\documentclass[11pt]{article}
\pdfoutput=1 
\usepackage{./aux/jheppub}

\usepackage[T1]{fontenc}

\usepackage{graphicx}
\usepackage{amsfonts,amsmath,amssymb}
\usepackage{tikz}
\usepackage{verbatim}
\usepackage{array,mathrsfs,yfonts,dsfont,bbm,colonequals}
\usepackage{amscd}
\usepackage{relsize}
\usepackage{suffix,mathtools,cancel}
\usepackage{bbm}
\usepackage{caption}

\usepackage{float}

\usepackage{textcomp}

\usepackage{stmaryrd}

\usepackage{colortbl}
\definecolor{mygray}{gray}{0.9}

\usepackage{multirow}

\usepackage{enumerate}

\usepackage{array}
\newcolumntype{P}[1]{>{\centering\arraybackslash}p{#1}}

\newcommand{\be}{\begin{eqnarray}}
\newcommand{\ee}{\end{eqnarray}}
\newcommand{\bea}{\begin{eqnarray}}
\newcommand{\eea}{\end{eqnarray}}

\newcommand{\bn}{\begin{enumerate}}
\newcommand{\en}{\end{enumerate}}

\def\CC{{\cal C}}

\def  \sl{\frak{sl}}  
  
\def\psl{\frak{psl}} 

\def\det{{\rm det}}

\def\dim{{\rm dim}}

\def \mf{\mathfrak}

\def \wt{\widetilde}

\def \zb{{\bar z}}

\newcommand{\beq}{\begin{equation}}
\newcommand{\eeq}{\end{equation}}

\newcommand\nn{\nonumber}

\newcommand{\cB}{\mathcal{B}}
\newcommand{\cC}{\mathcal{C}}
\newcommand{\cD}{\mathcal{D}}

\newcommand{\cG}{\mathcal{G}}
\newcommand{\cH}{\mathcal{H}}
\newcommand{\cI}{\mathcal{I}}
\newcommand{\cJ}{\mathcal{J}}
\newcommand{\cK}{\mathcal{K}}

\newcommand{\cM}{\mathcal{M}}
\newcommand{\cN}{\mathcal{N}}
\newcommand{\cO}{\mathcal{O}}
\newcommand{\cP}{\mathcal{P}}

\newcommand{\cR}{\mathcal{R}}

\newcommand{\cT}{\mathcal{T}}

\newcommand{\cV}{\mathcal{V}}
\newcommand{\cW}{\mathcal{W}}

\newcommand\qq{\mathbbmtt{Q}}

\numberwithin{equation}{section}

\def\NO#1#2{{\rm NO}(#1,#2)}

\def\bea{\begin{eqnarray}}
\def\eea{\end{eqnarray}}

\DeclarePairedDelimiterX\MeijerM[3]{\lparen}{\rparen}%
{\begin{smallmatrix}#1 \\ #2\end{smallmatrix}\delimsize\vert\,#3}

\newcommand\MeijerG[8][]{%
  G^{\,#2,#3}_{#4,#5}\MeijerM[#1]{#6}{#7}{#8}}

\WithSuffix\newcommand\MeijerG*[7]{%
  G^{\,#1,#2}_{#3,#4}\MeijerM*{#5}{#6}{#7}}

\def\cH{\mathcal{H}}

\def\cN{\mathcal{N}}
\def\cV{\mathbb{V}}

\def \beg#1{\begin{#1}} 
\def \be{\beg{equation}}
\def \bea{\beg{eqnarray}}
\def \eea{\end{eqnarray}}
\def \ee{\end{equation}}

\def \qq{\mathbbmtt{Q}\,}

\def \slf{\mf{sl}}

\def \restr#1#2{{\left.\kern-\nulldelimiterspace#1\vphantom{\big|}\right|_{#2}}}

\def \hf{\frac12}

\def \mf{\mathfrak}

\def \nn{\nonumber}

\def \wt{\widetilde}

\def \zb{{\bar z}}

\def \Cb{\mathbb{C}}

\def \Zb{\mathbb{Z}}

\def \BB{\mathcal{B}}
\def \CC{\mathcal{C}}
\def \DD{\mathcal{D}}

\def \FF{\mathcal{F}}
\def \GG{\mathcal{G}}

\def \OO{\mathcal{O}}

\def \QQ{\mathcal{Q}}

\def \VV{\mathcal{V}}

\title{
\Large VOAs labelled by complex reflection groups and $4d$ SCFTs
}

\preprint{YITP-SB-2018-25}

\author[a]{Federico Bonetti,\!}
\author[b]{Carlo Meneghelli,\!}
\author[c]{and Leonardo Rastelli}

\affiliation[a]{Department of Physics and Astronomy, Johns Hopkins University, 3400 North Charles Street, Baltimore, MD 21218, USA}
\affiliation[b]{Mathematical Institute, University of Oxford, Andrew Wiles Building, Radcliffe Observatory Quarter, Woodstock Road, Oxford, OX2 6GG, United Kingdom}
\affiliation[c]{C.~N.~Yang Institute for Theoretical Physics, Stony Brook University, Stony Brook, NY 11794, USA}

\abstract
{We define 
and study a class of ${\cal N}=2$ vertex  operator algebras $\mathcal{W}_{\mathcal{\mathsf{G}}}$  labelled by complex reflection groups. They are extensions of the ${\cal N}=2$ super Virasoro algebra
obtained by introducing additional generators, in correspondence with the invariants of the complex reflection group $\mathcal{\mathsf{G}}$.  If  $\mathcal{\mathsf{G}}$
 is a Coxeter group, the ${\cal N}=2$ super Virasoro algebra
enhances to the (small) ${\cal N}=4$ superconformal algebra.  With the exception of  $\mathcal{\mathsf{G}} = \mathbb{Z}_2$, which corresponds to just the  ${\cal N}=4$ algebra, these are non-deformable VOAs that exist only
for a specific negative value of the central charge. 
We describe a  free-field realization of  $\mathcal{W}_{\mathcal{\mathsf{G}}}$ in terms of rank$(\mathcal{\mathsf{G}})$  $\beta \gamma bc$ ghost systems,
generalizing a construction of Adamovic for the ${\cal N}=4$ algebra at $c = -9$. If $\mathcal{\mathsf{G}}$ is a Weyl group,  $\mathcal{W}_{\mathcal{\mathsf{G}}}$
is believed to coincide with the ${\cal N}=4$ VOA
that arises from the four-dimensional super Yang-Mills theory whose gauge  algebra has Weyl group $\mathcal{\mathsf{G}}$.
More generally, if $\mathcal{\mathsf{G}}$ is a crystallographic complex reflection group, $\mathcal{W}_{\mathcal{\mathsf{G}}}$ is conjecturally associated to an ${\cal N}=3$ $4d$ superconformal field theory.
The free-field realization allows to determine  the elusive ``$R$-filtration'' of  $\mathcal{W}_{\mathcal{\mathsf{G}}}$, and thus to recover the full Macdonald index of the parent $4d$ theory.}

\begin{document} 

\maketitle
\flushbottom



\section{Introduction and summary}

Any four-dimensional ${\cal N}=2$ superconformal field theory (SCFT) contains a subsector isomorphic to a vertex operator algebra (VOA) \cite{Beem:2013sza}.
This $4d/2d$ correspondence  (see \cite{Beem:2014rza, Lemos:2014lua, Lemos:2015orc, Cecotti:2015lab, 
Arakawa:2016hkg, 
Bonetti:2016nma, Song:2016yfd, Fredrickson:2017yka, Cordova:2017mhb, Song:2017oew, Buican:2017fiq, Beem:2017ooy, Pan:2017zie, Fluder:2017oxm, Choi:2017nur, Arakawa:2017fdq, Niarchos:2018mvl, Feigin:2018bkf, Creutzig:2018lbc} for some further developments)  promises  to become
an organizing principle for the whole landscape of ${\cal N}=2$ SCFTs. The aspiration is to combine the rigid associativity constraints of  VOAs with additional physical requirements, such as unitary of the $4d$ theory,
in order to constrain and ideally classify the set of ${\cal N}=2$ SCFTs.\footnote{A complementary approach to the classification problem of ${\cal N}=2$ SCFTs is by 
studying their {\it Coulomb branch} geometries. See 
\cite{Argyres:2018zay, Argyres:2018urp, Caorsi:2018zsq, Caorsi:2018ahl} for the state of the art of this  approach. 
}
To make inroads into this   program,
it  may be fruitful to start with theories that enjoy enhanced supersymmetry. The map of  \cite{Beem:2013sza} associates to
a generic ${\cal N}=2$ SCFTs   a conformal VOA whose Virasoro subalgebra has central charge $c = - 12 c_{4d}$, where $c_{4d}$ is the  (Weyl)$^2$ conformal anomaly coefficient.
If the $4d$ SCFT has ${\cal N}=3$ or ${\cal N}=4$ supersymmetry, the associated VOA is supersymmetric, containing an ${\cal N}=2$ or (small) ${\cal N}=4$ super Virasoro subalgebra, respectively.

With this broad motivation in mind, we have undertaken a systematic study of  the classes of ${\cal N}=2$ and ${\cal N}=4$ VOAs that arise from $4d$ SCFTs. 
In this paper we  describe a uniform construction of all the {\it known} examples (and, as we shall see, of additional examples, some of which are unlikely to have a four-dimensional interpretation).  
Our aim here is descriptive rather than taxonomic. The general classification program outlined above will require different methods, and is left for future work.

A central class of examples are  the VOAs  
associated to the ${\cal N}=4$ super Yang-Mills (SYM) theories.  The ${\cal N}=4$ $4d$  theory SYM$_{\frak{g}}$
 with gauge algebra $\frak{g}$ descends to
an ${\cal N}=4$ VOA $\chi[ {\rm SYM}_{\frak{g}}]$ 
 with central charge $c = - 3 \, {\rm dim}(\frak{g})$.
As the $4d$ theory has a Lagrangian description, 
one can apply the methods of \cite{Beem:2013sza}
to give a description of the associated VOA, 
as a subalgebra of  ${\rm dim}(\frak{g})$ copies of the
 $ \beta \gamma b c$ ghost system. The subalgebra is defined by passing to the cohomology of a certain nilpotent  operator $Q_{\rm BRST}$ built in terms of the 
 $\beta \gamma b c$ ghosts.
While giving in principle a complete definition of the VOA, this
description   is  cumbersome and redundant. The calculation of the requisite BRST cohomology is a difficult problem that so far has only be solved in examples, by brute force level-by-level calculation 
up to some maximum conformal weight. The simplest example is the VOA associated to ${\cal N}=4$ SYM theory with gauge algebra $\sl(2)$. There is strong evidence \cite{Beem:2013sza} that $\chi[ {\rm SYM}_{\sl{(2)}}]$
coincides with the small ${\cal N}=4$ superVirasoro algebra with central charge $c = -9$.  More generally, for a simple Lie algebra $\frak{g}\neq \sl(2)$,  $\chi[ {\rm SYM}_{\frak{g}}]$  is an extension  of Vir$_{{\cal N}=4}$ 
obtained by introducing additional  strong generators. The list of generators includes one {\it short} 
 superprimary of  Vir$_{{\cal N}=4}$   for each  Casimir invariant of $\frak{g}$.\footnote{Additional generators which are {\it long} superprimaries of Vir$_{{\cal N}=4}$ are needed for some choices of $\frak{g}$, as we shall discuss in detail below. (By a short/long superprimary we mean the superprimary of a short/long superconformal multiplet). }

A concrete description $\chi[ {\rm SYM}_{\frak{g}}]$  as a W-algebra ({\it i.e.},
in terms  of the singular OPE of its strong generators),  becomes more and more involved as the rank of $\frak{g}$ increases.  
We have found such a W-algebra description for a few low rank cases, see section \ref{sec:N4Examples} and 
\ref{sec:N=2Examples}, 
 but 
it seems very difficult to find such an explicit presentation in the general case.  Instead, the main subject of this paper 
is a proposal for a novel {\it free-field realization} of these VOAs,  much simpler and more explicit than the cohomological description of \cite{Beem:2013sza}. Our new 
free-field realization is in terms
 ${\rm rank}(\frak{g})$ copies of the $\beta \gamma b c$ ghost system.  There is a heuristic understanding of these free fields as corresponding to the low-energy degrees of freedom 
of  the $4d$ theory at a generic point on its Higgs branch of vacua. This physical picture will be discussed elsewhere \cite{Beem:2019tfp}:
it appears to be much more general, possibly valid for  all VOAs that arise from ${\cal N}=2$ SCFTs.

Our proposal generalizes to all $\frak{g}$ a  construction of Adamovic \cite{Adamovic:2014lra}, who exhibited a free-field realization of  Vir$_{{\cal N}=4}$  with $c = -9$ (in our framework, the  $\frak{g} = \sl(2)$ case)
 in terms of a single  $\beta \gamma b c$ system. The   Vir$_{{\cal N}=4}$  VOA with $c=-9$ contains a large set of null vectors, and a remarkable feature of Adamovic's construction is that they  identically vanish
 when expressed in terms of the free fields. In other terms, this is a construction of the {\it simple quotient} of the VOA. We find strong evidence that our generalization to  $\chi[ {\rm SYM}_{\frak{g}}]$
 shares the same property.

 A notable corollary of  our proposal is a compelling conjecture for the ``$R$-filtration'' of this class of VOAs.  As we review in detail below, any VOA that descends from a $4d$ ${\cal N}=2$ SCFT
 inherits a filtration associated to the $4d$  $R$-symmetry quantum number -- the details of the cohomological construction of  \cite{Beem:2013sza} imply that while $R$ is in general not preserved by the OPE, 
 it can at most {\it decrease}.  The $R$-filtration is of paramount importance in extracting four-dimensional physical information from the VOA, but it is completely hidden in an abstract W-algebra presentation. By contrast,
  our free-field realizations  come equipped with a natural filtration, which coincides with the $R$-filtration in all examples that we have been able to check.

 Curiously, we overshot our initial target,  finding a {larger} set of ${\cal N}=4$ VOAs than originally expected. We  found  free-field constructions for ${\cal N}=4$ VOAs
  ${\cal W}_\Gamma$ labelled by a {\it general}  Coxeter group $\Gamma$. So far, we have  described the situation
 when  $\Gamma$ is the Weyl group ${\rm Weyl}(\frak{g})$
 of a simple Lie algebra $\mathfrak g$, hence
 both Coxeter and \emph{crystallographic}.\footnote{ A group
is called crystallographic if it preserves a lattice \cite{2003math11012G,MR2358376}.}
Our basic contention is   that ${\cal W}_{{\rm Weyl}(\frak{g})} =\chi[ {\rm SYM}_{\frak{g}}]$. However, our construction
 goes through even if $\Gamma$ is not crystallographic,  in which case ${\cal W}_\Gamma$ does not have any obvious four-dimensional interpretation. Clearly, it cannot descend from a $4d$ SYM theory.
Even  if one is willing to entertain the possibility that the SYM theories do not exhaust the set of $4d$ ${\cal N}=4$ SCFTs,  the conventional wisdom is that none of them can give rise
to ${\cal W}_\Gamma$  if $\Gamma$ is a non-crystallographic Coxeter group.
 Indeed, as we explain below,
    the moduli space of vacua
 of  the putative parent $4d$ theory  would be  the orbifold $\mathbb{R}^{6n} /\Gamma$, where $n$ is the rank of $\Gamma$,
 but general consistency conditions on the low-energy effective  theory  restrict $\Gamma$ to be crystallographic \cite{Caorsi:2018zsq,Argyres:2018wxu}.\footnote{
 We are grateful to Philip Argyres  and Mario Martone for very insightful correspondence
 about the moduli space geometry of ${\cal N}=4$ and ${\cal N}=3$ SCFTs.}

\begin{figure}
  \centering
    \includegraphics[width=0.65\textwidth]{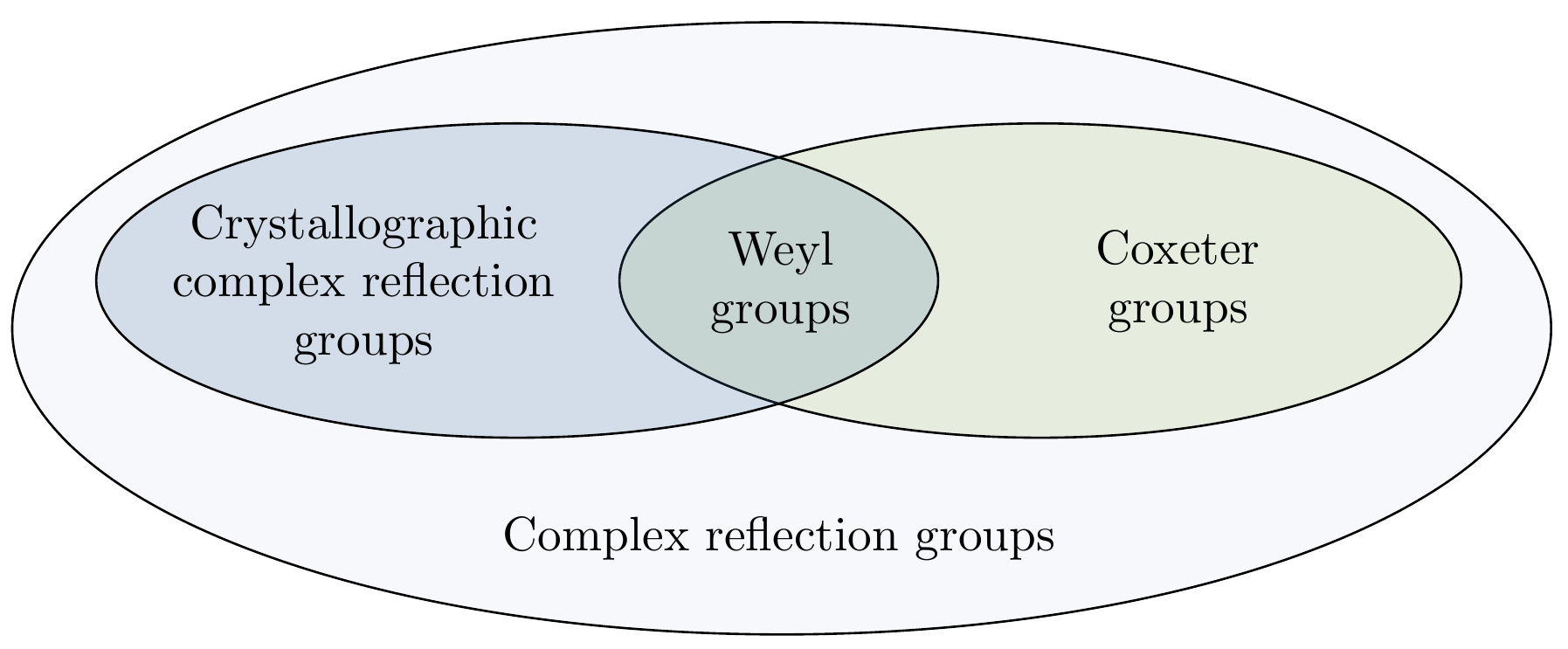}
      \caption{Euler-Venn diagram depicting the relations among
      complex reflection groups, crystallographic complex reflection groups,
      Coxeter groups, and Weyl groups. } \label{EulerVenn}
\end{figure}

This whole circle of ideas admits a  natural extension to a class  of VOAs
  $\mathcal{W}_\mathcal{\mathsf{G}}  \supset {\cal W}_\Gamma$, labelled by a general complex reflection group $\mathsf{G}$. These vertex algebras are extensions of the ${\cal N}=2$ superVirasoro algebra by additional generators, including
  one short superprimary for each of the fundamental invariants of $\mathsf{G}$. Their central charge is fixed in terms of the degrees of the primitive invariants of $\mathsf{G}$, see (\ref{centralcharge_intro}).
  We propose a free-field construction of the simple quotient of these algebras in terms of rank($\mathsf{G}$) $\beta \gamma bc$ ghost systems.
    If (and only if) $\mathsf{G} = \Gamma$ is a Coxeter group, its lowest fundamental invariant has degree two, and the corresponding short generator of the VOA
    induces an enhancement of the superconformal algebra from ${\cal N}=2$ to ${\cal N}=4$, so that we recover the  construction of  ${\cal W}_\Gamma$
     discussed above. If $\mathsf{G}$ is a crystallographic complex reflection group (which is not a Coxeter group),  $\mathcal{W}_\mathcal{\mathsf{G}}$ may descend via the map of  \cite{Beem:2013sza} from an ${\cal N}=3$ $4d$ SCFT (which is not  an ${\cal N}=4$ SCFT). 
     Many
      examples of ${\cal N}=2$ VOAs 
    associated to ${\cal N}=3$ SCFTs 
    have been described in the literature \cite{Nishinaka:2016hbw,Lemos:2016xke},      and we are able to identify each of them with $\mathcal{W}_\mathcal{\mathsf{G}}$ for some choice of $\mathsf{G}$.
An Euler-Venn diagram of complex reflection groups
is presented in figure \ref{EulerVenn}.      

    In summary, we have found a uniform construction for:
    \vspace{-0.2cm}
    \begin{enumerate}
     \item[(i)] All the VOAs that descend from currently  known ${\cal N}=3$ and ${\cal N}=4$ $4d$ SCFTs. They are labelled by Weyl groups in the
    ${\cal N}=4$ case and by  a subset of the crystallographic complex  reflection groups in the ${\cal N}=3$ case.
    \item[(ii)] Additional VOAs, labelled by the remaining crystallographic complex reflection groups, which are candidates for  $4d$ uplifts to ${\cal N}=3$ SCFTs, but whose $4d$ counterparts are currently unknown.
    \item[(iii)] VOAs 
    labelled by non-crystallographic Coxeter and complex reflection
    groups, which presumably do  not correspond to standard $4d$ theories.
    \end{enumerate}
    
     Perhaps the most interesting open question is whether abstract bootstrap methods can lead to a rigorous classification of all 
   ${\cal N} \geq 2$ VOAs that descend from ${\cal N} \geq 3$ $4d$ SCFTs. The bootstrap constraints would have to incorporate intrinsic $4d$ conditions, such as $4d$ unitarity. The simplest conjecture  generalizing
  all available data is that the complete list of such VOAs coincides with  $\mathcal{W}_\mathcal{\mathsf{G}}$ with $\mathcal{\mathsf{G}}$  crystallographic, as well as of the additional VOAs obtained from the above set by performing discrete quotients (see, {\it e.g.}, \cite{Argyres:2018wxu,Bourton:2018jwb}).
     We have made some preliminary progress in this direction, which we will report in an upcoming publication \cite{bootstrappaper}. We preview some our findings in the outlook subsection of this introduction.



 \subsection{Main results}
\label{mainresultsINTRO}

In this work we  study
a class of   
$\cN = 2$ VOAs labelled by
complex reflection groups. 
By an  $\cN = 2$
 VOA     we mean
 a VOA that contains the   $\cN = 2$ superconformal
algebra (SCA) as a subalgebra.
The   $\cN = 2$ SCA   is generated by the stress tensor,
together with an affine $\mathfrak{gl}(1)$ current
and two   supercurrents.
The global part of the   $\cN = 2$ SCA is 
$\mathfrak{osp}(2|2)$, and it is natural to organize the operator content
of our VOAs 
into representations of $\mathfrak{osp}(2|2)$.
We will discuss $\mathfrak{osp}(2|2)$ representations 
 in more detail in section \ref{smallN4recap}
and in appendix \ref{OPEandsl2cov}, but for the purposes of the present introduction,
it suffices to recall they
are labelled by two quantum numbers $(h,m)$,
which are the conformal dimension and the $\mathfrak{gl}(1)$ charge
of the highest weight state in the multiplet.
Representations with $h=\pm m$ obey a shortening condition,
and are referred to as  chiral, anti-chiral, respectively.
A representation with $h \neq |m|$ will be referred to as non-chiral.

The class of VOAs that we analyze in this work are
 extensions of the   $\cN = 2$ SCA obtained  by 
introducing additional strong generators.  The   additional
strong generators,
as well as several other interesting properties
of the VOAs under examination,
are intimately related to the theory of invariants of the
complex reflection group~$\mathsf G$.
Therefore, before we proceed, we need to
introduce some algebraic structures 
related to complex reflection groups.

Let $\mathsf G$ be a complex reflection group, regarded as a subgroup of 
$GL(V_{\mathsf G})$ with $V_{\mathsf G} \cong \mathbb C^{\text{rank}(\mathsf G)}$.
According to the Chevalley--Shephard--Todd theorem,
the ring of invariants $\mathbb C[V_\mathsf G]^\mathsf G$
is a freely generated polynomial ring.
The generators of the ring $\mathbb C[V_\mathsf G]^\mathsf G$
are usually referred to as the fundamental
invariants of $\mathsf G$.
Their number   equals the rank of $\mathsf G$,
and their degrees are denoted $p_\ell$,
with $\ell = 1, \dots, \text{rank}(\mathsf G)$.
For example, the rank-one complex reflection group 
$\mathsf G = \mathbb Z_p$ has a unique fundamental
invariant of degree $p_1 = p$. If we introduce the coordinate
$z$ on $V_{\mathbb Z_p} \cong \mathbb C$,
the action of $ \mathbb Z_p$ on $V_{\mathbb Z_p}$ is simply
$z \mapsto e^{2  \pi i/p} \, z$,
and the ring of invariants $\mathbb C[V_{\mathbb Z_p}]^{\mathbb Z_p}$
is freely generated by $z^p$.

For our purposes, we need to consider the canonical
symplectic variety 
associated to the action of $\mathsf G$ on $V_\mathsf G$.
More precisely, let us define the variety
 \begin{equation}\label{associatedvarietyDEF2}
 \mathscr{M}_{\mathsf{G}}\,=\,
 \frac{V^{}_{\mathsf{G}}\oplus
 V^*_{\mathsf{G}}}{ 
 \mathsf{G}}  \ ,
 \end{equation}
where $V^*_{\mathsf G}$ denotes the dual of the vector space 
$V_{\mathsf G}$. The space $V^{}_{\mathsf{G}}\oplus
 V^*_{\mathsf{G}}$ admits a canonical symplectic structure,\footnote{The
 symplectic structure $\omega$ is defined by  
 \beq
\omega(x_1 \oplus \xi_1 , x_2 \oplus \xi_2  ) = \xi_2(x_1) - \xi_1(x_2)  \ ,
\eeq 
where $x_{1,2} \in V_{\mathsf G}$, and $\xi_{1,2} \in V^*_{\mathsf G}$.
}
which is preserved by the action of $\mathsf G$.
We also define the ring associated to $ \mathscr{M}_{\mathsf{G}}$,
\beq
\mathscr R_{\mathsf G} = \mathbb C[ \mathscr{M}_{\mathsf{G}}]
= \mathbb C [ 
V^{}_{\mathsf{G}}\oplus
 V^*_{\mathsf{G}}
 ]^{\mathsf G}   \ .
\eeq
As opposed to the ring $\mathbb C[V_\mathsf G]^\mathsf G$,
the ring $\mathscr R_{\mathsf G}$ 
is not freely generated. For instance, in the case $\mathsf G = \mathbb Z_p$,
we may parametrize the $\mathsf G$ action on $V^{}_\mathsf G \oplus V^*_\mathsf G$
as $(z, \bar z) \mapsto (e^{2\pi i/p} \, z ,e^{-2\pi i/p} \, \bar z )$.
The ring $\mathscr R_{\mathsf G}$  is then generated by
the monomials
$\mathsf j = z \, \bar z$, $\mathsf w = z^p$, $\bar {\mathsf w} = \bar z^p$,
subject to the relation $\mathsf w \, \bar {\mathsf  w} = \mathsf j^p$.

As $\mathscr M_{\mathsf G}$ admits a natural
$GL(1) \times GL(1)$ action, 
 the ring $\mathscr R_\mathsf G$
can always be given a presentation in terms of generators
and relations, each possessing definite quantum
numbers $(h,m)$ under the $GL(1) \times GL(1)$
action. 
The generators with $h=m$ are in 1-to-1 correspondence
with the  fundamental invariants of 
$\mathsf G$. Our normalization of the quantum
numbers $h$, $m$ is such that
 $h=m=p_\ell/2$ for the $\ell$-th
fundamental invariant of $\mathsf G$.
For any complex reflection group $\mathsf G$,
there is a unique generator with $(h,m) = (1,0)$.\footnote{If
$e_\ell$, $\ell = 1, \dots, \text{rank}(\mathsf G)$
is a basis of $V_\mathsf G$, and $e^\ell$
denotes the dual basis of $V_\mathsf G^*$,
this generator has the form $\sum_{\ell=1}^{\rm{rank}(\mathsf G)} e^\ell \, e_\ell$.
Moreover, this generator is the moment map for the $GL(1)$ action
with quantum number $m$.
}
In the example of $\mathsf G =\mathbb Z_p$,
the quantum numbers of the generators $\mathsf j$, $\mathsf w$, $\bar {\mathsf w}$
are 
$(1,0)$, $(\frac p2,\frac p2)$, $(\frac p2, - \frac p2)$,
respectively.
We refer the reader to section \ref{subsec:Coxeter} for more
information on the general case.
 
We are now in a position to articulate our main proposal.
Given a complex reflection group $\mathsf G$, we claim 
that there exists an   $\cN = 2$ VOA, denoted $\cW_\mathsf G$,
satisfying     the   properties (i)-(vii) listed below.
Before giving the list of properties,
we would like to caution the reader that
we do not have a general existence 
proof for $\cW_\mathsf G$,
but we have nonetheless 
gathered a considerable amount of evidence 
in favor of our proposal.
In particular, in the case in which the complex
reflection group is the Weyl group
of a semisimple Lie algebra,
one can resort to the BRST construction of \cite{Beem:2013sza}
to demonstrate the existence of $\cW_\mathsf G$.
Beyond Weyl groups, we have a fully explicit
construction of $\cW_\mathsf G$
for the infinite series $\mathsf G = I_2(p)$
and $\mathsf G =\mathbb Z_p$,
as well as for all rank-three Coxeter groups,
and the rank-two complex reflection group $G(3,1,2)$
(whose definition is recalled  in section \ref{N2rank2Example}).
We now give the list of properties enjoyed by 
$\cW_\mathsf G$:
\begin{enumerate}[(i)]
\item $\cW_\mathsf G$ 
is a simple $\cN =2$ VOA  that is strongly generated
by a finite number of operators.\footnote{This
means that any element of $\cW_\mathsf G$
can be obtained by taking linear combinations of
normal ordered products of derivatives of the strong
generators. Furthermore, all null composites 
constructed in such a way from the strong generators
are modded out  in $\cW_\mathsf G$.
All strong generators are non-null operators.
}
All strong generators of $\cW_\mathsf G$
are organized in  
multiplets of $\mathfrak{osp}(2|2)$.

\item Chiral and anti-chiral multiplets
of strong generators of $\cW_\mathsf G$
come in pairs, which are in 1-to-1
correspondence with the fundamental invariants of $\mathsf G$.
The quantum numbers of the chiral/anti-chiral pairs
are $(h,m)= (p_\ell/2, \pm p_\ell/2)$, with $\ell =1, \dots,\mathrm{rank}(\mathsf G)$,
where $p_\ell$ are the degrees of the fundamental invariants of
$\mathsf G$.

\item For each generator of the ring $\mathscr R_\mathsf G$
with $GL(1) \times GL(1)$ quantum numbers $(h,m)$,
there is an $\mathfrak{osp}(2|2)$ multiplet
of strong generators of $\cW_\mathsf G$,
labelled by the same quantum numbers.

Because of point (ii), this observation
is trivial if $h = \pm m$, 
but it is non-trivial for generators of 
$\mathscr R_\mathsf G$ with $h \neq |m|$,
which  are mapped to
non-chiral multiplets of strong generators
 of $\cW_\mathsf G$.\footnote{In
all examples we analyzed, 
the entire   set
of non-chiral strong generators of $\cW_\mathsf G$
is obtained in this way from the ring $\mathscr R_\mathsf G$.
We do not have conclusive evidence, however, that
this is always the case for all choices of~$\mathsf G$.
  }
The unique
generator of $\mathscr R_\mathsf G$
with $(h,m) = (1,0)$ is mapped to the
set of generators of the $\cN =2$ SCA.  
\item The central charge of $\cW_\mathsf G$ is
given by
\beq\label{centralcharge_intro}
c = -3 \sum_{\ell=1}^{\text{rank}(\mathsf G)} (2 \, p_\ell -1) \ ,
\eeq
where $p_\ell$ are the degrees of the fundamental invariants of
$\mathsf G$.

\item  $\mathcal{W}_{\mathsf G}$ can be realized as a subalgebra of
 $\text{rank}(\mathsf G)$ copies of
a standard $\beta\gamma b c$ system.
We may write 
\begin{equation}\label{WGromfreefields}
\mathcal{W}_{\mathsf G}\,\subset\,
\mathbb{M}_{\beta\gamma b c}^{(\mathsf G)}:=\,
 \bigotimes_{\ell=1}^{\text{rank}(\mathsf G)}\, M_{\beta\gamma b c}^{(p_{\ell})}\,.
\end{equation}
The symbol $M_{\beta\gamma b c}^{(p_{\ell})}$
denotes a standard $\beta\gamma b c$ system,
endowed with its canonical $\cN = 2$ superconformal
structure labelled by the quantum number $p_\ell$.
More explicitly, each tensor factor 
 $M_{\beta\gamma b c}^{(p_{\ell})}$
comes with a free-field realization of the $\cN = 2$ SCA,
in which the quantum numbers of the free fields in  
 $M_{\beta\gamma b c}^{(p_{\ell})}$
are determined by $p_\ell$, see table \eqref{betagammabc_hmr}. 
For $\mathsf G=\mathbb{Z}_2$,  the free-field realization 
\eqref{WGromfreefields} of $\cW_\mathsf G$ 
 coincides with the one given in 
 \cite{Adamovic:2014lra}.
 
Crucially, we claim that the above free-field realization
is a realization of $\cW_\mathsf G$ regarded 
as simple quotient of the span of the strong
generators. More explicitly, we propose that 
all null states built from the strong generators are automatically identically zero 
in the free-field realization.
This has been proven for $\mathsf G = \mathbb Z_2$ in~\cite{Adamovic:2014lra}.

\item  
Thanks to the free-field realization of point (v) above,
we have a natural way to associate a Poisson algebra to $\cW_\mathsf G$.
This is achieved by means of the map
\beq
\mathcal P \;\; : \;\; \cW_\mathsf G \;\; \rightarrow \;\; \mathbb M^{(\mathsf G) {\rm cl}}_{\beta
\gamma} \ ,
\eeq
where $\mathbb M^{(\mathsf G) {\rm cl}}_{\beta
\gamma}$ denotes the Poisson algebra of polynomials
in the indeterminates $\beta_\ell$, $ \gamma_\ell$,
with $\ell =1, \dots, \text{rank}(\mathsf G)$,
with each pair $(\beta_\ell, \gamma_\ell)$
regarded as a pair of canonically conjugate variables.
The image of an operator $\cO$ in $\cW_\mathsf G$
under the map $\mathcal P$ 
is obtained starting from the free-field realization of $\cO$
and setting to zero all derivatives and all factors of $b_\ell$, $c_\ell$,
$\ell =1, \dots, \text{rank}(\mathsf G)$. More details on the map 
$\mathcal P$ are found in section \ref{sec:freefieldsclassPOISSON}.
The image of $\cW_\mathsf G$ under the map $\mathcal P$ is
a Poisson subalgebra  $\cP(\cW_\mathsf G)$ of
$\mathbb M^{(\mathsf G) {\rm cl}}_{\beta
\gamma}$. We propose that   $\cP(\cW_\mathsf G)$ 
is isomorphic to the ring of invariants $\mathscr R_\mathsf G$,
regarded as a Poisson algebra,
\beq
\cP(\cW_\mathsf G) \cong \mathscr R_\mathsf G \ .
\eeq

\item Following the general
construction of \cite{Arakawa2012}, 
one can define the vector subspace $C_2(\cW_\mathsf G) \subset \cW_\mathsf G$
and the commutative 
Zhu algebra $\cW_\mathsf G/ C_2(\cW_\mathsf G)$.
The latter is naturally endowed with the structure of a Poisson algebra.
Crucially,   $\cW_\mathsf G/ C_2(\cW_\mathsf G)$ needs not
be a reduced ring, i.e.~$\cW_\mathsf G/ C_2(\cW_\mathsf G)$
may contain nilpotent elements.
One can mod out nilpotent elements of $\cW_\mathsf G/ C_2(\cW_\mathsf G)$
and obtain a reduced ring, denoted $(\cW_\mathsf G/ C_2(\cW_\mathsf G) )_{\rm red}$.
The latter is still a Poisson algebra.
We propose the following Poisson algebra isomorphism,
\beq
(\cW_\mathsf G/ C_2(\cW_\mathsf G) )_{\rm red} \cong \mathscr R_\mathsf G \ .
\eeq 
In particular, this means that the associated variety
to $\cW_\mathsf G$, in the sense of  \cite{Arakawa2012},
is precisely the variety $\cM_\mathsf G$ defined in \eqref{associatedvarietyDEF2}.

\end{enumerate}

In the case of a complex reflection group that is also a Coxeter group,
the structure of the corresponding VOA is richer.
In what follows, we reserve the symbol $\Gamma$ for a Coxeter group.

First of all, supersymmetry is enhanced from $\cN =2$
to small $\cN = 4$. The small $\cN = 4$ SCA is generated by the stress
tensor, an $\mathfrak{sl}(2)$ triplet of  affine currents,
and two $\mathfrak{sl}(2)$ doublets of supercurrents.
The explicit embedding of the $\cN = 2$ SCA into the small 
$\cN = 4$ SCA is given in \eqref{N2inN4preliminaries}.
The global part of the small $\cN = 4$ SCA is $\mathfrak{psl}(2|2)$.
Multiplets of $\mathfrak{psl}(2|2)$ are labelled by a pair
$(h,j)$, where $h$ is the conformal dimension
and $j$ is the $\mathfrak{sl}(2)$ spin of the highest
weight state. Multiplets with $h = j$ obey a shortening condition,
and are referred to as short in what follows.
Generic multiplets with $h >j$ are referred to as long multiplets.
We refer the reader to section \ref{smallN4recap} and appendix \ref{OPEandsl2cov}
for more information.

Recall that 
Coxeter groups are characterized within
complex reflection groups by the property of possessing
exactly one fundamental invariant of degree two.
By virtue of point (ii) above,
it follows that the VOA $\cW_\Gamma$ associated to a 
Coxeter group $\Gamma$
has exactly one chiral/anti-chiral pair of strong generators
with $h = \pm m = 1$.
This chiral/anti-chiral pair
combines with the set of generators of the   
$\cN = 2$ SCA to give a short multiplet of $\mathfrak{psl}(2|2)$
that encompasses all generators of the small
$\cN = 4$ SCA.

This phenomenon extends to all chiral/anti-chiral
pair of multiplets of strong generators of $\cW_\Gamma$.
More precisely, each chiral/anti-chiral pair in $\cN = 2$
language is paired with  non-chiral $\cN =2$ multiplets
to give a short multiplet of $\mathfrak{psl}(2|2)$. 
Crucially, not all $\cN = 2$ non-chiral
multiplets are necessarily paired with 
chiral/anti-chiral pairs in the enhancement to small
$\cN =4$ supersymmetry.
As a result, the VOA $\cW_\Gamma$ generically
admits long multiplets of strong generators.

The supersymmetry enhancement at the level of the VOA
is mirrored by an enhancement of the isometry group
of the variety $\mathscr M_\mathsf G$.
Indeed,  if $\mathsf G$ is a Coxeter group $\Gamma$,
the variety \eqref{associatedvarietyDEF2} admits an alternative presentation,
\beq
\label{MGammavarietyintro}
\mathscr{M}_{\Gamma}\,=\,
 \frac{\mathbb{C}^2 \otimes V^{\mathbb R}_{\Gamma}}{ 
 \Gamma} \ ,
\eeq
where $\Gamma$ acts trivially on the $\mathbb C^2$ factor.
In the above expression, 
we have implicitly identified the Coxeter group
$\Gamma$ with a subgroup of $O(V^{\mathbb R}_{\Gamma})$,
where
$V^{\mathbb R}_{\Gamma} \cong \mathbb R^{\mathrm{rank}(\Gamma)}$ is a \emph{real}
vector space.\footnote{If $\mathsf G$ is a complex reflection group,
and $g\in \mathsf G$,
the action of $g$ on $V_\mathsf G \oplus V^*_{\mathsf G}$ is
of the form
\beq
(z, \bar z) \mapsto (M \, z , M^{T,-1} \, \bar z) \ ,
\eeq
where $z \in V_\mathsf G$, $\bar z \in V_\mathsf G^*$, and $M \in GL(V_\mathsf G) \cong GL({\rm rank}(\mathsf G), \mathbb C)$.
If $\mathsf G$ is a Coxeter group $\Gamma$,
for any $g \in \Gamma$ the matrix $M$ can be taken to be real and orthogonal, $M^{T,-1} = M$,
making it manifest that $M$ acts on the $GL(2)$ doublet $(z,\bar z)$.
}
The presentation \eqref{MGammavarietyintro}
makes it manifest that the $GL(1) \times GL(1)$ action
is enhanced to a $GL(2) \cong GL(1) \times SL(2)$ action.
It follows that the ring
$\mathscr R_\Gamma$
associated to $\mathscr{M}_{\Gamma}$
admits a presentation in terms of generators
and relations with a definite $GL(1)$ weight $h$
and $SL(2)$ spin $j$.
The simplest example of Coxeter group is $\Gamma = \mathbb Z_2$.
In this case, the generators $\mathsf j$, $\mathsf w$, $\bar {\mathsf w}$
(introduced above for $\mathbb Z_p$)
all have the same weight and
form a triplet of $SL(2)$.
The relation $\mathsf w \, \bar {\mathsf w} = \mathsf j^2$
is a singlet of $SL(2)$.

Let us summarize how points (i)-(iii) 
above can be rephrased in the   Coxeter case.

\begin{enumerate}[(i)${}^\prime$]
\item $\cW_\Gamma$ 
is a simple VOA with small $\cN =4$
supersymmetry  that is strongly generated
by a finite number of operators.
All strong generators of $\cW_\Gamma$
are organized in  
multiplets of~$\mathfrak{psl}(2|2)$.

\item Short multiplets
of strong generators of $\cW_\Gamma$
are in 1-to-1
correspondence with the fundamental invariants of $\Gamma$.
The quantum numbers of the short multiplets
of strong generators are
$h = j = p_\ell/2$, with $\ell =1, \dots,\mathrm{rank}(\mathsf G)$,
where $p_\ell$ are the degrees of the fundamental invariants of
$\Gamma$. The short multiplet containing the generators
of the small $\cN =4$ SCA is in correspondence
with the fundamental invariant of $\Gamma$ of degree two.

\item 
Upon expressing generators and relations of the ring
$\mathscr R_\Gamma$ in a $GL(2)$ covariant way,
for each $GL(2)$ multiplet of generators of   $\mathscr R_\Gamma$
with   quantum numbers $(h,j)$,
there is a $\mathfrak{psl}(2|2)$ multiplet
of strong generators of $\cW_\Gamma$,
labelled by the same quantum numbers.

Because of point $\rm{(ii)}'$ above, this observation
is trivial if $h = j$, 
but it is non-trivial for generators of 
$\mathscr R_\Gamma$ with $h \neq j$,
which  are mapped to
long multiplets of strong generators
 of $\cW_\Gamma$.

 \end{enumerate}

The points (iv) to (vii) in the list of properties
in the complex reflection case are specialized to the
Coxeter case with obvious modifications.
In particular, we still have a free-field realization of $\cW_\Gamma$,
in which all null states are automatically zero.

\subsection{Connection with $4d$ physics}

The class of VOAs labelled by 
complex reflection groups that
we have described in the previous section
constitutes a rich and novel 
set of VOAs, worth studying in its own right.
We now come back to our original
motivation for the study of these $2d$ algebraic
structures, which is rooted in the analysis of $4d$ SCFTs.
We propose that the class of VOAs $\cW_\mathsf G$
associated to
complex reflection groups $\mathsf G$ provides a unified
framework to describe VOAs originating from $4d$ SCFT with $\cN \ge 3$
via the map of \cite{Beem:2013sza}.

In a generic $4d$ $\cN = 2$ SCFT,
there is no obvious relation between the geometries
of the Coulomb and Higgs branches.
If the theory has $\cN \ge 3$ supersymmetry, however,
the geometry of the Higgs branch is modeled
after the geometry of the Coulomb branch.
This is a consequence of the fact that both branches
are subspaces of the full moduli space of the theory,
which is highly constrained by $\cN \ge 3$ supersymmetry.

Let us explain this point in more detail.
The moduli space of a $4d$ $\cN = 3$ theory
is parametrized by the expectation value of scalars
in $\cN= 3$ vector multiplets, which are identical to $\cN = 4$
vector multiplets. 
Therefore, the  moduli space of a $4d$ $\cN \ge 3$
has real dimension $6\mathsf r$,   
where $\mathsf r$ is the rank of the SCFT,
and is locally flat.
Upon selecting an $\cN = 2$ subalgebra of the $\cN \ge 3$
superconformal symmetry,
the moduli space geometry is reformulated in terms of a Coulomb
branch and a Higgs branch.
Let us assume that the Coulomb branch 
can be written globally as a quotient of $\mathbb C^{\mathsf r}$ by a discrete
group, $\mathbb C^{\mathsf r} / \mathsf{G}$. 
The analysis of
\cite{Caorsi:2018zsq,Argyres:2018wxu}
 then reveals that $\mathsf G$
must be a \emph{crystallographic} complex reflection group,
whose rank equals the rank $\mathsf r$ of the SCFT.
Moreover, if the Coulomb branch is $\mathbb C^\mathsf r/\mathsf G$,
$\cN \ge 3$ supersymmetry implies that 
the Higgs branch must coincide
with the variety \eqref{associatedvarietyDEF2} (recall $V_{\mathsf G} \cong \mathbb 
C^{\mathsf r}$).

%


Let us now consider the VOA associated to the
SCFT. According to the conjecture of \cite{Beem:2017ooy},
the associated variety to the VOA must coincide with the Higgs branch of the
$4d$ theory,
which is the variety \eqref{associatedvarietyDEF2}.
Moreover, we know from \cite{Beem:2013sza} that the $2d$ central
charge $c_{ 2d}$ must be given by $c_{ 2d} = -12 \, c_{ 4d}$,
where $c_{\rm 4d}$ is the $\text{(Weyl)}^2$ trace anomaly
coefficient. On the other hand, in any $4d$ $\cN \ge 3$ SCFT,
the two trace anomaly coefficients are equal,
$a_{4d} = c_{4d}$ \cite{Aharony:2015oyb}. Furthermore, we can
use the Shapere-Tachikawa formula \cite{Argyres:2007tq,Shapere:2008zf}
to relate them to the dimensions of the Coulomb
branch generators $D_\ell$, $\ell = 1, \dots, \mathsf r$,
\beq
c_{\rm 4d} = \frac 14 \sum_{\ell = 1}^{\mathsf r} (2 \, D_\ell - 1) \ .
\eeq
By assumption, the Coulomb branch is
$\mathbb C^{\mathsf r}/\mathsf G$,
which implies that the dimensions of the Coulomb generators
coincide with the degrees of the fundamental invariants
of $\mathsf G$, $D_\ell = p_\ell$, $\ell =1, \dots, \mathsf r$.
It follows that the central charge of the VOA is given
by the formula \eqref{centralcharge_intro}.

The above considerations provide strong evidence
that, if we start with a $4d$ $\cN \ge 3$ SCFT
with Coulomb branch $\mathbb C^{\mathsf r}/\mathsf G$,
with $\mathsf G$ a crystallographic complex
reflection group of rank $\mathsf r$,
the associated VOA is precisely the VOA $\cW_\mathsf G$
described in section \ref{mainresultsINTRO}.
Conversely, it seems natural to expect that,
for any crystallographic complex
reflection group $\mathsf G$,
the VOA $\cW_\mathsf G$ should originate
from a $4d$ $\cN \ge 3$ SCFT.

In table \ref{crystal_compl_refl}
we list all irreducible crystallographic complex reflection groups.\footnote{For
the character fields of irreducible crystallographic complex reflection groups,
we refer the reader to table 3 of \cite{Caorsi:2018zsq}.}
If $\mathsf G$ is 
a crystallographic
complex reflection group that is also Coxeter, i.e.~a Weyl group,
the identification of the parent $4d$ theory for the VOA $\cW_{\mathsf G}$
is straightforward:
it is simply $4d$ $\cN = 4$ SYM with the appropriate
gauge algebra $\mathfrak g$, such that $\mathsf G = {\rm Weyl}(\mathfrak g)$.
If we consider a crystallographic complex reflection
group that is not Coxeter, the putative $4d$ parent
theory in the sense of \cite{Beem:2013sza}
should be a genuine $\cN = 3$ SCFT.
For some of the entries in  table 
\ref{crystal_compl_refl}
the parent $\cN = 3$ theory has been identified in \cite{Garcia-Etxebarria:2015wns,Nishinaka:2016hbw,Aharony:2016kai}.\footnote{More
precisely, 
the rank-one cases associated to $\mathbb Z_k$, $k\in\{3,4,6\}$,
have been analyzed in \cite{Nishinaka:2016hbw},
while the S-fold construction of \cite{Garcia-Etxebarria:2015wns,Aharony:2016kai}
provides the parent $4d$ $\cN = 3$ theory in the cases
$G(3,3,n)$, $G(3,1,n)$, $G(4,4,n)$,
$G(4,1,n)$, $G(6,6,n)$ (see table (2.13) in \cite{Aharony:2016kai}).
Additional $\cN = 3$ theories have been
introduced in \cite{Garcia-Etxebarria:2016erx},
but not enough information about their Coulomb branches
is available to establish a clear connection to
this work.
} It would be interesting to confirm or rule out
the existence of parent $\cN = 3$ theories for all other entries
of table \ref{crystal_compl_refl}.

Any VOA arising from a $4d$ SCFT via the construction of
\cite{Beem:2013sza} is equipped with the so-called ``$R$-filtration''
\cite{Beem:2017ooy}, which originates from the $\mathfrak{sl}(2)_R$
symmetry of the parent $4d$ $\cN \ge 2$ SCFT.
While natural from a $4d$ perspective,
 the $R$-filtration does not seem to be intrinsic to the VOA
in any obvious way. For instance, given an abstract presentation
of the VOA in terms of its strong generators and the singular
OPEs among them, it is not clear in general
how to recover the $R$-filtration. 
This is a pressing problem, since 
the $R$-filtration is instrumental in achieving 
a detailed understanding of the map
from  $4d$ to $2d$ operators,
which in turn is pivotal in many applications of the VOA
technology to $4d$ physics.

In this work, we propose a simple solution to the problem of the $R$-filtration
for all VOAs that admit a $4d$ origin via the map of \cite{Beem:2013sza},
and at the same time can be identified with one of the
VOAs $\cW_\mathsf G$ for some complex reflection group 
$\mathsf G$. In this case we can utilize the free-field
construction to define a filtration of $\cW_\mathsf G$,
which will be referred to as $\cR$-filtration.
This filtration is specified in a simple way by assigning
weights to the free field ``letters'' $\beta_\ell$, $\gamma_\ell$,
$b_\ell$, $c_\ell$, $\ell = 1, \dots, \rm{rank}(\mathsf G)$,
see \eqref{tableRweightsbetagammabc}.
We propose the identification of this novel $\cR$-filtration
with the sough-for $R$-filtration,
and we perform several tests of this proposal.

The identification of   $R$-filtration
and $\cR$-filtration allows us to recover the 
Macdonald limit of the superconformal index of the parent $4d$
theory from the corresponding VOA.
This is achieved with the refinement the vacuum character of the VOA
by the $\cR$-filtration.

\newpage

\begingroup
\renewcommand{\arraystretch}{1.15}

\begin{table}

\noindent
\begin{minipage}[t]{0.49 \textwidth} 
 \begin{tabular}[t]{|    p{2.5cm} | P{0.8cm}  | P{3.0cm} |        } 
 \hline
 \multicolumn{3}{|c|}{ \small \emph{Non-Coxeter groups}} 
 \\ \hline  \hline
\multicolumn{1}{|c|}{ \small Group}    &
\small \centering Rank
 &  \small    Degrees
 \\ 
 \hline\hline
\rowcolor{mygray} \small $G(3,1,n)$ & \small   $n$  &
\small $3,6,...,3n$
\\ \hline
\rowcolor{mygray} \small $G(3,3,n) , \,  n \ge 3$ & \small $n$  &
\small  $3,6,...,3(n-1);n$
\\ \hline
\rowcolor{mygray} \small $G(4,1,n)$ & \small $n$  &
\small $4,8,...,4 n$
\\ \hline
\small $G(4,2,n)$ & \small $n$  &
\small $4,8,...,4 (n-1);2n$
\\ \hline
\rowcolor{mygray} \small $G(4,4,n) , \, n \ge 3$ & \small $n$  &
\small $4,8,...,4 (n-1);n$
\\ \hline
\small $G(6,1,n)$ & \small $n$  &
\small $6,12,...,6 n$
\\ \hline
\small $G(6,2,n)$ & \small $n$  &
\small $6,12,...,6 (n-1);3n$
\\ \hline
\small $G(6,3,n)$ & \small $n$  &
\small $6,12,...,6 (n-1);2n$\
\\ \hline
\rowcolor{mygray} \small $G(6,6,n) , \, n \ge 3$ & \small $n$  &
\small $6,12,...,6 (n-1);n$
\\ \hline
\rowcolor{mygray} \small $\mathbb Z_k , \; k \in \{3,4,6\}$
& 
\small $1$ 
&
\small $k$
\\ \hline
\small $G_4$ & \small $2$ & \small $4,6$
\\ \hline
\small $G_5$ & \small $2$ & \small $6,12$
\\ \hline
\small $G_8$ & \small $2$ & \small $8,12$
\\ \hline
\small $G_{12}$ & \small $2$ & \small $6,8$
\\ \hline
\small $G_{24}$ & \small $3$ & \small $4,6,14$
\\ \hline
\small $G_{25}$ & \small $3$ & \small $6,9,12$
\\ \hline
\small $G_{26}$ & \small $3$ & \small $6,12,18$
\\ \hline
\small $G_{29}$ & \small $4$ & \small $4,8,12,20$
\\ \hline
\small $G_{31}$ & \small $4$ & \small $8,12,20,24$
\\ \hline
\small $G_{32}$ & \small $4$ & \small $12,18,24,30$
\\ \hline
\small $G_{33}$ & \small $5$ & \small $4,6,10,12,18$
\\ \hline
\small $G_{34}$ & \small $6$ & \small $6,12,18,24,30,42$
\\ \hline
 \end{tabular}
\end{minipage}  
\begin{minipage}[t]{0.49 \textwidth} 

 \begin{tabular}[t]{| p{1.4cm} | P{0.8cm} | P{0.8cm} | P{3.cm} |     } 
 \hline
\multicolumn{4}{|c|}{ \small \emph{Coxeter groups}} 
 \\ \hline\hline
   \multicolumn{1}{|c|}{\small Group} & \small $\mathfrak g$ &
\small Rank & \small Degrees
 \\ \hline \hline
\rowcolor{mygray}\small $S_n$ & \small $ \mathfrak a_{n-1}$  & \small $n-1$  & \small $2,3,...,n$  
\\ \hline
\rowcolor{mygray} \small $G(2,1,n)  $ & \small $\mathfrak b_{n}, \mathfrak c_n$ & \small $n$ &
\small $2,4,...,2n$
\\ \hline
\rowcolor{mygray} \small $G(2,2,n) $ & \small $\mathfrak d_{n}$ & \small $n$ &
\small $2,4,...,2(n-1);n$
\\ \hline
\rowcolor{mygray} \small $G(6,6,2) $ & \small $\mathfrak g_2$ & \small $2$ &
\small $2,6$
\\ \hline
\rowcolor{mygray} \small $G_{28}  $ & \small $\mathfrak f_4$ & \small $4$ &
\small $2,6,8,12$
\\ \hline
\rowcolor{mygray} \small $G_{35}  $ &\small  $\mathfrak e_6$ & \small $6$ &
\small $2,5,6,8,9,12$
\\ \hline
\rowcolor{mygray} \small $G_{36} $ & \small $\mathfrak e_7$ & \small $7$ &
\small $2,6,8,10,12,14,18$
\\ \hline
\rowcolor{mygray} \small $G_{37} $ & \small $\mathfrak e_8$ & \small $8$ &
\parbox[c]{2.3cm}{ \centering \small \vspace{1.5mm}$2,8,12,14,$\\$18,20,24,30$\vspace{1mm}  }
\\ \hline
 \end{tabular}
\end{minipage}
\caption{ 
Irreducible crystallographic complex reflection groups,
partitioned into non-Coxeter and Coxeter groups.
For each group, we give the rank and the degrees of the fundamental
invariants. 
A crystallographic Coxeter group
is a Weyl group: in this case we also include the corresponding Lie
algebra(s). 
The notations $G(m,p,n)$ and $G_n$ refer to the original
list by Shephard and Todd. 
The symbol $S_n$ denotes the symmetric group
of permutations of $n$ objects.
Unless otherwise stated, it is understood that $n \ge 2$.
The specifications $n \ge 3$ exclude 
$G(3,3,2) \cong {\rm Weyl}(\mathfrak a_2)$,
$G(6,6,2) \cong {\rm Weyl}(\mathfrak g_2)$,
and $G(4,4,2)$, which is conjugate in $U(2)$ to
${\rm Weyl}(\mathfrak b_2)$.
The shaded entries correspond to
crystallographic complex reflection
groups that govern the Coulomb branch
geometry of known $4d$ $\cN \ge 3$
SCFTs.
} 
\label{crystal_compl_refl}
\end{table}
\endgroup




 \linepenalty=1000

 \subsection{Outlook}
  
The main goal of this work is
to provide a unified description of a large class of 
supersymmetric VOAs,
which includes all known VOAs originating
from $4d$ SCFTs with $\cN \ge 3$ via the map of \cite{Beem:2013sza}.
A full classification of all VOAs with $\cN = 2$
or small $\cN = 4$ supersymmetry is a more ambitious task,
which would   require different tools. One way to address the classification problem 
consists in posing and trying to solve suitable supersymmetric
VOA bootstrap problems.  We plan to report
progress in this direction in an upcoming paper \cite{bootstrappaper}.

A simple example of VOA bootstrap problem is the following.
Consider a VOA with small $\cN = 4$ supersymmetry,
and suppose it is strongly generated
by the generators of the small $\cN = 4$ SCA,
together with additional strong generators,
organized in a single short $\mathfrak{psl}(2|2)$
multiplet with prescribed quantum numbers $h=j=p/2$.\footnote{The
highest weight state of this short $\mathfrak{psl}(2|2)$
multiplet is always assumed to be Grassmann even.}
We may then ask:
for a given $p \ge 3$, for which values of the central charge $c$
does the VOA exist?
To address this question,
one tries to determine the singular OPEs of the 
strong generators in such a way that all axioms
of a \emph{bona fide} VOA are satisfied.
The outcome of this analysis is that
the VOA exists in three cases:
\beq
\begin{array}{rcl}
\text{(A)} :  &  & \text{$c = -6(p+1)$, for any integer $p \ge 3$}; \\[2mm]
\text{(B)} : & & \text{$c = -6 (\tfrac 12 \, p + 1)$, for even $p \ge 4$}; \\[2mm]
\text{(C)}:  &  &\text{$c = 3(p-2)$, for odd $p \ge 3$}. 
\end{array}
\eeq
The interpretation of case (A) is clear: 
the VOA is the  VOA $\cW_{I_2(p)}$ associated to the Coxeter
group $I_2(p)$, which is discussed in detail 
in section \ref{I2p} below. The VOAs of cases (B) and 
(C) cannot be identified with any VOA $\cW_\Gamma$
with $\Gamma$ Coxeter group.
We do have, however, an interpretation for case (B) in terms
of a \emph{quotient} of the VOA $\cW_{I_{2}(p/2)}$
associated to the Coxeter group $I_2(p/2)$.
More precisely, the VOA $\cW_{I_{2}(p/2)}$
admits a $\mathbb Z_2$ automorphism,
and we propose the identification
of the VOA of case (B) with $\cW_{I_2(p/2)}/\mathbb Z_2$.
A first trivial check of this proposal
is the value of the central charge;
more non-trivial checks
can be performed  by analyzing
null states in the VOA.
As far as the VOA of case (C) is concerned,
we restrict ourselves to the simple observation that,
since its central charge is positive,
it cannot originate from any unitary $4d$ SCFT
via the map of \cite{Beem:2013sza}.

Another simple VOA bootstrap problem we can address is the following.
Consider a small $\cN = 4$ VOA, which is by assumption
strongly generated by the generators of the small $\cN =4$
SCA, together with additional generators,
organized in two short $\mathfrak{psl}(2|2)$
multiplets with given quantum numbers $h = j = p_1/2$
and $h = j = p_2/2$.\footnote{Also in this case the
highest weight state of these short $\mathfrak{psl}(2|2)$
multiplets are   assumed to be Grassmann even
for any choice of $p_1$, $p_2$.} Once again, 
we imagine to fix $p_1 \le p_2$,
and we investigate if there is any value of $c$
for which the VOA exists. 
We find that the VOA  exists only 
in four cases:
\beq
\begin{array}{rclcl}
\text{(a)} : && (p_1, p_2) = (3,4) \ , && c=-36 \ ; \\[2mm]
\text{(b)} : && (p_1, p_2) = (4,6)  \ ,  && c=-54 \ ; \\[2mm]
\text{(c)} : && (p_1, p_2) = (6,10) \ , && c=-90 \  ; \\[2mm]
\text{(d)} : && (p_1, p_2) = (4,6) \ , && c=-36 \ .
\end{array}
\eeq
We can interpret all these four cases in terms of VOAs
associated to a Coxeter group.
The VOAs in the cases (a), (b), (c)
are the VOAs $\cW_{A_3}$,
$\cW_{B_3}$,
$\cW_{H_3}$,  respectively.
The VOA of case (d) is the quotient $\cW_{A_3}/\mathbb Z_2$.

As a final comment, we have studied similar VOA
bootstrap problems involving more strong generators.
The picture emerging from the bootstrap analysis
is compatible with the expectation that the  VOAs 
labelled by Coxeter groups introduced in this work 
exhaust the complete list of (small) $\mathcal{N}=4$ W-algebras
with certain ``good'' properties.
The 
fundamental question of 
identifying such properties 
will be  addressed in
\cite{bootstrappaper}.






 \linepenalty=1000

\section{Preliminaries}
\label{sec:prelliminaries}

In the first part of this section, we briefly review some basic features of the
$\cN = 2$ and small $\cN = 4$ SCAs and of their representation theory.
In the second part of this section, we collect some standard 
material on Coxeter and complex reflection groups. In particular, 
we describe their ring of invariants  in the cases
of interest for applications in the the rest of the paper.

Unless otherwise stated, all Lie (super)algebras
in this work are understood to be Lie (super)algebras over 
the complex numbers.

\subsection{$\cN = 2$ and small $\cN = 4$ SCAs}
\label{smallN4recap}
The small $\cN = 4$  super-Virasoro algebra is generated 
by  affine $\mathfrak{sl}(2)$ currents $J^{0,\pm}$,
 a stress tensor $T$, and four fermionic  operators $\widetilde G^{\pm}$ and   $G^{\pm}$.
 The bosonic sub-VOA has OPEs
 \begin{align}
   J^{0}(z_1)J^{0}(z_2)& \sim
   \frac{2\,k}{(z_{1}-z_2)^2}\,,\\
    J^{0}(z_1)J^{\pm}(z_2)& \sim
   \frac{ \pm 2\,J^{\pm}}{(z_{1}-z_2)}\,,\\
   J^{+}(z_1)J^{-}(z_2)& \sim
   \frac{-k}{(z_{1}-z_2)^2}+ \frac{-J^0}{(z_{1}-z_2)}\,,\\
 T(z_1)\,T(z_2)&\sim
  \frac{c/2}{(z_{1}-z_2)^4}+ \frac{2\,T(z_2)}{(z_{1}-z_2)^2}+ \frac{\partial T(z_2)}{(z_{1}-z_2)}\,,
 \end{align}
the  OPE between $T$ and $J$ express the fact that 
 $J^{0,\pm}$ are Virasoro primaries of conformal dimension $h=1$, see 
 appendix \ref{OPEandsl2cov}.
The level and the central charge are related as 
\beq
c=6k\,.
\eeq
The fermionic generators are Virasoro and affine Kac-Moody (AKM) primaries 
with weights $h=\tfrac{3}{2}$ and $j=\tfrac{1}{2}$. Their OPE takes the form
\beq\label{GGtildeOPE}
G^I(z_1) \, \widetilde G^J(z_2)  \sim \frac{2 \, k \, \epsilon^{IJ}}{(z_1-z_2)^3}+ 
\frac{2 \, J^{IJ}(z_2)}{(z_1-z_2)^2} + \frac{  \epsilon^{IJ} T(z_2)+ \partial J^{IJ}(z_2)}{z_1-z_2}\,,
\eeq
where $\epsilon^{+-}=1$, $J^{\pm \pm}=J^{\pm}$ and $J^{+-}=J^{-+}= \tfrac{1}{2}J^0$.
The remaining OPE among fermionic generators are regular.
The small $\cN = 4$ SCA possesses an $SL(2)$ outer automorphism that rotates  $G$ and $\widetilde G$ as a doublet and acts trivially on the bosonic generators. 
We denote by\footnote{This $r$ should not to be confused with $\mathsf r=\text{rank}(\Gamma)$.}
 $GL(1)_r$ the corresponding Cartan generator, normalized 
 as $r[G]=\tfrac{1}{2}$, $r[\widetilde{G}]=-\tfrac{1}{2}$. 
In order to avoid keeping track of
$\mathfrak{sl}(2)$ indices we use a standard index free notation 
by introducing the auxiliary variable $y$ and write
\beq
J(y) = J^+ + J^0 \, y + J^- \, y^2 \ , \qquad
G(y) = G^+ + G^- \, y \ , \qquad
\widetilde G(y) = \widetilde G^+ + \widetilde G^- \, y \ ,
\eeq
where we omitted the explicit $z$-dependence of the operators.
Because of the auxiliary variable $y$, we often refer to the $\mathfrak{sl}(2)$
R-symmetry of the small $\cN = 4$ SCA as $\mathfrak{sl}(2)_y$.
In contrast, the conformal algebra $\mathfrak{sl}(2)$ on the Riemann sphere with coordinate $z$
will be referred to as $\mathfrak{sl}(2)_z$.
More details are given in appendix \ref{OPEandsl2cov}.

The $\cN = 2$ SCA can be defined as the sub-VOA of the small  $\cN = 4$ SCA  generated by
\beq\label{N2inN4preliminaries}
\cJ = J^0 \ , \qquad
\cG = G^- \ , \qquad
\widetilde \cG = \widetilde G^+\,,
\qquad
\mathcal{T}=T \ .
\eeq
The remaining generators of the small $\cN = 4$ SCA, namely $(J^+,G^+)$ and $(J^-,\widetilde{G}^-)$,
are $\mathcal{N}=2$ chiral  and anti-chiral multiplets respectively.

The global part of the small $\cN = 4$ and $\cN = 2$ SCA are  
 $\mathfrak{psl}(2|2)$ and $\mathfrak{osp}(2|2)$ respectively.\footnote{In fact, the
 small $\cN = 4$ and $\cN = 2$ SCAs
can be obtained by quantum Drinfeld-Sokolov  reduction 
of the Lie superalgebras $\mathfrak{psl}(2|2)$ and $\mathfrak{osp}(2|2)$,
respectively,
see \cite{KAC2004400}.
This is consistent with the fact that 
the small $\cN = 4$ and $\cN = 2$ SCAs exist for any value 
of the central charge $c$.
} 
The representations of 
 $\mathfrak{psl}(2|2)$ and $\mathfrak{osp}(2|2)$
relevant for this work are summarized in table \ref{table_reps}.
The bosonic subalgebra of $\mathfrak{psl}(2|2)$ is
$\mathfrak{sl}(2)_z \oplus \mathfrak {sl}(2)_y$, and representations
are labelled by the conformal dimension $h$ and the (half-integer) spin
$j$ of the superconformal primary. The bosonic subalgebra of $\mathfrak{osp}(2|2)$
is $\mathfrak{sl}(2)_z \oplus \mathfrak{gl}(1)$, 
and representations
are labelled by the conformal dimension $h$ and the (half-integer) 
$\mathfrak{gl}(1)$ charge $m$ of the superconformal primary.

%
%
%
%

\begingroup
\renewcommand{\arraystretch}{1.4}

\begin{table}[h!]
\centering
 \begin{tabular}{|c | c | c |c|} 
 \hline
algebra & 
quantum numbers of s.c.p.~$X$
 &  shortening conditions
 & notation
 \\ 
 \hline\hline
\multirow{2}{*}{$\mathfrak{psl}(2|2)$}  & $h > j$ & $-$ & $\mathfrak L_{(h,j)}$\\
 \cline{2-4}
    &  $h=j$ & $G^\uparrow X = \widetilde G^\uparrow X = 0$  &
   $\mathfrak S_h$  \\
  \cline{2-4} \hline\hline
  \multirow{3}{*}{$\mathfrak{osp}(2|2)$}    &  $h > |m|$ & $-$  &
  $\mathfrak X_{(h,m)}$  \\
   \cline{2-4}
   &  $h=+m$ & $\widetilde {\cG} \cdot X = 0$  &   $\mathfrak C_h$ \\
   \cline{2-4}
   &  $h=-m$ & $  {\cG} \cdot X = 0$  &   $\overline {\mathfrak C}_h$ \\
   \hline
\end{tabular}
\caption{
Representations of $\mathfrak{psl}(2|2)$ and $\mathfrak{osp}(2|2)$
that appear in this work. We use $X$ to the denote the
superconformal primary, or s.c.p.~for short.
The symbol $G^\uparrow X$ denotes the
descendant of $X$ with weight $h+\tfrac 12$ and spin $j+\tfrac 12$ obtained
by acting once with the supercharge $G$.
The notation $\cG \cdot X$ stands for
the descendant of $X$ with weight $h+ \tfrac 12$ and
charge $m + \tfrac 12$ obtained
by acting once with the supercharge $\cG$.
Similar remarks apply to $\widetilde G^\uparrow X$,
$\widetilde \cG \cdot X$.
More details on our notation can be found 
in appendix~\ref{OPEandsl2cov}.
} 
\label{table_reps}
\end{table}

\endgroup

Finally we recall the definition of super-Virasoro primary.
The operator $\cO$ is a super-Virasoro primary if
the OPE of any super-Virasoro generator with $\cO$
does not contain any pole of order  higher than one, 
with the obvious exception of the order-two pole 
in the $T \cO$ OPE, which
encodes the conformal weight.
The first order poles in the OPEs
of the super-Virasoro generators with $\cO$ 
encode the action of the global part of the SCA.
See  appendix \ref{OPEandsl2cov} for more details.

\vspace{0.5cm}
\noindent
\emph{Notation:} 
Occasionally we will use the following notation for poles in the OPE:
\beq \label{bracket_notation}
A(z_1)B(z_2)=\sum_n \frac{\{AB\}_{n}(z_2)}{(z_1-z_2)^n}\,.
\eeq
We will also use $(A\,B)_n$ to represent be the
 completion of $\{A\,B\}_n$ to a quasiprimary, {\it i.e.}~an 
$\mathfrak{sl}(2)_z$ primary.
The object  $(A\,B)_n$  is obtained from 
$\{A\,B\}_n$ by adding $z$-derivatives of higher-order poles in the $AB$ OPE, the explicit 
formula is given in \eqref{quasiPrimary_extraction}.
See {\it e.g.}~\cite{Thielemans:1994er} for details.
Concerning  the $\mathfrak{sl}(2)_y$ structures, we will use the notation $(A\,B)^j$
 for the spin $j$ projection of the relevant product of $A$ with $B$, 
 see appendix \ref{OPEandsl2cov} for more details.
Finally, given a quasiprimary $X$ of $\mathfrak{sl}(2)_y$ with spin $j$, we introduce the 
shorthand notation for its supersymmetric descendant
\beq\label{SUSYdescNOT}
G^\downarrow X = (G \, X)^{j -\frac 12}_1 \,.
\eeq
This is the quasiprimary in the order-one pole of the OPE of $G$  with $X$,
projected onto the component with spin $j-\frac 12$. Similar remarks
apply to the operations $G^\uparrow $ and  $\widetilde G^{\uparrow,\downarrow}$.

\subsection{Coxeter groups, complex reflection groups, rings of invariants}
\label{subsec:Coxeter}

We will now describe the symplectic varieties
  \eqref{MGammavarietyintro}
  and \eqref{associatedvarietyDEF2} and the associated ring of functions in more detail.
 The case of Coxeter groups is described first, the generalization to complex reflection groups is presented in the end of this section.
 Let us start by fixing a  basis of $V_{\Gamma}$ to be $z_1,\dots,z_{\mathsf{r}}$, 
 where $\mathsf{r}=\text{rank}(\Gamma)$.
 The action of the  (finite) Coxeter group $\Gamma$ on the real vector space $V_{\Gamma}$ 
 is  generated by reflections.
 This can be taken to be the definition of Coxeter group.
  Their full list is given by the Weyl groups of finite dimensional semisimple Lie algebras $ABCDEFG$
  together with $I_2(p)$, $H_3$,  $H_4$, see table~\ref{tableinvariants} for more details. 
The expression for the central charge given in \eqref{centralcharge_intro}
can be rewritten as
\beq\label{cmassagedform}
-\frac{c}{3}=
\sum_{\ell=1}^{\text{rank}(\Gamma)}\,(2 p_{\ell}-1)=
|\Phi_{\Gamma}|+\text{rank}(\Gamma)\,
\stackrel{\text{Weyl}}{=}\text{dim}(\mathfrak{g}_{\Gamma})\,,
\eeq
where $|\Phi_{\Gamma}|$ is the cardinality of the root system associated to $\Gamma$.

\begingroup
\renewcommand{\arraystretch}{1.3}
\begin{table}[h!]
\centering
 \begin{tabular}{|c | c | c |c|} 
 \hline
$\Gamma$ & 
$p_1,\dots, p_{\mathsf{r}}$
 &  $(h,j)$ quantum numbers of long generators*
 & $-\frac{c}{3}$
 \\ 
 \hline\hline
$A_{n-1}$  & $2,3,\dots,n$ & $-$ & $n^2-1$\\ 
 \hline
 $B_n$  &  $2,4,\dots,2n$ & $-$  & $n(2n+1)$  \\
 \hline
  $D_n$  &  $2,4,\dots,2(n-1);n$ & $(\tfrac{n+2}{2},\tfrac{n-4}{2}), \dots$  &  $n(2n-1)$  \\
 \hline
 $E_6$  &  $2,5,6,8,9,12$ & $(4,0),(\tfrac{9}{2},\tfrac{3}{2}), (6,3)$  &   $78$ \\
 \hline
  $E_7$  &  $2,6,8,10,12,14,18$ & $(5,1), (6,3), (7,3), (8,5)$  & $133$ \\
 \hline
   $E_8$  &  $2,8,12,14,18,20,24,30$ & $(6,0), (7,3), (9,6), (9,4),(9,3),(10,6), (10,0)$  &  $248$ \\
 \hline
    $F_4$  &  $2,6,8,12$ & $(4,0), (6,3)$  &  $52$\\
 \hline
  $H_3$  &  $2,6,10$ & $-$  &  $33$ \\
 \hline
  $H_4$  &  $2,12,20,30$ & $(6,0), (10,6), (10,0)$  &  $124$\\
 \hline
   $I_2(p)$  &  $2,p$ &$-$  & $2(p+1)$ \\
 \hline
\end{tabular}
\captionof{table}{ 
The notation for Coxeter groups is such that when restricting to Weyl groups one has the identification
 $A=\text{Weyl}(\mathfrak{a})$,  $B=\text{Weyl}(\mathfrak{b})$ and so on.
 Moreover, the Weyl group $G_2$ appears as $G_2=I_2(6)$. 
 Recall that
 $\text{Weyl}(\mathfrak{b}_n)=\text{Weyl}(\mathfrak{c}_n)$. 
 The asterisk * above indicates that only  long generators whose conformal weight $h$
 is smaller than the lightest relation are listed.
 These  are the one that can be extracted unambiguously from the 
 associated Hilbert series but there might be more generators. 
 The \dots in the $D_n$ series  are given explicitly for  $4\leq n \leq 9$
  in equation  \eqref{ALLplethysticLOGS}.
See appendix \ref{appMOLIEN} for more details.} 
\label{tableinvariants}
\end{table}
\endgroup

%
 The main ring of interest in the following is the ring of invariant polynomials in two sets of  variables $z_i^{\pm}$
\beq\label{wannabeHiggsBrChiralRing}
\mathscr{R}_{\Gamma}\,=
\,\mathbb{C}[z_1^+,\dots,z_{\mathsf{r}}^+,z_1^-,\dots,z_{\mathsf{r}}^-]^{\Gamma}
=\mathbb{C}[ \mathscr{M}_{\Gamma}]\,,
\eeq
where the Coxeter group acts independently on  $z_i^{+}$ and $z_i^{-}$.
This ring carries the action of $GL(2)= SL(2) \times GL(1) $, where $z^{\pm}$ transform as a doublet  under $SL(2)$
and have the same $GL(1)$ weight. 
This ring is thus  graded  by $m$, the eigenvalues of the Cartan of  $SL(2)$, and  $h$  with $m[z_{i}^{\pm}]=\pm \tfrac{1}{2}$, $h[z_{i}^{\pm}]=\tfrac{1}{2}$.

The ring $\mathscr{R}_{\Gamma}$ has  an alternative description in terms of 
generators and relations.
Let us present the simplest example of $\Gamma=\text{Weyl}(\mathfrak{a}_1)=\mathbb{Z}_2$,
 whose action is generated by  $\sigma \cdot z_{1}^{\pm}=-\, z_{1}^{\pm}$. 
In this case the invariants are $j^{\pm}=z_1^{\pm}z_1^{\pm}$ and 
 $j^{0}=2 z_1^{+}z_1^{-}$ with the obvious relation  $j^+j^-=\tfrac{1}{4} j^0 j^0$.
For groups $\Gamma$ of higher rank giving an explicit description of this type 
is more involved but in principle straightforward. 
 We do so in the low rank examples of $I_2(p)$, $A_3$, $B_3$, $H_3$ 
 and comment on the general case in appendix \ref{appMOLIEN}.
One can develop an opinion about the set of generators and relations  by considering the Hilbert series of
 $\mathscr{R}_{\Gamma}$ that we will now review.

\paragraph{Hilbert series.}
The (refined) Hilbert series of $\mathscr{R}_{\Gamma}$ is defined as 
\beq\label{HS}
\mathsf{HS}_{\Gamma}(\tau,x)\,=\,
\text{Tr}_{\mathscr{R}_{\Gamma}}\big(
\tau^{2h}x^{2m}\big)\,.
\eeq
In the case of ring of invariants as \eqref{wannabeHiggsBrChiralRing}
 the Hilbert series can be computed by  averaging over $\Gamma$ 
 the Hilbert series  of the freely generated ring $\mathbb{C}[z^+,z^-]$.
 This is the content of the Molien formula
\beq\label{Molien}
\mathsf{HS}_{\Gamma}(\tau,x)\,=\,\mathsf{Molien}_{\Gamma}(\tau,x)\,:=\,
\frac{1}{|\Gamma|}\sum_{g\in\Gamma}\,
\frac{1}{\det_{\mathbb{C}^2\otimes V_{\Gamma}}(1-h\otimes g)}\,,
\qquad 
h=\tau\,\begin{pmatrix}
x & \!0 \\
0 & \,x^{-1}
\end{pmatrix}\,,
\eeq
where $|\Gamma|$ is the order of $\Gamma$.
The simplest example is given by
\beq\label{MolienZ2text}
\mathsf{Molien}_{\mathbb{Z}_2}(\tau,x)\,=\,
\frac{1-\tau^4}{(1-\tau^2\,x^{-2})(1-\tau^2)(1-\tau^2\,x^{+2})}\,.
\eeq
In this formula the denominator can be interpreted as the contribution of the generators $j^+,j^-,j^0$ defined above and the subtraction of the terms $\tau^4$ in the numerator corresponds to the relation
$j^+j^-=\tfrac{1}{4} j^0 j^0$.
While it is not possible in general to extract the set of generators and relations, 
together with their $h,m$ quantum numbers, from  the series \eqref{HS} alone,
one can show that certain generators and relations must be present, see {\it e.g.}~\cite{stanley1979}.
This can be done efficiently by using the so-called plethystic logarithm.
The expressions are collected in appendix \ref{appMOLIEN} and the resulting set of 
generators  is given in table~\ref{tableinvariants}.
For convenience of the reader we collect the expression for  Molien generating functions for all Coxeter groups in an ancillary Mathematica file.
It should be noticed that the generators of $\mathscr{R}_{\Gamma}$ can be divided in two groups:
one  consisting of elements with  quantum numbers $h=j$, which will be referred to as short generators, and 
the other with  quantum numbers $h>j$, which will be referred to as long generators.
We do not give a complete list of long generators for general $\Gamma$ but list generators whose existence can be shown unambiguously
 using the Molien series \eqref{Molien}. The result is collected in table~\ref{tableinvariants}.
\paragraph{Symplectic structure.}

The ring $\mathscr{R}_{\Gamma}$ possesses the important property of being a Poisson algebra
 with Poisson bracket
\beq\label{zsPB}
\{ z_i^I , z_j^J \}_{\rm PB} = \eta_{ij} \, \epsilon^{IJ} \ , \qquad
i,j = 1,2,\dots, \mathsf{r}  \ , \quad
I,J = \pm \ , 
\eeq
where $\epsilon^{IJ}$ is antisymmetric and normalized as $\epsilon^{+-}=1$ while 
$\eta_{ij}$  is symmetric and non-degenerate.
This implies that 
\eqref{MGammavarietyintro}
is a symplectic variety.
It is straightforward to compute the Poisson bracket of the generators of $\mathscr{R}_{\Gamma}$ using \eqref{zsPB}.
In the simple example of $\Gamma=\mathbb{Z}_2$ this gives the Lie algebra of $SL(2)$. For higher rank this procedure  will produce an extension of it. 
A priori, there is no canonical choice for the set of generators,
but there is a distinguished one originating from the associated VOA.
 We remark that when long generators are present they can be generated by 
 Poisson brackets of short generators, 
  as the example of $D_4$ given in section \ref{sec:D4Example} illustrates.

\paragraph{Complex reflection groups.}
The case of complex reflection groups is very similar so we will be brief. 
The first important difference is that the $GL(1) \times SL(2)$ symmetry 
of the Coxeter case is reduced to $GL(1) \times GL(1)$.
The Molien formula is a slight generalization of  \eqref{Molien} to 
\beq\label{MolienG}
\mathsf{Molien}_{\mathsf{G}}(\tau,x)\,=\,
\frac{1}{|\mathsf{G}|}\sum_{g\in\mathsf{G}}\,
\frac{1}{\det_{V^{}_{\mathsf{G}}}(1-\tau\,x\, g)\,\det_{V^*_{\mathsf{G}}}(1-\tau\,x^{-1}\, g)}
\ .
\eeq
Finally the generalization of Poisson brackets \eqref{zsPB}  uses the canonical pairing between 
$V^{}_{\mathsf{G}}$ and its dual $V^*_\mathsf G$.

\vspace{0.5cm}
\noindent
\emph{Remark:}
One might refer to \eqref{wannabeHiggsBrChiralRing} as the Higgs branch chiral rings.
It contains a subring defined as the  graded component  with $h=m$.
The latter can be referred to as Coulomb branch chiral ring and is a freely generated  polynomial ring.
This is the content of a famous theorem of Chevalley, Shephard and Todd states that the ring of invariants
 $\mathbb{C}[V]^{\mathsf{G}}$ 
is a freely generated polynomial ring if and only if $\mathsf{G}$ acts as a  complex reflection group on $V$.
These finite groups have been classified by  Shephard and Todd, see {\it e.g.}~\cite{reflectiongroups}.



\section{Free-field realizations}
\label{sec:free_fields}

This section is devoted to our proposal for a free-field realization
of the VOAs associated to Coxeter and complex reflection groups.

\subsection{Realization of the $\cN  =2$ SCA
 in terms of $\beta\gamma bc$ systems}
 \label{sec:N2SCAFree}
To begin with, we present a free-field realization
of the $\cN = 2$ SCA. 
According to our proposal, the relevant free fields 
consist of $\mathsf{r} = {\rm rank} (\mathsf{G})$ copies of a free $\beta\gamma b c$
system. The relevant non-trivial OPEs are simply
\beq\label{defberagammbc}
\beta_{\ell_1} (z_1) \, \gamma_{\ell_2} (z_2) = - \frac{\delta_{\ell_1\ell_2}}{z_{12}} + \text{reg.} \ , \qquad
b_{\ell_1}(z_1) \, c_{\ell_2}(z_2) = \frac{\delta_{\ell_1\ell_2}}{z_{12}} + \text{reg.}  \ ,
\eeq
where $\ell_1,\ell_2$ run from $1$ to $\mathsf{r}$ and $\delta_{\ell_1\ell_2}$ denotes the Kronecker delta.
The generators of the $\cN = 2$ SCA 
take a simple expression in terms of the free fields,
consisting of a direct sum of terms, one for each copy of the
$\beta\gamma bc$ system.
More precisely,
\begin{align}\label{N2freefireld}
\cJ & =  \sum_{\ell = 1}^{\mathsf{r}} \Big[ p_\ell \, \beta_\ell \, \gamma_\ell 
+ (p_\ell -1) \, b_\ell \, c_\ell \Big]  \ , \nn \\
\cG & = \sum_{\ell = 1}^{\mathsf{r}} b_\ell \, \gamma_\ell \ ,
\qquad  
\widetilde \cG  =\sum_{\ell = 1}^{\mathsf{r}} \Big[
 p_\ell \, \beta_\ell \, \partial c_\ell 
+ (p_\ell -1) \, \partial \beta_\ell \, c_\ell
\Big]
\ , \nn \\
\cT & = \sum_{\ell = 1}^{\mathsf{r}} \Big[
- \frac 12 \, p_\ell \, \beta_\ell \, \partial \gamma_\ell
+ \left ( 1 - \frac 12 \, p_\ell \right) \, \partial \beta_\ell \, \gamma_\ell
- \frac 12 (p_\ell +1) \, b_\ell \, \partial c_\ell
+ \frac 12 (1 - p_\ell) \, \partial b_\ell \, c_\ell
\Big]
 \ .
\end{align}
The quantities $p_\ell$ are the degrees of the invariants of the
Coxeter group.
The central charge of the $\cN =2$ SCA is given by
\beq
c =- 3 \, \sum_{\ell = 1}^{\mathsf{r}} (2 \, p_\ell -1) \,,
\eeq
compare to 
\eqref{centralcharge_intro}.
The conformal weights $h$, the $\mathfrak {gl}(1)$ charges $m$ 
and the $\mathfrak {gl}(1)_r$ charge $r$ defined below \eqref{GGtildeOPE}
of the free fields are summarized in table \eqref{betagammabc_hmr}.
The charge $m$ is normalized in such a way that
$h = m$ for chiral primary operators.
Notice that, although the combined $\beta\gamma bc$ system
contains states of negative conformal dimension,
all states have a non-negative ``twist'' $h-m$.
Furthermore, the space of states with given twist
and charge is finite-dimensional.

\begingroup
\renewcommand{\arraystretch}{1.3}
\begin{equation}
 \begin{tabular}{| c | c  | c| c  |c || c |} 
 \hline
  & $h$ & $m$ &$h-m$ & $h+m$ & $r$ \\
 \hline\hline
 $\beta_\ell$  &  $\frac{1}{2} \, p_\ell $ &   $\frac{1}{2} \, p_\ell$
 &0 & $p_\ell$  & $\,\,0$ \\ 
   \hline
  $b_\ell$  &  $\frac{1}{2} (p_\ell + 1)  $ &   $\frac{1}{2} (p_\ell - 1)$
  &1& $p_\ell$  & $+\tfrac{1}{2}$ \\
 \hline
  $c_\ell$  &  $-\frac{1}{2} (p_\ell - 1)$ &   
  $-\frac{1}{2} (p_\ell - 1)$&0& $1-p_\ell$  & $-\tfrac{1}{2}$ \\
 \hline
   $\gamma_\ell$  &  $1 -  \frac 12 \, p_\ell $ &   $- \frac 12 \,
   p_\ell$&1 & $1-p_\ell$ & $\,\,0$  \\
 \hline
    $\partial$  &  $1$ &   $0$& $1$ & $1$   &  $\,\,0$ \\
    \hline
\end{tabular}
\label{betagammabc_hmr}
\end{equation}
\endgroup
\vspace{1mm}

A comment on
normal-ordered products in the $\beta \gamma bc$ system is in order.
In all equations in \eqref{N2freefireld}, the juxtaposition
of free fields is understood as their $\{ \cdot \; \cdot \}_0$
normal-ordered product, see \eqref{bracket_notation}.
 More generally,
let $X_1$, \dots, $X_n$ stand for any of the free fields $\beta$,
$\gamma$, $b$, $c$, or any $z$-derivative thereof.
A natural object is the nested normal-ordered product
\beq
: X_1  \, X_2 \dots X_n :   \;\; = \{ X_1 \{ X_2 \{ \dots \{ X_{n-1} \, X_n \}_0
\dots \}_0 \}_0 \ .
\eeq
Since we are considering a free theory,
and since the $X_i$ are (derivatives of) free fields,
in the normal-ordered product $: X_1  \, X_2 \dots X_n:$
we are free to permute the factors $X_i$, up to Grassmann signs,
exactly 
as we would do in a supercommutative algebra.
For instance
\beq
: \beta \, \partial \beta \, \gamma \, \gamma \, \gamma : \;\; = 
\{ \beta \, \{ \partial \beta \, \{ \gamma \, \{ \gamma \, \gamma
\}_0 \}_0 \}_0 \}_0
= \{ \gamma \, \{ \partial \beta \, \{ \gamma \, \{ \gamma \, \beta
\}_0 \}_0 \}_0 \}_0
= \;\; : \gamma \, \partial \beta \, \gamma \, \gamma \, \beta :  \ .
\eeq
This allows to write compactly $:  \beta \, \partial \beta \, \gamma^3 :$
without ambiguities.
Notice that such manipulations are not allowed in a generic VOA.
For the sake of brevity, we omit the colons from the normal-ordered
products in the rest of this work.

\subsection{Realization of the remaining generators in the $\cN = 4$ case}
\label{sec:remainingGENN4}

Let us first discuss the case of the $\cN = 4$ VOA $\mathcal{W}_{\Gamma}$ associated to a
Coxeter group $\Gamma$. 
As outlined in section \ref{mainresultsINTRO}, the set of strong generators of this VOA 
 includes elements transforming in short representations 
$\mathfrak{S}_{\frac{p_{\ell}}{2}}$ of the global conformal algebra $\mathfrak{psl}(2|2)$, 
see table \ref{table_reps}.
Among these, 
there is a distinguished invariant of degree two which corresponds to the generators of the 
small $\mathcal{N}=4$ super-Virasoro VOA. 
This will be labelled by $\ell=1$ so that $p_{1}=2$.
The remaining short generators are denoted  as\footnote{
The notation for their super-descendants is given in \eqref{Sncontent}.} $W_{\ell}$.
A notable feature of the proposed free-field realization is that 
the highest weight states of the short generators  are identified with $\beta$s,
 see \eqref{defberagammbc}, more precisely
\begin{subequations}
\label{betabN2pair}
\begin{align}
\label{JpGpdoublet}
J^+ &= \beta_1\,,
\qquad\,\,\,\,\,\,\,\,\,\,\,\,\,\,\,\,\,\,
G^+ \,: =\, \{ G^- \, J^+\}_1\,\,\,\,\,\,\,\,\,=\,b_1\,, \\
\label{Whighestweight}
W_\ell {}^{\rm h.w.} &= \beta_\ell \ , \qquad \,\,\,\,\,\,
G_{W_\ell}{}^{\rm h.w.} \,:=\, \{ G^- \,W_\ell {}^{\rm h.w.} \}_1 = \, b_\ell\,,
\end{align}
\end{subequations}
$\ell = 2, \dots , \text{rank}(\Gamma)$. Two remarks are in order.
First notice that the $(h,m)$ weights assignment of these object is consistent by construction, 
see table \eqref{betagammabc_hmr}. The $\mathcal{N}=2$ subalgebra,
 embedded as specified by \eqref{N2inN4preliminaries},  is realized  
as in \eqref{N2freefireld}. The second remark is that each pair $(\beta_{\ell},b_{\ell})$
 forms an $\mathcal{N}=2$ chiral multiplet as \eqref{betabN2pair} indicates.

The next generator that needs to be constructed is $J^-$,
the  $\widehat{\mathfrak{sl}(2)}$ affine Kac-Moody current with 
weight $m=-1$, see section~\ref{smallN4recap}.
Once this operator is constructed one can build the whole 
$\mathcal{N}=4$ super-conformal multiplets to which \eqref{betabN2pair} belong by
taking appropriate poles in the  OPE with $J^-$ and its $\mathcal{N}=2$ partner 
$\widetilde{G}^-$. 
Next, one takes the OPE of the generators obtained in this way.
If $\mathcal{W}_{\Gamma}$ does not have long generators, see table~\ref{tableinvariants},
the latter OPE needs to close on the generators that have already been constructed.
If long generators are present, they will be defined
 by the failure of these OPE to close on the short generators. 
An example of this mechanism, which is already at play at the classical level when closing the Poisson brackets of the short generators of the ring $\mathscr{R}_{\Gamma}$,  is given for the example of $D_4$
in section \ref{sec:D4Example}.

In order to construct $J^-$, we build an Ansatz and impose necessary conditions that this operator must obey. The construction goes as follows:
\begin{itemize}
\item[{\bf 1.}] Construct the most general Ansatz for $J^-$ in terms of
the $\beta\gamma bc$ free fields.
 There is always a finite number of
terms in the Ansatz.
This is easy to verify by recalling that $J^-$ has weights $h=1$, $h-m=2$ 
and by staring at weight assignments of the constituent free fields
given  in table~\ref{betagammabc_hmr}.
\item[{\bf 2.}] Impose
that the small $\cN = 4$ algebra closes, this is equivalent to:
\begin{itemize} 
\item[{\bf a.}] 
Linear constraints: $J^-$  is an anti-chiral
 $\cN=2$ super-Virasoro primary of weight $h=1$
  and its  OPE with  $J^+$ closes on $J^0$.
  \item[{\bf b.}] 
   Non-linear  constraints:
   $J^-$  has a regular OPE with itself.
   \end{itemize}
\item[{\bf 3.}] Impose that the short generators $W_\ell {}^{\rm h.w.}$ given in \eqref{betabN2pair}
must be super-Virasoro primary, this implies:
\beq
\{ J^- \, \beta_\ell \}_{n \ge 2} = 0  \ , \qquad
\ell = 2, \dots, \text{rank}(\Gamma) \ .
\eeq
\item[{\bf 4.}] Impose that the short generators $W_\ell$  have non-zero norms. 
\item[{\bf 5.}] Impose that the  VOA closes on the strong generators.
\end{itemize}
It is convenient to write the Ansatz for $J^-$ in the form
\beq\label{JisJminplusJnorms}
J^-=J^-_{\rm min} +J^-_{\rm norms}\,,
\eeq
where
\beq \label{Jmindef}
J^-_{\rm min} := k \, \partial \gamma_1
+ \beta_1 \, (\gamma_1)^2
+ \gamma_1 \, b_1 \, c_1
+ \gamma_1 \, \widehat{\mathcal{J}}
- c_1 \, \widehat{\mathcal{G}}\,.
\eeq 
The hat on $\mathcal{J}$ and $\mathcal{G}$ signals the 
omission of all terms built with $\beta_1$, $\gamma_1$,
$b_1$, $c_1$ in  \eqref{N2freefireld}. 
For all $\Gamma$ different from  $A_1$ the factor $J^-_{\rm norms}$ has to be non-zero.
As the name suggests,  it must be there in order for point {\bf 4}.~above to be satisfied.
To illustrate this point, let us explain what happens if we set 
$J^-_{\rm norms}=0$ in \eqref{JisJminplusJnorms}.
It is easy to verify that in this case the $\mathcal{N}=4$ super-Virasoro 
subalgebra is correctly reproduced.
Next, let us construct $W^{\text{min}}_{\ell}(z,y)$ by summing up the  $\mathfrak{sl}(2)_y$
descendants  of \eqref{Whighestweight} defined by the action of $J^-_{\rm min}$.
A little computation shows that 
\beq
W^{\text{min}}_\ell(y) = (1 + y \, \gamma_1)^{p_\ell} \, \beta_\ell
-  (1 + y \, \gamma_1)^{p_\ell-1} \, c_1 \, b_\ell \ , \qquad
\ell = 2, \dots, \text{rank}(\Gamma) \ .
\eeq
These operators have obviously regular OPE among themselves,  in particular they have zero norm.
This explains the necessity of adding $J^-_{\rm norms}$.
By using an Ansatz of the form \eqref{JisJminplusJnorms} in the steps above one quickly verifies that 
$J^-_{\rm norms}$ does not include $\gamma_1$ and $c_1$.

Before describing various examples of this construction in the next section, let us make a few remarks:

\vspace{0.3cm}
\noindent
\emph{Remark 1:} 
The  generators of the
 $\mathcal{N}=2$ subalgebra in \eqref{N2freefireld} are  invariant under 
 the transformation\footnote{Notice that if there are two generators with the same weight, like $D_4$ for which 
$(p_1,p_2,p_3,p_4)=(2,4,4,6)$, the corresponding $GL(1) \times GL(1)$ is enhanced to $GL(2)$.
} 
 \beq\label{severalU1sonbetagammas}
\big(\beta_{\ell},\gamma_{\ell},b_{\ell},c_{\ell}\big)\,\mapsto\,
\big(\lambda_{\ell}^{}\,\beta_{\ell},\lambda_{\ell}^{-1}\gamma_{\ell},\lambda_{\ell}^{}\,b_{\ell},\lambda_{\ell}^{-1}c_{\ell}\big)\,,
\qquad
\lambda_{\ell}\,\in\, GL(1) \,,
\eeq
for $\ell=2,\dots,\text{rank}(\Gamma)$.  
We claim that $J^-$ is uniquely determined by the steps above up to this ambiguity.
\vspace{0.3cm}
\\ \noindent
\emph{Remark 2:} For low ranks $\text{rank}(\Gamma)=1,2,3$ 
the steps ${\bf 1.}$-${\bf4.}$
 are sufficient to 
determine $J^-$ up to the action \eqref{severalU1sonbetagammas} and condition ${\bf 5.}$ holds automatically.
For higher ranks condition ${\bf 5.}$  needs to be used as well. 
As the example  of $D_4$ illustrates, see section \ref{sec:D4Example},  there is a subset of these 
conditions that is easy to implement and is sufficient to fully determine $J^-$.
\vspace{0.3cm}
\\ \noindent
\emph{Remark 3.} It is natural to ask what happens if one applies 
the procedure presented above to a set of weights $p_1=2, p_2, \dots, p_{r}$ 
that do not correspond to a Coxeter group. 
We observed experimentally, by looking at  rank 3 examples with 
$2\leq p_2 \leq p_3 \leq 10$, that the norms mentioned above 
are non-zero if and only if the weights are the one associated to a 
Coxeter group, which in rank $3$ are $A_3,B_3,H_3$, see table~\ref{tableinvariants}.


\subsection{Realization of the remaining generators in the $\cN = 2$ case}
\label{sec:N2freeFIELDprescription}

The realization of the $\cN = 2$ VOA associated to a complex reflection
group $\mathsf{G}$ is qualitatively similar.
The generators of the  $\cN = 2$  SCA
algebra are given in \eqref{N2freefireld} 
and the additional chiral generators have the form
\beq
W_\ell = \beta_\ell \ ,
\qquad
\mathcal{G}_{W_{\ell}} \,: =\, \{\mathcal{G} \, W_\ell\}_1\,=\,b_\ell\,,
 \qquad \ell = 1, \dots, \text{rank}(\mathsf{G}) \,.
\eeq
As outlined in section \ref{mainresultsINTRO} the 
complete set of generators 
can be found from the set of generators of the ring $\mathscr{R}_{\mathsf{G}}$.
 This include in particular anti-chiral operators with conformal weights $\tfrac{p_{\ell}}{2}$, which will 
 be denoted as $\overline{W}_\ell$.
Concerning the construction of the remaining generators we propose the following strategy:
\begin{itemize}
\item[{\bf 1.}] Make an Ansatz for the remaining generators
in terms of free fields
and impose that they are  $\cN = 2$ super-Virasoro primary with the correct $(h,m)$ weights.
As in the $\cN=4$ case there is a finite number of terms in the Ansatz.
\item[{\bf 2.}]  Impose that the  VOA closes on the strong generators.
This gives both linear and non-linear conditions on the coefficients of the Ansatz.
It is practically convenient to  first solve the linear constraints.
\end{itemize}
We will present all rank $1$ examples as well as a rank $2$ example
  of this procedure in section~\ref{sec:N=2Examples}.

\subsection{Free-field realization and classical Poisson structure}
\label{sec:freefieldsclassPOISSON}

We will now show how the Poisson algebra  \eqref{wannabeHiggsBrChiralRing}, \eqref{zsPB}
 can be obtained starting from the free-field realization of $\mathcal{W}_{\Gamma}$.
Let $\mathbb{M}^{(\Gamma)}_{ \beta\gamma bc }$ denote the
free $\beta \gamma bc$ system associated to $\Gamma$, consisting
of   $\mathsf r$ copies of a single $\beta\gamma bc$ system.
Let $\mathbb{M}^{(\Gamma)\rm cl}_{\beta\gamma}$ denote the classical Poisson
algebra comprised by all polynomials
in the variables $\beta_\ell$, $\gamma_\ell$, $\ell = 1, \dots, \mathsf{r}$,
with Poisson bracket
\beq
\{ f  , g \}_{\rm PB} =\sum_{\ell = 1}^{\mathsf{r}}
 \Big[
 \partial_{\gamma_\ell} f \, \partial_{\beta_\ell} g
-\partial_{\beta_\ell} f \, \partial_{\gamma_\ell} g
\Big] \ .
\eeq
We can define a linear map
\beq\label{MAPP}
\mathcal P \; : \; \mathbb{M}^{(\Gamma)}_{ \beta\gamma bc} \; \rightarrow \;
\mathbb{M}^{(\Gamma)\rm cl}_{\beta\gamma} \ ,
\eeq
according to the following prescription. Any element of  
$\mathbb{M}^{(\Gamma)}_{ \beta\gamma bc}$ can be cast as a linear combination
of nested normal ordered products of derivatives of free fields.
The image under $\mathcal P$ of such an object is simply 
obtained by dropping all terms with derivatives and/or
fermionic free fields $b_\ell$, $c_\ell$,
and by replacing all normal-ordered products with
regular products in the 
algebra of polynomials
in the variables $\beta_\ell$, $\gamma_\ell$.
One may verify that $\mathcal P$  is well-defined.
Furthermore, the map $\mathcal P$ satisfies
\begin{align}
\mathcal P (  \{ X_1 \, X_2  \}_0 ) &= \mathcal P(X_1) \, \mathcal P(X_2)
 \ ,\nn \\
 \mathcal P (  \{ X_1 \, X_2  \}_1 ) &= \{ \mathcal P(X_1) , \mathcal P(X_2)
 \}_{\rm PB} \ .
\end{align}
On the  right hand side of the first relation, the product is the commutative
and associative product in the algebra of polynomials
in $\beta_\ell$, $\gamma_\ell$.
Since the VOA  $\cW_\Gamma$
 associated to a given Coxeter   group $\Gamma$
is realized as a subalgebra of   $\mathbb{M}^{(\Gamma)}_{ \beta\gamma bc}$,
we can apply the map $\mathcal P$ to any element of $\cW_\Gamma$,
thus defining the Poisson algebra  $\mathcal P (W_\Gamma)$.
We claim the following 
 isomorphism of 
Poisson algebras,
\beq\label{Pisomorphism}
\mathcal{P}(\mathcal{W}_{\Gamma})\,\simeq\,\mathscr{R}_{\Gamma}\,,
\eeq
where $\mathscr{R}_{\Gamma}$ is defined in 
  \eqref{wannabeHiggsBrChiralRing} and its symplectic structure is given in \eqref{zsPB}.
The explicit form of the isomorphism  \eqref{Pisomorphism}  will be given in some 
 examples in section \ref{sec:N4Examples}.
The case of complex reflection groups is identical.

In close analogy to the above discussion about the map $\cP$,
we can also define the map
 \beq\label{PprimeDEF}
 \mathcal{P}'\,:\,\mathbb{M}^{(\Gamma)}_{\beta\gamma b c}\,\rightarrow\,
 \mathbb{M}_{\beta\gamma b c}^{(\Gamma)\text{cl}} \   .
 \eeq
In the above expression, $\mathbb{M}_{\beta\gamma b c}^{(\Gamma)\text{cl}}$
denotes the Poisson superalgebra of functions of the classical
Grassmann even variables $\beta_\ell$, $\gamma_\ell$
and Grassmann odd variables $b_\ell$, $c_\ell$.
The usual properties of the Poisson bracket in a  Poisson algebra 
hold, up to the obvious modifications due to Grassmann signs.
In particular, the Poisson bracket on 
 $\mathbb{M}_{\beta\gamma b c}^{(\Gamma)\text{cl}}$
 is entirely specified by
\beq
\{ \beta_{\ell_1} , \gamma_{\ell_2} \}_{\rm PB} = - \delta_{\ell_1 \ell_2}
\ , \qquad
\{ b_{\ell_1} , c_{\ell_2} \}_{\rm PB} =  \delta_{\ell_1 \ell_2} \ .
\eeq
The map $\cP'$ is defined as follows. Given any object in
the VOA $\mathbb{M}^{(\Gamma)}_{\beta\gamma b c}$,
presented as a polynomial of normal orders of derivatives
of the free fields, its image under $\cP'$ is obtained
by setting to zero all $z$-derivatives
and by replacing the normal ordered product of the VOA
with the associative, supercommutative product in the Poisson
superalgebra  $\mathbb{M}_{\beta\gamma b c}^{(\Gamma)\text{cl}}$.
The map $\cP'$ will be useful in section \ref{sec:HL}
in relation to the discussion of the Hall-Littlewood ring.



\section{Examples of $\cN = 4$ VOA $\mathcal{W}_{\Gamma}$ }
\label{sec:N4Examples}

In this section we present the proposed free-field construction of  $\mathcal{W}_{\Gamma}$ in some examples.
 We start by reviewing the rank one case $\Gamma=A_1$ following  \cite{Adamovic:2014lra}.
 Next we present all rank two and three cases, namely $I_2(p), A_3, B_3,H_3$, and some aspects of the interesting example of $D_4$. All algebraic manipulations were performed on a laptop using the Mathematica package introduced in~\cite{Thielemans:1994er}. The analysis of higher rank Coxeter groups along the lines of this paper will require the use of more  computational power and/or packages that deal more efficiently wit $\beta \gamma bc$ free fields.

\subsection{Rank $1$: $\Gamma=A_1$}
\label{sec:A1}
The VOA associated to the Coxeter group $\Gamma =A_1$
is simply the small $\cN= 4$ SCA with central charge $c= -9$.
Our proposed free-field realization 
reduces in this case to the free-field realization
studied in \cite{Adamovic:2014lra}. All generators of the small $\cN=4$
SCA are expressed in terms of a single $\beta \gamma bc$
system. 
In this simple case, the object
$J^-_{\rm min}$ introduced in \eqref{Jmindef}
is actually sufficient to obtain
the desired free-field realization of all
the generators of the small $\cN = 4$ SCA.
For the convenience of the reader,
we summarize here all the relevant   formulae,
\begin{align}\label{allAdamovic}
J^+  & =  \beta \ , \nn \\
J^0 & = b  \, c  + 2 \, \beta  \, \gamma 
 \ , \nn \\
J^- & = 
\beta \, \gamma \, \gamma
+ \gamma \, b \, c
- \tfrac 32 \, \partial \gamma  \ , \nn \\
G^+ & = b \ , \nn \\
G^- & =  b \, \gamma  \ , \nn \\
\widetilde G^+ & = 
c  \, \partial \beta   + 2 \, \partial c  \, \beta   \ , \nn \\
\widetilde G^- &  =  
- b \, \partial c \, c
+ 2 \, \beta \, \gamma \, \partial c
+ \partial \beta \, \gamma \, c
- \tfrac 32 \, \partial^2 c  \ , \nn \\
T & =  - \tfrac 32 \, b  \, \partial c 
- \beta \, \partial \gamma
- \tfrac 12 \, \partial b \, c\ .
\end{align}
The ring $\mathscr R_{A_1}$
defined in \eqref{wannabeHiggsBrChiralRing} is 
generated by the $SL(2)$ triplet
\beq
j(y) =  j^+ + y \, j^0 + y^2 \, j^- =  z_1(y) \, z_1(y) \ , \qquad
z_1(y) = z_1^+ + y \, z_1^- \ ,
\eeq
subject to the relation
\beq
j^+ \, j^-  - \tfrac 14 \, (j^0)^2 = 0 \ .
\eeq
The corresponding composite operator in the small
$\cN = 4$ SCA is the $\mathfrak{psl}(2|2)$ primary
operator
\beq
\mathfrak L_{2,0} = (JJ)_0^0  + \tfrac 13 T
=  \tfrac 23 \, \{ J^+ \, J^- \}_0 - \tfrac 16 \, \{J^0 \, J^0\}_0
+ \tfrac 13 \partial_z J^0 + \tfrac 13 T   \ ,
\eeq
where $(JJ)_0^0$ denotes the quasiprimary completion
of the normal ordered product of two $J$'s,
projected onto the spin-0 component,
see appendix \ref{OPEandsl2cov} for more details on the notation.
It is straightforward to check that the composite
operator $\mathfrak L_{2,0}$ is identically zero
in the free-field realization \eqref{allAdamovic}.
This comes as no surprise, since it has been proven in \cite{Adamovic:2014lra}
that  \eqref{allAdamovic} is a  free-field realization 
of the simple quotient of the small $\cN = 4$ SCA at $c = -9$,
implying that all super-Virasoro descendants
of the identity operator that are null for $c = -9$
are automatically zero in the free-field realization.

It is worth recalling that in 
\cite{Adamovic:2014lra} it is also proven that the
simple quotient of the small $\cN = 4$ SCA at $c = -9$
can be characterized as the kernel of a suitable
screening operator $\mathbb S$ acting on the free $\beta \gamma bc$ system.
In order to write down the screening operator,
 we first have to express $\beta$ and $\gamma$
in terms of chiral bosons $\chi$, $\phi$,
\beq \label{beta_gamma_chiral_bosons}
\beta = e^{\chi + \phi} \ , \qquad
\gamma = \partial \chi \, e^{- \chi - \phi} \ .
\eeq
The chiral bosons non-trivial OPEs are
\beq \label{chiral_boson_OPEs}
\chi(z_1) \, \chi(z_2) = + \log z_{12} + \text{reg.} \ , \qquad
\phi(z_1) \, \phi(z_2) = - \log z_{12} + \text{reg.} \ .
\eeq
Using this notation, the screening operator  and the screening current read
\beq \label{screening_current}
\mathbb S=\int dz\,\mathsf J (z)\,,
\qquad
\mathsf J = b \, e^{- \frac 12  (\chi + \phi)} \ .
\eeq
The screening current $\mathsf J$ has conformal dimension 1 and $J^0$-eigenvalue 0.
It acts on the $\beta \gamma bc$ system via 
the order-one pole in the OPE,
\beq
X \mapsto
\mathbb S\cdot X \,=\,\{ \mathsf J \, X\}_1 \ ,
\eeq
where $X$ is any operator in the $\beta \gamma bc$ system.
It is also worth pointing out that $\mathsf J$ can be written as
a supersymmetry descendant of a chiral operator $\mathsf K$
with dimension $1/2$ and $J^0$-eigenvalue $1/2$,
\beq
\mathsf J = \{ G^- \, \mathsf K \}_1 \ , \qquad
\mathsf K = e^{\frac 12 (\chi + \phi)} \ .
\eeq

\subsection{Rank $2$: $\Gamma=I_2(p)$}
\label{I2p}

This section is devoted to a detailed description of
the free-field realization of the $\cN = 4$ VOA associated to the
Coxeter group $I_2(p)$, $p\ge 3$.

\subsubsection{The Coxeter groups  $I_2(p)$ and associated rings}
The group $I_2(p)$ is the symmetry group of the regular
$p$-gon on the plane.
As a Coxeter group,
it is the subgroup of $O(2, \mathbb R)$
generated by two reflections with respect to two lines forming
an angle $\pi/p$. Equivalently, we may regard it as generated by a
reflection $\sigma$ and a rotation $\rho$ by an angle $2\pi/p$,
\beq
\sigma = \begin{pmatrix}
1 &  0 \\
0 & -1
\end{pmatrix} \ , \qquad
\rho  = \begin{pmatrix}
\cos \frac{2\pi}{p} & - \sin \frac{2\pi}{p} \\
\sin \frac{2\pi}{p} & \cos \frac{2\pi}{p}
\end{pmatrix} \ .
\eeq
The action on $I_2(p)$ on $\mathbb R^2$ is extended naturally to
$\mathbb C^2$. It is then convenient to perform a change of basis,
and introduce coordinates $z^{1,2}$ in $\mathbb C^2$
such that
\beq \label{I2p_action}
\sigma \; : \; \begin{pmatrix}
z_1 \\ z_2
\end{pmatrix} \; \mapsto \;
\begin{pmatrix}
z_2 \\ z_1
\end{pmatrix} \ , \qquad
\rho \; : \; \begin{pmatrix}
z_1 \\ z_2
\end{pmatrix} \; \mapsto \;
\begin{pmatrix}
e^{+2\pi i/p} \, z_1 \\ e^{-2\pi i /p } \, z_2
\end{pmatrix} \ .
\eeq
The original space $\mathbb R^2 \subset \mathbb C^2$ is recovered
via the reality condition $z_2 = (z_1)^*$.
In terms of $z_{1,2}$ the invariants of $I_2(p)$ take the simple form
\beq \label{Coulomb_I2p_invariants}
\cI_2  = z_1 \, z_2 \ , \qquad \cI_p =  z_1 ^p +  z_2 ^p \ .
\eeq
The Coxeter group $I_2(p)$ has the property of being crystallographic only for $p = 3,4,6$.
This is equivalent to the fact that the plane $\mathbb{R}^2$ admits a 
tessellations   by regular $p$-gons only for $p = 3,4,6$, i.e.~triangles, squares and hexagons.
 These are also the value for which 
it coincides with a Weyl group, namely 
\beq\label{I2pisWeyl}
I_2(3) \cong \text{Weyl}(\mathfrak a_2) \ , \qquad
I_2(4) \cong\text{Weyl}(\mathfrak b_2) \cong \text{Weyl}(\mathfrak c_2) \ , \qquad
I_2(6) \cong \text{Weyl}(\mathfrak g_2) \ .
\eeq

Let us describe the ring $\mathscr{R}_{I_2(p)}$ defined in \eqref{wannabeHiggsBrChiralRing}.
The set of generators in this case is obtained by promoting the generators \eqref{Coulomb_I2p_invariants}
to $SL(2)$ multiplets
\beq\label{jwgenerators}
j(y)=z_1(y)z_2(y)\,,
\qquad
w(y)=z_1(y)^p+z_2(y)^p\,,
\eeq
where $z_i(y)=z_i^++y\,z_i^-$.
While \eqref{Coulomb_I2p_invariants} are algebraically independent the generators $j(y)$ and $w(y)$
satisfy the following relations
\beq\label{I2prelations}
(j\,w\,)\big|_{\frac{p}{2}-1}=0\,,
\qquad
(w\,w)\big|_{p-2m}+c_{p,m} \,(j^2\big|_0)^m\,j^{p-2m}\big|_{p-2m}=0
\eeq
$m=1,\dots,\left \lfloor{\tfrac{p}{2}}\right \rfloor$ and the notation $\big{|}_*$ denotes the projection 
onto the $SL(2)$ spin $*$ component. 
Using the same  normalization of $SL(2)$  projections 
as in appendix \ref{OPEandsl2cov},
the coefficients $c_{p,m}$ take the form
\beq
c_{p,m}=-\frac{2\left(\tfrac{3}{2}\right)^m\,
\left(\Gamma(p+1)\right)^2\Gamma(m-p-\tfrac{1}{2})
}{
\Gamma(2m+1)\Gamma(1-2m+p)\Gamma(1-m+p)\Gamma(2m-p-\tfrac{1}{2})
} \,.
\eeq
We have obtained an alternative description of the ring $\mathscr{R}_{I_2(p)}$ given in \eqref{wannabeHiggsBrChiralRing} as the ring generates by the $y$-components of \eqref{jwgenerators}
subject to the relations \eqref{I2prelations}.
 As we will show shortly, these relations are the image of null states 
 in the chiral algebra $\mathcal{W}_{I_2(p)}$.

Finally, let us comment on the Poisson algebra structure of $\mathscr{R}_{I_2(p)}$.
It is given by \eqref{zsPB} with $\eta= \begin{psmallmatrix}0&-1\\-1&0\end{psmallmatrix}$. This choice of $\eta$ ensures
\beq
\{ j(y_1) \,, \, j(y_2) \}_{\rm PB} = 2 \, y_{12} \, \big( 1 + \tfrac 12 \, y_{12} \, \partial_{y_2}  \big) j(y_2) \ ,
\eeq
which is the Poisson counterpart of the $JJ$ OPE,
with the same normalization conventions.

\subsubsection{Free-field realization of all generators}
We will now apply the procedure outlined in section \ref{sec:free_fields} to
obtain a free-field realization of the  VOA $\mathcal{W}_{I_2(p)}$.
Since $I_2(p)$ has rank 2, we need two copies of the $\beta\gamma bc$
system, denoted $\beta_1$, $\beta_2$, and so on. We associate the label
$1$ to the invariant of degree $2$,
\beq
p_1 = 2 \ , \qquad
p_2 = p \ .
\eeq
As explained in section \ref{sec:remainingGENN4} the only quantity that has to be constructed is
$J^-_{\text{norms}}$ in  \eqref{JisJminplusJnorms}. Following the steps given below \eqref{JpGpdoublet}
one obtains the unique solution
\beq\label{Jminus_norms_for_I2p}
J^- _{\text{norms}} =\Lambda\,(\beta_1)^{p-2}\,\gamma_2
\left(\beta_1\gamma_2+(p-1)b_1c_2\right)\,. 
\eeq
The normalization $\Lambda$ could be scaled to one by the transformation \eqref{severalU1sonbetagammas} but it is instructive to keep it as a parameter.
The remaining generators of the small $\cN = 4$ SCA take the form given in 
\eqref{N2inN4preliminaries},
\eqref{N2freefireld} and  \eqref{JpGpdoublet} together with $\widetilde{G}^-=\{G^-\,J^-\}_{1}$.
For convenience of the reader, we summarize here the expression
of all generators of the small $\cN = 4$ SCA in terms of the
two $\beta \gamma bc$ systems:
\begin{align}\label{allVIrpartW2p}
J^+  & =  \beta_1 \ , \nn \\
J^0 & = b_1 \, c_1 + 2 \, \beta_1 \, \gamma_1
+ (p-1) \, b_2 \, c_2 + p \, \beta_2 \, \gamma_2 \ , \nn \\
J^- & = 
b_1 \, c_1 \, \gamma_1
+ \beta_1 \, \gamma_1 \, \gamma_1
+(p-1) \, \gamma_1 \, b_2 \, c_2
+ p \, \gamma_1 \, \beta_2 \, \gamma_2
-  c_1 \, b_2 \, \gamma_2
- (p+1) \, \partial \gamma_1 \nn \\
&
+ \Lambda  \,  (\beta_1)^{p-1} \, ( \gamma_2)^2 
+ (p-1) \, \Lambda \, b_1 \, (\beta_1)^{p-2} \, c_2 \, \gamma_2 \ , \nn \\
G^+ & = b_1 \ , \nn \\
G^- & =  b_1 \, \gamma_1 + b_2 \, \gamma_2 \ , \nn \\
\widetilde G^+ & = c_1 \, \partial \beta_1  + 2 \, \partial c_1 \, \beta_1
+ (p-1) \, c_2 \, \partial \beta_2 + p \, \partial c_2 \, \beta_2 \ , \nn \\
\widetilde G^- &  = 
- b_1 \, \partial c_1 \, c_1
- c_1 \, b_2 \, \partial c_2
+ c_1 \, \partial \beta_1 \, \gamma_1
+ c_1 \, \partial \beta_2 \, \gamma_2
+ (p-1) \, \partial c_1 \, b_2 \, c_2
+ 2 \, \partial c_1 \, \beta_1 \, \gamma_1
+ p \, \partial c_1 \, \beta_2 \, \gamma_2 \nn \\
& + (p-1) \, \gamma_1 \, c_2 \, \partial \beta_2
+ p \, \gamma_1 \, \partial c_2 \,   \beta_2
- (p+1) \, \partial^2 c_1 \nn \\
& - (p-1) \, \Lambda \,  b_1 \, (\beta_1)^{p-2} \,
 \partial c_2 \, c_2
+ 2 \, \Lambda\,   (\beta_1)^{p-1} \, \partial c_2 \, \gamma_2
+ (p-1) \, \Lambda  \, \partial \beta_1 \, 
(\beta_1)^{p-2} \, c_2 \, \gamma_2 \ , \nn \\
T & =  - \frac 32 \, b_1 \, \partial c_1
 - \frac 12 \, \partial b_1 \, c_1 
- \beta_1 \, \partial \gamma_1
 - \frac{p+1}{2} \, b_2 \, \partial c_2
  - \frac{p-1}{2} \, \partial b_2 \,   c_2
 - \frac p2 \, \beta_2 \, \partial \gamma_2
  - \left(\frac p2 -1 \right)\,  \partial \beta_2 
  \,   \gamma_2 \ .
\end{align}
Given the small $\cN = 4$ SCA in terms of free fields as above, 
we  proceed building the additional short generator $W$ 
of the VOA associated to $I_2(p)$.
 According to
our general prescription, we simply set
\beq
W^{\rm h.w.} = \beta_2 \ ,
\eeq  
where h.w.~stand for highest weight and
refers to the component of $W$ with charges $h=m=p/2$.
The whole $\mathfrak{psl}(2|2)$ short supermultiplet
 $\mathbb{W}=\{W,G_W,\widetilde{G}_W,T_W\}$,
see appendix \ref{OPEandsl2cov} for our notation,
 is
generated from $W^{\rm h.w.}$.
We have now entirely specified our free-field
realization. We refrain from giving the 
expressions for the other components of $W$,
 since their complexity grows  quickly.

The next step is to verify that the set of strong generators of the
 VOA $\mathcal{W}_{I_2(p)}$ is given by  $\mathbb{W}=\{W,G_W,\widetilde{G}_W,T_W\}$
 together with  the generators of the small $\mathcal{N}=4$ super-Virasoro algebra 
 $\mathbb{J}=\{J,G,\widetilde{G}, T\}$. To do so we need to close the OPE on this set of generators.
 The $\mathbb{J}$-$\mathbb{J}$ and
$\mathbb{J}$-$\mathbb{W}$ OPEs
take the required form by construction.
The 
   $\mathbb{W}$-$\mathbb{W}$ OPEs
are fixed in terms of the $WW$ OPE.
  By means of a direct computation we verified that  
  \beq
  \label{WWOPEI2p}
  W\,\times\,W\sim\,g_{WW}\,\,[\text{id}]\,,
  \qquad
  g_{WW} = \frac{(2p)!}{p! \, p^2}  \, \Lambda \,,
  \eeq
  where the notation $[\text{id}]$ 
stands for the small $\cN =4$ super-Virasoro
family of the identity operator. The first few terms are given by
  \beq\label{idPIECEN4}
[\text{id}]
\,=\,
\text{id} -\tfrac{6p}{c}\, (J -\tfrac{1}{6} T)
+\tfrac{18 p(p-1)}{c(c-6)}\, (JJ)^2_0
-\tfrac{9p(p+1)}{c(c+9)}\,((JJ)^0_0+ \tfrac 13 T)+
\dots\
  \eeq  
    with $c = -6(p+1)$.
  It is worth remarking that the stress tensor $T$
  appears as $\mathfrak{psl}(2|2)$ descendants of $J$
  and as completion of $(JJ)^0_0$ to a  $\mathfrak{psl}(2|2)$  primary.
 Note that all factors $z_{12}$, $y_{12}$,
as well as all $\partial_z$, $\partial_y$ operators,
are implicit, since they can be unambiguously
restored exploiting $\mathfrak{sl}(2)$
covariance. This compact notation for OPEs is
described in more detail in appendix \ref{OPEandsl2cov},
and is also utilized below in other examples.
The formula \eqref{WWOPEI2p}
has been tested explicitly up to $p=7$.
As anticipated, the free parameter $\Lambda$ entering \eqref{Jminus_norms_for_I2p} 
is crucial in order to obtain a viable realization of the full VOA.

\subsubsection{Null states}
Our proposed free-field realization is conjectured to 
enjoy the highly non-trivial property that all null states in the
abstract VOA are realized manifestly as zero.
In this section we check  that the null states in the VOA $\mathcal{W}_{I_2(p)}$
associated to the   ``Higgs Branch" relations  \eqref{I2prelations} are identically zero in the 
free field realization. We expect that these nulls generated the maximal ideal of $\mathcal{W}_{I_2(p)}$,
but at the moment we do not have a complete proof of this fact.

To begin with, let us consider the long composite operator linear in the
extra generator $W$,
\beq \label{W_linear_null}
\mathfrak L_{\frac p2 + 1, \frac p2 -1} = (J \, W)_0^{\frac p2 -1} + \frac{1}{p+1} \,
T_W \ .
\eeq
The first term  in the expression above is the quasiprimary 
completion of the normal ordered product $J W$  projected 
on the $\mathfrak{sl}(2)_y$ spin $\frac p2 -1$ component.
The second term contains
the supersymmetry descendant of $W$, namely
$T_W = - G^\downarrow \, \widetilde G^\downarrow \, W$, see \eqref{SUSYdescNOT} for the notation.
This term is necessary in order for \eqref{W_linear_null} to be a $\mathfrak{psl}(2|2)$ primary.
Being linear in the new generator $W$, the operator  $\mathfrak L_{\frac p2 + 1, \frac p2 -1}$
can be defined for any value of the central charge without any reference to the 
free field realization and by construction is a $\cN =4$ super-Virasoro 
descendant of $W$ itself. 
If the central charge takes the special value $c= -6(p+1)$ the operator
$\mathfrak L_{\frac p2 + 1, \frac p2 -1}$  becomes an $\cN =4$ super-Virasoro primary operator.
Being a primary and a descendant at the same time it must be null.
In  our free-field realization the operator $\mathfrak L_{\frac p2 + 1, \frac p2 -1}$ vanishes identically.
We have thus recovered the VOA counterpart of the first  ``Higgs Branch" relation in \eqref{I2prelations}.
Let us turn to the second set of relations in  \eqref{I2prelations}.
Let us consider the composite operators
\beq \label{WW_nulls}
\mathfrak L^{WW}_{p, j} = (W \,W)_0^{j} + \text{(super-Virasoro primary completion)} \ , \quad j < p \ , \quad 
\text{$p -j$ even} \ .
\eeq
In other words, $\mathfrak L^{WW}_{p, j}$ is defined to be the
$\cN = 4$ super-Virasoro primary completion of the normal ordered product 
$WW$
projected onto the component with $\mathfrak{sl}(2)_y$ spin $j$.
The requirement that $p-j$ be even stems from Bose symmetry.
One can verify that this definition is well-posed, in the sense that,
making only use of the OPEs of the abstract VOA, one can check
that there exists a unique super-Virasoro primary operator
starting with $(W \,W)_0^{j}$.
Having unambiguously defined the composite 
$\mathfrak L^{WW}_{p, j}$ in the abstract VOA,
we can resort to our free-field realization and verify, in a few examples,
that this object is indeed identically vanishing.
This finding is in perfect agreement with the bootstrap analysis of 
\cite{bootstrappaper}.

\subsubsection{Classical limit: relation between $\beta\gamma$ and $z^{\pm}$}
Using the map $\mathcal P$ defined in section \ref{sec:freefieldsclassPOISSON},
we can define the classical objects
associated to the generators $J$ and $W$,
\beq\label{JclWcl}
J_{\rm cl} := \mathcal P(J) \ , \qquad
W_{\rm cl} := \mathcal P(W) \ ,
\eeq
more explicitly
\beq
J_{\rm cl}^+=\beta_1\,,
\quad 
J_{\rm cl}^0=2\beta_1\gamma_1+p\beta_2\gamma_2\,,
\quad 
J_{\rm cl}^-=(\beta_1\gamma_1+p\,\beta_2\gamma_2)\gamma_1+
\Lambda\,(\beta_1)^{p-1}(\gamma_2)^2\,,
\eeq
and $W_{\rm cl}=\beta_2+\text{descendants}$. 
Recall that, as explained in  section \ref{sec:freefieldsclassPOISSON},
 $\beta \gamma$ are now commuting variables. 
The combinations \eqref{JclWcl} satisfy the same relations  
\eqref{I2prelations} as
\eqref{jwgenerators}.
This implies that 
$J_{\rm cl},
W_{\rm cl}$ 
provide a  realization of the ring $\mathscr{R}_{I_2(p)}$
as a subring of $\mathbb{C}[\beta_1,\gamma_1,\beta_2,\gamma_2]$.
It is instructive to determine 
$\beta_{1,2}$, $\gamma_{1,2}$ in terms of the quotient variables
$z^{\pm}_{1,2}$ by equating \eqref{jwgenerators} with \eqref{JclWcl}.
This gives the remarkably simple expressions
\begin{align} \label{canonical_transf}
\beta_1 &= 
\, z_1^+ \, z_2^+ \ , &
\gamma_1 &= \frac{(z_1^+)^{p-1} \, z_1^- - (z_2^+)^{p-1} \, z_2^-}{(z_1^+)^p
- (z_2^+)^p} \ , \nn \\
\beta_2 &= (z_1^+)^p
+ (z_2^+)^p\ , &
\gamma_2 & =\frac{1}{p}\, \frac{
z_1^+ \, z_2^- - z_2^+ \, z_1^-
 }{   (z_1^+)^p
- (z_2^+)^p} \ , &
\Lambda & = p^2  \ .
\end{align}
Notice that $\beta_{1,2}$, $\gamma_{1,2}$
are rational functions that are invariant under the action \eqref{I2p_action}
of  $I_2(p)$. The Poisson brackets \eqref{zsPB} with 
$\eta= \begin{psmallmatrix}0&-1\\-1&0\end{psmallmatrix}$
imply the expected Poisson brackets
\beq
\{ \beta_{\ell_1} \ , \gamma_{\ell_2} \}_{\rm PB} =- \delta_{\ell_1,\ell_2} \ , \qquad 
\{ \beta_{\ell_1} \ , \beta_{\ell_2} \}_{\rm PB}
 =\{ \gamma_{\ell_1} \ , \gamma_{\ell_2} \}_{\rm PB} =0\,,
\qquad
\ell_1,\ell_2 =1,2 \ .
\eeq
The minus sign in the first equation is a consequence
of our conventions for the $\beta\gamma$ OPEs,
see \eqref{defberagammbc}.


\subsubsection{Comments on the screening operator}

It is natural
to ask if $\cW_{I_2(p)}$ can be identified with the kernel of a
suitable screening operator acting on the free field VOA
 $\mathbb M^{(I_2(p))}_{\beta \gamma bc}$.
A simpler version of this problem 
is obtained using the map $\cP'$ of section
\ref{sec:freefieldsclassPOISSON}.
More precisely, we aim at identifying $\cP'(\cW_{I_2(p)})$
with the kernel of a suitable object $\mathsf J_{\rm cl}$ in 
the classical Poisson superalgebra 
$\mathbb M^{(I_2(p))\rm cl}_{\beta \gamma bc}$.
The object $\mathsf J_{\rm cl}$ acts 
via Poisson bracket.

The object  $\mathsf J_{\rm cl}$
can be presented as
\beq\label{JclfromK}
\mathsf J_{\rm cl} =
 b_1 \, (\beta_1)^{- \frac 12}
\,
\Big[  \tfrac 12  \, F(x) - \tfrac 12 \, p \, x \, F'(x)
\Big]
+ b_2 \, (\beta_1)^{\frac 12 - \frac p2} \, F'(x)
= \{ \cP'(G^-) , \mathsf K_{\rm cl} \}_{\rm PB} \ ,
\eeq
where in the last step we introduced the auxiliary object
\beq \label{K_classical}
\mathsf K_{\rm cl} = (\beta_1)^{\frac 12} \, F(x) \ ,  \qquad
x := (\beta_1)^{-\frac p2} \, \beta_2 \ .
\eeq
The function $F(x)$
is required to be a solution to the differential equation
\beq
\label{diffeqscreening}
\left(p^2 \, x^2-4 \, \Lambda \right)
 F''(x)+p^2\, x \,
  F'(x) -  F(x)  
   = 0 \ .
\eeq
We have checked that, by virtue of the above equation,
one has
\beq
\{ \mathsf J_{\rm cl} , \cP'(X) \}_{\rm PB} = 0 \qquad
\text{for} \qquad
X \in \{ J^+, J^0, J^-, G^+, G^-, \widetilde G^+, \widetilde G^-, T, W^{\rm h.w.} \} \ .
\eeq
This is enough to guarantee that  $\cP'(\cW_{I_2(p)})$
lies inside the kernel of $\mathsf J_{\rm cl}$ acting of the Poisson
superalgebra  
$\mathbb M^{(I_2(p))\rm cl}_{\beta \gamma bc}$.
It seems natural to conjecture that 
$\cP'(\cW_{I_2(p)})$
is actually the entirety of the kernel of  $\mathsf J_{\rm cl}$,
but we do not have a proof of this fact.

In order to promote the results of the previous paragraphs
from the level of the Poisson algebra to the level of the full VOA,
we have to be able to make sense of expressions
like \eqref{K_classical} in the context of the VOA.
This is possible by expressing $\beta_1$, $\beta_2$
in terms of chiral bosons and making use of vertex operators.
We refrain, however, from pursuing this direction further.

\subsection{Rank $3$:  $\Gamma=A_3, B_3, H_3$}

We will now present the free field construction of $\mathcal{W}_{\Gamma}$
 for $\text{rank}(\Gamma)=3$.
There are only three examples in this case, namely  $\Gamma=A_3, B_3, H_3$.
The structure of these VOA is analyzed in less details compared to the rank $2$
 series discussed in the previous section.
A few remarks are in order. 
Notice that in all three cases the degrees are $(2,3,4)$, $(2,4,6)$, $(2,6,10)$
so that $p_3=2p_2-2$.
Moreover all super-Virasoro primary operators of the form
$(W_{\ell_1}W_{\ell_2}+\dots)_{\mathfrak{L}}$ are null except for 
$(W_{2}W_{3}+\dots)_{\mathfrak{L}_{(\frac{p_2+p_3}{2},\frac{p_2+p_3}{2}-1)}}$.
Finally, as in the rank two series, 
the classical limit of  $J,W_{2},W_{3}$ obtained by applying the map $\mathcal{P}$
defined in section \ref{sec:freefieldsclassPOISSON},
gives a  realization of the rings $\mathscr{R}_{A_3}$,
$\mathscr{R}_{B_3}$
 and $\mathscr{R}_{H_3}$
as subrings of $\mathbb{C}[\beta_1,\gamma_1,\beta_2,\gamma_2,\beta_3,\gamma_3]$.
This properties is not manifest but has been checked by verifying that all the relations 
 are satisfied\footnote{
 Alternatively one can find the analogues 
  of \eqref{canonical_transf} for $\Gamma=A_3, B_3, H_3$.
 }.

\vspace{0.3cm}
\noindent
\emph{Notation:} In order to make the equations easier to read we will label 
 $W$ generators as well as $\beta \gamma b c$ by their weight with the gothic suffix  
 $\mathfrak{p}\in \{\mathfrak{3},\mathfrak{4},\mathfrak{5},\mathfrak{6},\mathfrak{7},\dots\}$.

\subsubsection{Example:  $\Gamma=A_3$}

\paragraph{Free field realization}
By following the steps described in section \ref{sec:free_fields} we find a unique solution for 
$J^-_{\text{norms}}$ up to the rescaling \eqref{severalU1sonbetagammas}. 
Its explicit form is rather long so we present only its classical limit:
\beq
\mathcal{P}(J^-_{\text{norms}})=
\Lambda_2 \big(\Lambda_1\beta_{\mathfrak{2}}^2-\beta_{\mathfrak{4}}\big)\gamma_{\mathfrak{3}}^2
-\tfrac{16\,\Lambda_1}{3}\,\beta_{\mathfrak{2}}\beta_{\mathfrak{3}}
\gamma_{\mathfrak{3}}\gamma_{\mathfrak{4}}+
\tfrac{\Lambda_1}{12\Lambda_2}
 \big(20\Lambda_1\Lambda_2\, \beta_{\mathfrak{2}}^3
 +51\, \beta_{\mathfrak{3}}^2
 +28\Lambda_2\, \beta_{\mathfrak{2}}\beta_{\mathfrak{4}}
  \big)\gamma_{\mathfrak{4}}^2\,,
\eeq
where $\mathcal P$ is defined in section \ref{sec:freefieldsclassPOISSON}.
Given $J^-_{\text{norms}}$ we can construct  the remaining 
strong
generators as descendants of $\beta_{\mathfrak{2}}$ and $\beta_{\mathfrak{3}}$.
The parameters $\Lambda_1, \Lambda_2$ are related to the normalization appearing below as
$g_{\mathfrak{3}\mathfrak{3}}=\tfrac{85}{2} \Lambda_1 \Lambda_2$, 
$g_{\mathfrak{4}\mathfrak{4}}=595 \Lambda_1^2$.

\paragraph{Closing the OPE.}
Let us present the OPE of strong generators in this case:
\begin{subequations}
\begin{align}
W_{\mathfrak{3}}\,\times\,W_{\mathfrak{3}}
&\sim
g_{\mathfrak{3}\mathfrak{3}}\,[\text{id}]+
\lambda_{\mathfrak{3}\mathfrak{3}}^{\mathfrak{4}}\,
[W_{\mathfrak{4}}]\\
W_{\mathfrak{3}}\,\times\,W_{\mathfrak{4}}
&\sim
\,\lambda_{\mathfrak{3}\mathfrak{4}}^{
\mathfrak{3}}\,[W_{\mathfrak{3}}]\\
\label{OPEW4W4intypeA3}
W_{\mathfrak{4}}\,\times\,W_{\mathfrak{4}}
&\sim
g_{\mathfrak{4}\mathfrak{4}}\,[\text{id}]+
\lambda_{\mathfrak{4}\mathfrak{4}}^{\mathfrak{4}}\,[W_{\mathfrak{4}}]+
\lambda_{\mathfrak{4}\mathfrak{4}}^{(\mathfrak{3}\mathfrak{3})}\,
[(W_{\mathfrak{3}})^2_{\mathfrak{S}}]
\end{align}
\end{subequations}
where 
\beq
\tfrac{
\sqrt{g_{\mathfrak{4}\mathfrak{4}}}}{g_{\mathfrak{3}\mathfrak{3}}}\,
\lambda_{\mathfrak{3}\mathfrak{3}}^{\mathfrak{4}}=
\tfrac{1}{\sqrt{g_{\mathfrak{4}\mathfrak{4}}}}\,
\lambda_{\mathfrak{3}\mathfrak{4}}^{\mathfrak{3}}=
-4\sqrt{\tfrac{7}{85}}\,,
\qquad
\tfrac{1}{\sqrt{g_{\mathfrak{4}\mathfrak{4}}}}\,
\lambda_{\mathfrak{4}\mathfrak{4}}^{\mathfrak{4}}=
\tfrac{11}{3}\sqrt{\tfrac{5}{119}}\,,
\qquad
\tfrac{
g_{\mathfrak{3}\mathfrak{3}}}{g_{\mathfrak{4}\mathfrak{4}}}
\lambda_{\mathfrak{4}\mathfrak{4}}^{(\mathfrak{3}\mathfrak{3})}=
\tfrac{17}{28}\,,
\eeq
and
\beq
\label{W3W3composite_A3}
(W_{\mathfrak{3}})^2_{\mathfrak{S}}\,=\,
(\beta_{\mathfrak{3}})^2
-\tfrac{8 \,g_{\mathfrak{3}\mathfrak{3}}}{23\sqrt{ g_{\mathfrak{4}\mathfrak{4}}}}
\sqrt{\tfrac{7}{85}}\,\beta_{\mathfrak{2}}\beta_{\mathfrak{4}}
+\tfrac{8 \,g_{\mathfrak{3}\mathfrak{3}}}{4845}
(\beta_{\mathfrak{2}})^3
+\text{descendants}\,,
\eeq
is the completion of $(\beta_{\mathfrak{3}})^2$ to a super-Virasoro primary.
Its norm is given by $\tfrac{1344}{437}g^2_{\mathfrak{3}\mathfrak{3}}$.
As usual,
 $[X]$ denotes the contribution from the $\mathcal{N}=4$ super-Virasoro family of the 
primary  $X$.
 Notice that there  is a null state of type $\mathfrak{L}_{(3,1)}$ 
of the schematic form $W_{\mathfrak{3}}W_{\mathfrak{3}}+J W_{\mathfrak{4}}+J^3+\dots$.
This is a primary operator that, if not null, could  appear in the 
 right hand side of  \eqref{OPEW4W4intypeA3}.
The quantum numbers of the relations among the generators of $\mathscr{R}_{A_3}$ are
\beq
\mathfrak{L}_{(3,1)}\,,
\quad
\mathfrak{L}_{(\frac{7}{2},\frac{3}{2})}\,,
\quad
\mathfrak{L}_{(\frac{7}{2},\frac{1}{2})}\,,
\qquad
\mathfrak{L}_{(4,2)}\,,
\quad
\mathfrak{L}_{(4,0)}\,.
\eeq
They all  correspond to  null operators in the VOA.
Notice that all  the operators of the type
$(W_{\mathfrak{p}_1}W_{\mathfrak{p}_2}+\dots)_{\mathfrak{L}}$ are null except for 
$(W_{\mathfrak{3}}W_{\mathfrak{4}}+\dots)_{\mathfrak{L}_{(\frac{7}{2},\frac{5}{2})}}$.

\subsubsection{Example:  $\Gamma=B_3$}

\paragraph{Free field realization}
As before, 
following the recipe given in   section \ref{sec:free_fields} we find a unique solution for 
$J^-_{\text{norms}}$. In the  classical limit it reads
\begin{subequations}
\begin{align}
\mathcal{P}(J^-_{\text{norms}})=\,\,&
\big( 
u_1\,\beta_{\mathfrak{2}}^3
+u_2\,\beta_{\mathfrak{2}}\beta_{\mathfrak{4}}
+u_3\,\beta_{\mathfrak{6}}
\big)\gamma_{\mathfrak{4}}^2+\\
+&\big( 
u_4\,\beta_{\mathfrak{2}}^4
+u_5\,\beta_{\mathfrak{2}}^2\beta_{\mathfrak{4}}
+u_6\,\beta_{\mathfrak{4}}^2+
u_7\,\beta_{\mathfrak{2}}\beta_{\mathfrak{6}}
\big)\gamma_{\mathfrak{4}}\,\gamma_{\mathfrak{6}}+\\
+&\big( 
u_8\,\beta_{\mathfrak{2}}^5
+u_9\,\beta_{\mathfrak{2}}^3\beta_{\mathfrak{4}}
+u_{10}\,\beta_{\mathfrak{2}}\beta_{\mathfrak{4}}^2+
u_{11}\,\beta_{\mathfrak{2}}^2\beta_{\mathfrak{6}}
+u_{12}\,\beta_{\mathfrak{4}}\beta_{\mathfrak{6}}
\big)\gamma_{\mathfrak{6}}^2
\end{align}
\end{subequations}
The explicit form of the coefficients $u_k$ is not very illuminating so we omit it.
With this ingredient we can produce all the strong generators.

\paragraph{Closing the OPE.}
Let us present the OPE of strong generators in this case, setting the normalizations to one,
\begin{subequations}
\begin{align}
W_{\mathfrak{4}}\,\times\,W_{\mathfrak{4}}
&\sim
[\text{id}]+
\lambda_{\mathfrak{4}\mathfrak{4}}^{\mathfrak{4}}\,
[W_{\mathfrak{4}}]
+
\lambda_{\mathfrak{4}\mathfrak{4}}^{\mathfrak{6}}\,
[W_{\mathfrak{6}}]
\\
\label{OPEW4W6intypeB3}
W_{\mathfrak{4}}\,\times\,W_{\mathfrak{6}}
&\sim
\,\lambda_{\mathfrak{4}\mathfrak{6}}^{\mathfrak{4}}
\,[W_{\mathfrak{4}}]+
\lambda_{\mathfrak{4}\mathfrak{6}}^{\mathfrak{6}}
\,[W_{\mathfrak{6}}]+
\lambda_{\mathfrak{4}\mathfrak{6}}^{(\mathfrak{4}\mathfrak{4})}\,
[(W_{\mathfrak{4}})^2_{\mathfrak{S}}]
\\
\label{OPEW6W6intypeB3}
W_{\mathfrak{6}}\,\times\,W_{\mathfrak{6}}
&\sim
[\text{id}]+
\lambda_{\mathfrak{6}\mathfrak{6}}^{\mathfrak{4}}
\,[W_{\mathfrak{4}}]
+
\lambda_{\mathfrak{6}\mathfrak{6}}^{\mathfrak{6}}
\,[W_{\mathfrak{6}}]
+
\lambda_{\mathfrak{6}\mathfrak{6}}^{(\mathfrak{4}\mathfrak{4})}\,
[(W_{\mathfrak{4}})^2_{\mathfrak{S}}]
+
\lambda_{\mathfrak{6}\mathfrak{6}}^{(\mathfrak{6}\mathfrak{4})}\,
[(W_{\mathfrak{6}}W_{\mathfrak{4}})_{\mathfrak{S}}]
\end{align}
\end{subequations}
where
\beq
\lambda_{\mathfrak{4}\mathfrak{4}}^{\mathfrak{4}}
= -\tfrac{143}{3\sqrt{2415}}\,,
\quad
\lambda_{\mathfrak{4}\mathfrak{4}}^{\mathfrak{6}}
=
\lambda_{\mathfrak{4}\mathfrak{6}}^{\mathfrak{4}}
= \tfrac{46}{5}\sqrt{\tfrac{11}{609}}\,,
\quad 
\lambda_{\mathfrak{4}\mathfrak{6}}^{(\mathfrak{4}\mathfrak{4})}=
-12\sqrt{\tfrac{5}{7337}}\,,
\eeq
\beq
\lambda_{\mathfrak{4}\mathfrak{6}}^{\mathfrak{6}}=
\lambda_{\mathfrak{6}\mathfrak{6}}^{\mathfrak{4}}=
\tfrac{65}{58}\sqrt{\tfrac{35}{69}}\,,
\qquad
\lambda_{\mathfrak{6}\mathfrak{6}}^{\mathfrak{6}}=
-\tfrac{27}{29}\sqrt{\tfrac{21}{319}}\,,
\lambda_{\mathfrak{6}\mathfrak{6}}^{(\mathfrak{4}\mathfrak{4})}=\tfrac{285}{319}\,,
\qquad
\lambda_{\mathfrak{4}\mathfrak{4}}^{(\mathfrak{6}\mathfrak{4})}=2\sqrt{\tfrac{55}{667}}\,,
\eeq
and $(W_{\mathfrak{4}})^2_{\mathfrak{S}}$, $(W_{\mathfrak{6}}W_{\mathfrak{4}})_{\mathfrak{S}}$
are defined in a similar way to \eqref{W3W3composite_A3} so that they are $\mathcal{N}=4$
super-Virasoro primaries.
Notice that there are nulls of type $\mathfrak{L}_{(4,2)}$ and  $\mathfrak{L}_{(4,0)}$
which are the unique super-Virasoro  primary completions of $(W_{\mathfrak{4}})^2_{\mathfrak{L}}$.
These are  primary operators that, if not null, could  appear in the 
 right hand side of  \eqref{OPEW4W6intypeB3}.
Similarly,  there are nulls of type $\mathfrak{L}_{(5,3)}$, $\mathfrak{L}_{(5,2)}$,  $\mathfrak{L}_{(5,1)}$ 
relevant for the OPE \eqref{OPEW6W6intypeB3} of the schematic form 
$(W_{\mathfrak{4}}W^{}_{\mathfrak{6}})_{\mathfrak{L}}$.
The quantum numbers of the relations are
\beq
\mathfrak{L}_{(4,2)}\,,
\quad
\mathfrak{L}_{(4,0)}\,,
\quad
\mathfrak{L}_{(5,3)}\,,
\quad
\mathfrak{L}_{(5,2)}\,,
\quad
\mathfrak{L}_{(5,1)}\,,
\quad
\mathfrak{L}_{(6,4)}\,,
\quad
\mathfrak{L}_{(6,2)}\,,
\quad
\mathfrak{L}_{(6,0)}\,.
\eeq
As in the case of $A_3$ 
all  the super-Virasoro primary operators of the form 
$(W_{\mathfrak{p}_1}W_{\mathfrak{p}_2}+\dots)_{\mathfrak{L}}$ are null except for 
$(W_{\mathfrak{6}}W_{\mathfrak{4}}+\dots)_{\mathfrak{L}_{(5,3)}}$.

\subsubsection{Example:  $\Gamma=H_3$}

\paragraph{Free field realization}
Also in this case  the procedure outlined in   section \ref{sec:free_fields} gives a unique solution for 
$J^-_{\text{norms}}$. In the  classical limit it reads
\begin{subequations}
\begin{align}
\mathcal{P}(J^-_{\text{norms}})=\,\,&
\big( 
u_1\,\beta_{\mathfrak{2}}^5
+u_2\,\beta_{\mathfrak{2}}^2\beta_{\mathfrak{6}}
+u_3\,\beta_{\mathfrak{10}}
\big)\gamma_{\mathfrak{6}}^2+\\
+&\big( 
u_4\,\beta_{\mathfrak{2}}^7
+u_5\,\beta_{\mathfrak{2}}^4\beta_{\mathfrak{6}}
+u_6\,\beta_{\mathfrak{2}}\beta_{\mathfrak{6}}^2+
u_7\,\beta_{\mathfrak{2}}^2\beta_{\mathfrak{10}}
\big)\gamma_{\mathfrak{6}}\,\gamma_{\mathfrak{10}}+\\
+&\big( 
u_8\,\beta_{\mathfrak{2}}^9
+u_9\,\beta_{\mathfrak{2}}^6\beta_{\mathfrak{6}}
+u_{10}\,\beta_{\mathfrak{2}}^3\beta_{\mathfrak{6}}^2+
u_{11}\,\beta_{\mathfrak{6}}^3
+u_{12}\,\beta_{\mathfrak{2}}^4\beta_{\mathfrak{10}}
+u_{13}\,\beta_{\mathfrak{2}}\beta_{\mathfrak{6}}\beta_{\mathfrak{10}}
\big)\gamma_{\mathfrak{10}}^2\,,
\end{align}
\end{subequations}
where we omit the explicit form of the coefficients $u_k$.

\paragraph{Closing the OPE.}
Let us present the OPE of strong generators in this case:
\begin{subequations}
\begin{align}
W_{\mathfrak{6}}\,\times\,W_{\mathfrak{6}}
&\sim
[\text{id}]+
\lambda_{\mathfrak{6}\mathfrak{6}}^{\mathfrak{6}}\,
[W_{\mathfrak{6}}]
+
\lambda_{\mathfrak{6}\mathfrak{6}}^{\mathfrak{10}}\,
[W_{\mathfrak{10}}]
\\
\label{OPEW6W10intypeH3}
W_{\mathfrak{6}}\,\times\,W_{\mathfrak{10}}
&\sim
\,\lambda_{\mathfrak{6},\mathfrak{10}}^{\mathfrak{6}}
\,[W_{\mathfrak{6}}]+
\lambda_{\mathfrak{6},\mathfrak{10}}^{\mathfrak{10}}
\,[W_{\mathfrak{10}}]+
\lambda_{\mathfrak{6},\mathfrak{10}}^{(\mathfrak{6}\mathfrak{6})}\,
[(W_{\mathfrak{6}})^2_{\mathfrak{S}}]
\\
\label{OPEW10W10intypeH3}
W_{\mathfrak{10}}\,\times\,W_{\mathfrak{10}}
&\sim
[\text{id}]+
\lambda_{\mathfrak{10},\mathfrak{10}}^{\mathfrak{6}}
\,[W_{\mathfrak{6}}]
+\lambda_{\mathfrak{10},\mathfrak{10}}^{\mathfrak{10}}
\,[W_{\mathfrak{10}}]+
\lambda_{\mathfrak{10},\mathfrak{10}}^{(\mathfrak{6}\mathfrak{6})}\,
[(W_{\mathfrak{6}})^2_{\mathfrak{S}}]\,+\\
\nonumber
\,\,\,\,\,\,\,\,&\,\,\,\,\,\,\,\,\,\,\,\,\,\,\,\,\,\,\qquad\quad\,\,\,\,\,\,\,\,\,\,\,\,\,\,
+
\lambda_{\mathfrak{10},\mathfrak{10}}^{(\mathfrak{6},\mathfrak{10})}
\,[(W_{\mathfrak{6}}W_{\mathfrak{10}})_{\mathfrak{S}}]
+
\lambda_{\mathfrak{10},\mathfrak{10}}^{(\mathfrak{6}\mathfrak{6}\mathfrak{6})}\,
[(W_{\mathfrak{6}})^3_{\mathfrak{S}}]\,,
\end{align}
\end{subequations}
where
\beq
\lambda_{\mathfrak{6}\mathfrak{6}}^{\mathfrak{6}}=
\tfrac{57}{2}\sqrt{\tfrac{35}{10582}}\,,
\qquad 
\lambda_{\mathfrak{6}\mathfrak{6}}^{\mathfrak{10}}=
\lambda_{\mathfrak{6}\mathfrak{10}}^{\mathfrak{6}}=
\tfrac{37}{2}\sqrt{\tfrac{4845}{551122}}
\,,
\qquad 
\lambda_{\mathfrak{6},\mathfrak{10}}^{\mathfrak{10}}=
\lambda_{\mathfrak{10},\mathfrak{10}}^{\mathfrak{6}}=
-\tfrac{1189}{94}\sqrt{\tfrac{55}{6734}}\,,
\eeq
\beq
\lambda_{\mathfrak{6},\mathfrak{10}}^{(\mathfrak{6}\mathfrak{6})}=\,
\tfrac{817}{2}\sqrt{\tfrac{41}{11795637}}\,,
\qquad
\lambda_{\mathfrak{10},\mathfrak{10}}^{\mathfrak{10}}\,=\,
\tfrac{2747357}{1974}\sqrt{\tfrac{55}{48548838}}\,,
\qquad
\lambda_{\mathfrak{10},\mathfrak{10}}^{(\mathfrak{6}\mathfrak{6})}\,=\,
\tfrac{479167}{295630}\,,
\eeq
\beq
\lambda_{\mathfrak{10},\mathfrak{10}}^{(\mathfrak{6},\mathfrak{10})}\,=\,
-\tfrac{232}{3}\sqrt{\tfrac{287}{1685091}}\,,
\qquad
\lambda_{\mathfrak{10},\mathfrak{10}}^{(\mathfrak{6}\mathfrak{6}\mathfrak{6})}\,=\,
\tfrac{15416}{35853}\sqrt{\tfrac{286}{1295}}\,.
\eeq
In this case the defining relations  for the ring $\mathscr{R}_{H_3}$
are of type 
\beq
\big{\{}\mathfrak{L}_{(6,j)}\big{\}}_{j\in \{0,2,4\}}\,,
\qquad
\big{\{}\mathfrak{L}_{(8,j)}\big{\}}_{j\in \{2,3,4,5,6\}}\,,
\qquad
\big{\{}\mathfrak{L}_{(10,j)}\big{\}}_{j\in \{0,2,4,6,8\}}\,.
\eeq
Notice that the super-Virasoro primary operators of the form 
$(W_{\mathfrak{6}})^3_{\mathfrak{L}}$  are null as a consequence of  
$(W_{\mathfrak{6}})^2_{\mathfrak{L}}$ being null. 


\subsection{A rank $4$ example:  $\Gamma=D_4$}
\label{sec:D4Example}

In this example we will encounter two new features:
(1) the ring of invariants $\mathscr{R}_{D_4}$ and, according to our proposal, 
the VOA $\mathcal{W}_{\Gamma}$ has long generators,
(2) the form of $J^-$ is not uniquely determined by the first four steps 
given in section \eqref{sec:remainingGENN4} and the fifth condition needs
 to be incorporated.

\paragraph{Generators of the ring $\mathscr{R}_{D_4}$.}
The action of the Weyl group of type $D_4$ on $\mathbb{R}^4$ is generated by
\begin{align}
s_1:&\quad (z_1,z_2,z_3,z_4)\,\mapsto\, (-z_1,z_2,z_3,z_4)\\
s_2:&\quad (z_1,z_2,z_3,z_4)\,\mapsto\, 
\tfrac{1}{2}
(z_{12}^+-z_{34}^+,z_{12}^++z_{34}^+,-z_{12}^-+z_{34}^-,-z_{12}^--z_{34}^-)
\\
s_3:&\quad (z_1,z_2,z_3,z_4)\,\mapsto\, (z_1,z_2,-z_3,z_4)\\
s_4:&\quad (z_1,z_2,z_3,z_4)\,\mapsto\, (z_1,z_2,z_3,-z_4)
\end{align}
where $z_{ij}^{\pm}=z_i\pm z_j$.
In this basis,
a choice of generators for the ring of invariants
$\mathbb{C}[z_1,\dots,z_4]^{D_4}$ 
is given by
\begin{subequations}
\label{IinvariantsD4}
\begin{align}
\mathcal{I}_2&=\tfrac{1}{2}(z_1^2+z_2^2+z_3^2+z_4^2)\,,
\qquad
\mathcal{I}_6=
(z_1^2z_2^2-z_3^2z_4^2)(z_1^2+z_2^2-z_3^2-z_4^2)\\
&\qquad 
\mathcal{I}_4^{(1)}=(z_2^2-z_3^2)(z_1^2-z_4^2)\,,
\qquad
\mathcal{I}_4^{(2)}=(z_1^2-z_3^2)(z_2^2-z_4^2)\,.
\end{align}
\end{subequations}
These generator are algebraically independent. 
As discussed in section \eqref{subsec:Coxeter} we need to analyze the ring \eqref{wannabeHiggsBrChiralRing} in which the Coxeter group acts on two copies of   $z$, called $z^{\pm}$. Some of the generators of this ring are immediately identified starting from \eqref{IinvariantsD4} and promoting each $z_i$ to $z_i(y)=z_i^++y z_i^-$. These are the so-called short generators.
 As opposed to all the examples encountered so far, namely $A_1,I_2(p), A_3,B_3,H_3$,
this is the first example in which these are not all the generators of $\mathscr{R}_{D_4}$.
The missing generator is given by
\beq
w_{(3,0)}\,:=\,X_{123}+X_{134}-X_{124}-X_{234}\,,
\qquad
X_{ijk}:=\langle ij \rangle \langle ik \rangle\langle j k \rangle\,,
\eeq
where $\langle ij\rangle=\epsilon_{IJ}z_i^Iz_j^J$.
It is rather clear that this invariant cannot be written  as a composite of the short generators.
What is less obvious is that there is no additional generator. We claim that it is the case.
This fact can be checked by matching the Hilbert series computed from  the 
 proposed description of $\mathscr{R}_{D_4}$ in terms of generators 
 and relations with the Molien series obtained from the quotient description.
 Alternatively one can verify that the set of generators we propose closes under the Poisson bracket  \eqref{zsPB}. We followed  the second strategy.

Finally let us explain  how the triality automorphism of $D_4$ acts on the space of invariants.
Its action in terms of the $z_{i=1,\dots,4}$ variables can be defined as the group   $S_3$ of permutations of $z_1,z_2,z_3$.
It is a simple exercise to verify that, up to a simple redefinition of $\mathcal{I}_6$,
the invariants of  degree $2$ and $6$ transform trivially under $S_3$, 
the invariants of  degree $4$ transform in a two dimensional representation\footnote{
In the basis above, this is generated by $ \begin{psmallmatrix}0&1\\1&0\end{psmallmatrix}$  and
$ \begin{psmallmatrix}1&-1\\0&-1\end{psmallmatrix}$.}
and the long invariant
$w_{(3,0)}$ transform in the non-trivial one dimensional representation 
corresponding to the sign of the permutation.

\paragraph{Free field realization.}
This is the first example in which the first four steps described in  section 
\eqref{sec:remainingGENN4} to construct $J^-$ do not give a unique result.
There is a number or relatively simple conditions that the free field realization must satisfy
that are easy to add.
The first one is the following. 
Consider the OPE $W_{\ell_1} W_{\ell_2}$.
The term with next to extremal $\mathfrak{sl}(2)_y$ spin $j=\tfrac{1}{2}(p_{\ell_1}+p_{\ell_2})-1$
in the singular part of these OPE can appear only in the first order pole. 
By a simple quantum number analysis this term must be a short operator.
We require that this object is indeed a composite operator of the short generators that have been postulated in the free field realization.
The second condition is given by focusing on the first two most singular terms in the $WW$ OPE.
Their form is fixed by super-Virasoro symmetry to be
  \beq
W_{\ell_1}(z_1)W_{\ell_2}(z_1)\,=\,
\delta_{\ell_1,\ell_2}\,
\left(\text{id}
-\tfrac{6\,p_{\ell_1}}{c}\,J\right)
+\dots\,
  \eeq
 This form of the OPE is added as an extra requirement.
  We found experimentally that these two conditions allow to completely fix the free field realization.
  It is possible that in more complicated examples more conditions need to be added, but we expect 
  that the Ansatz we propose is sufficient.

In this case we obtain\footnote{Recall that $D_4$ has two fundamental invariants of degree $4$. We use the suffix $\pm$ to distinguish them. This sign encodes the transformation property of the free fields under a $\mathbb{Z}_2$ symmetry of the VOA $\mathcal{W}_{D_4}$. }
\begin{subequations}
\begin{align}
\mathcal{P}(J^-_{\text{norms}})=\,\,&
\Lambda_1\,\beta_{\mathfrak{2}}^3
\Big((\gamma_{\mathfrak{4}}^+)^2+(\gamma_{\mathfrak{4}}^-)^2\Big)
+
4\sqrt{\tfrac{\Lambda_1}{3}}\,\beta_{\mathfrak{2}}
\Big(\beta_{\mathfrak{4}}^+
\left((\gamma_{\mathfrak{4}}^-)^2-(\gamma_{\mathfrak{4}}^+)^2\right)
+2\beta_{\mathfrak{4}}^-\gamma_{\mathfrak{4}}^+\gamma_{\mathfrak{4}}^-
\Big)+
\\
+&\,
\Big(
\tfrac{16\Lambda_2}{\Lambda_1}
\beta_{\mathfrak{2}}\left((\beta_{\mathfrak{4}}^+)^2+(\beta_{\mathfrak{4}}^-)^2\right)
+\Lambda_2\beta_{\mathfrak{2}}^5
\Big)
\gamma_{\mathfrak{6}}^2
+
28\sqrt{\tfrac{\Lambda_2}{15}}
\beta_{\mathfrak{2}}^2
\left(\beta_{\mathfrak{4}}^+\gamma_{\mathfrak{4}}^++
\beta_{\mathfrak{4}}^-\gamma_{\mathfrak{4}}^-\right)\gamma_{\mathfrak{6}}+
\\
&\qquad\qquad+\,
\tfrac{1}{\sqrt{15\Lambda_2}}
\beta_{\mathfrak{6}}
\Big(5\Lambda_1 
\left((\gamma_{\mathfrak{4}}^+)^2+(\gamma_{\mathfrak{4}}^-)^2\right)
-22\Lambda_2
\beta_{\mathfrak{2}}^2
\gamma_{\mathfrak{6}}^2
\Big)
+
\\
&\qquad\qquad+\,16\sqrt{\tfrac{\Lambda_2}{5\Lambda_1}}
\Big(
\left((\beta_{\mathfrak{4}}^-)^2-(\beta_{\mathfrak{4}}^+)^2\right)
\gamma_{\mathfrak{4}}^++
2\beta_{\mathfrak{4}}^+\beta_{\mathfrak{4}}^-\gamma_{\mathfrak{4}}^-
\Big)
\gamma_{\mathfrak{6}}\,.
\end{align}
\end{subequations}
The parameters $\Lambda_1$, $\Lambda_2$ are related to the norms of the short generators as
$g_{\mathfrak{4}\mathfrak{4}}=1680\Lambda_1$,
$g_{\mathfrak{6}\mathfrak{6}}=665280\Lambda_2$.
\paragraph{Closing the OPE}
The extra generators in this case are $W^{\pm}_{\mathfrak{4}},W_{\mathfrak{6}}^{}$ and $W_{(3,0)}$
and
their OPEs take the form
\begin{subequations}
\begin{align}
W^{\pm}_{\mathfrak{4}}\,\times\,W^{\pm}_{\mathfrak{4}}
&\sim
g_{\mathfrak{4}\mathfrak{4}}\,[\text{id}]
\pm
\lambda_{\mathfrak{4}\mathfrak{4}}^{\mathfrak{4}^+}\,
[W^+_{\mathfrak{4}}]
+
\lambda_{\mathfrak{4}\mathfrak{4}}^{\mathfrak{6}}\,
[W_{\mathfrak{6}}]
\\
\label{OPEW4plusW4minusintypeD4}
W^+_{\mathfrak{4}}\,\times\,W^-_{\mathfrak{4}}
&\sim
\,\lambda_{\mathfrak{4}^+\mathfrak{4}^-}^{\mathfrak{4}^-}
\,[W^-_{\mathfrak{4}}]
+\,
[W_{(3,0)}]
\\
\label{OPEW4plusW6intypeD4}
W^+_{\mathfrak{4}}\,\times\,W_{\mathfrak{6}}
&\sim
\,\lambda_{\mathfrak{4}\mathfrak{6}}^{\mathfrak{4}}
\,[W^+_{\mathfrak{4}}]
+\,\lambda_{\mathfrak{4}\mathfrak{6}^{}}^{(\mathfrak{4}\mathfrak{4})}
[(W_{\mathfrak{4}})^{2,+}_{\mathfrak{S}}]
+\dots
\\
\label{OPEW4minusW6intypeD4}
W^-_{\mathfrak{4}}\,\times\,W_{\mathfrak{6}}
&\sim
\,\lambda_{\mathfrak{4}\mathfrak{6}}^{\mathfrak{4}}
\,[W^-_{\mathfrak{4}}]
+\,\lambda_{\mathfrak{4}\mathfrak{6}^{}}^{(\mathfrak{4}\mathfrak{4})}
[(W_{\mathfrak{4}})^{2,-}_{\mathfrak{S}}]
+\dots
\\
\label{OPEW6W6intypeD4}
W_{\mathfrak{6}}\,\times\,W_{\mathfrak{6}}
&\sim
g_{\mathfrak{6}\mathfrak{6}}\,
[\text{id}]
+
\lambda_{\mathfrak{6}\mathfrak{6}}^{\mathfrak{6}}
\,[W_{\mathfrak{6}}]
+
\lambda_{\mathfrak{6}\mathfrak{6}}^{(\mathfrak{4}\mathfrak{4})_C}\,
[(W_{\mathfrak{4}})^{2,0}_{\mathfrak{S}}]
+\dots
\end{align}
\end{subequations}
where
\beq
-\tfrac{1}{\sqrt{g_{\mathfrak{4}\mathfrak{4}}}}
\lambda_{\mathfrak{4}\mathfrak{4}}^{\mathfrak{4}^+}=
\tfrac{1}{\sqrt{g_{\mathfrak{4}\mathfrak{4}}}}
\lambda_{\mathfrak{4}^+\mathfrak{4}^-}^{\mathfrak{4}^-}=
\tfrac{6}{\sqrt{35}}\,,
\qquad
\tfrac{\sqrt{g_{\mathfrak{6}\mathfrak{6}}}}{g_{\mathfrak{4}\mathfrak{4}}}
\lambda_{\mathfrak{4}\mathfrak{4}}^{6}=
\sqrt{\tfrac{11}{7}}\,,
\qquad
\tfrac{1}{\sqrt{g_{\mathfrak{6}\mathfrak{6}}}}\,
\lambda_{\mathfrak{4}\mathfrak{6}}^{\mathfrak{4}}
=\sqrt{\tfrac{11}{7}}\,,
\eeq
\beq
\sqrt{\tfrac{g_{\mathfrak{4}\mathfrak{4}}}{g_{\mathfrak{6}\mathfrak{6}}}}
\lambda_{\mathfrak{4}\mathfrak{6}^{}}^{(\mathfrak{4}\mathfrak{4})}=
\tfrac{8}{3\sqrt{55}}\,,
\qquad
\tfrac{1}{\sqrt{g_{\mathfrak{6}\mathfrak{6}}}}
\lambda_{\mathfrak{6}\mathfrak{6}}^{\mathfrak{6}}=
-\tfrac{2}{3}\sqrt{\tfrac{7}{11}}\,,
\qquad
\tfrac{g_{\mathfrak{4}\mathfrak{4}}}{g_{\mathfrak{6}\mathfrak{6}}}
\lambda_{\mathfrak{6}\mathfrak{6}}^{(\mathfrak{4}\mathfrak{4})}=\tfrac{8}{9}
\eeq
and
\begin{align}
(W_{\mathfrak{4}})^{2,+}_{\mathfrak{S}}&=
(\beta^-_{\mathfrak{4}})^2-(\beta^+_{\mathfrak{4}})^2-
\tfrac{2\sqrt{g_{\mathfrak{4}\mathfrak{4}}}}{57\sqrt{35}}\,
\beta_{\mathfrak{2}}^2\beta_{\mathfrak{4}}^+
+\text{desc.}
\\
(W_{\mathfrak{4}})^{2,-}_{\mathfrak{S}}&=
2\,\beta^+_{\mathfrak{4}}\beta^-_{\mathfrak{4}}-
\tfrac{2\sqrt{g_{\mathfrak{4}\mathfrak{4}}}}{57\sqrt{35}}\,
\beta_{\mathfrak{2}}^2\beta_{\mathfrak{4}}^-
+\text{desc.}
\\
(W_{\mathfrak{4}})^{2,0}_{\mathfrak{S}}&=
(\beta^+_{\mathfrak{4}})^2+(\beta^-_{\mathfrak{4}})^2
+
\tfrac{1}{10}
\sqrt{\tfrac{11}{7}}
\tfrac{
g_{\mathfrak{4}\mathfrak{4}}
}{\sqrt{g_{\mathfrak{6}\mathfrak{6}}}}
\,\beta_{\mathfrak{2}}\beta_{\mathfrak{6}}
-\tfrac{g_{\mathfrak{4}\mathfrak{4}}}{28560}\,
\beta_{\mathfrak{2}}^4
+\text{desc.}
\end{align}
Notice that we omitted the OPEs of $W_{(3,0)}$ with the remaining generators 
as well as its explicit form in terms of free fields.
The $\dots$ in the last three OPEs indicate additional contributions.
For example in  the OPEs \eqref{OPEW4plusW6intypeD4} and \eqref{OPEW4minusW6intypeD4}
operators of type $\mathfrak{L}_{(4,3)}$, $\mathfrak{L}_{(4,2)}$ and $\mathfrak{L}_{(4,1)}$ 
of the schematic form $W_{\mathfrak{4}}W_{\mathfrak{4}}$ could appear.
Let us also collect the quantum numbers of the null operators  with small conformal weight
as  read off from the Hilbert series of $\mathscr{R}_{D_4}$: 
\beq\label{NULLsD4}
\mathfrak{L}_{(4,2)}\,,
\quad
\mathfrak{L}_{(4,1)}\,,
\quad
2\,\mathfrak{L}_{(4,0)}\,,
\quad
2\,\mathfrak{L}_{(5,3)}\,,
\quad
2\,\mathfrak{L}_{(5,2)}\,,
\quad
2\,\mathfrak{L}_{(5,1)}\,.
\eeq
The corresponding null operators have the schematic form 
$W_{\mathfrak{4}}W_{\mathfrak{4}}$  and $W_{\mathfrak{4}}W_{\mathfrak{6}}$.

\vspace{1cm}
\noindent
\emph{Remark:} The OPEs above have an $S_3\subset O(2)$ symmetry that acts non-trivially
 only on
 the generators
$W^{\pm}_{\mathfrak{4}}$ and is generated by the reflections $s_1,s_2$ as 
\beq\label{S3forD4}
\begin{pmatrix}
\beta_{\mathfrak{4}}^+\\
\beta_{\mathfrak{4}}^-
\end{pmatrix}
\mapsto
s_k\,\begin{pmatrix}
\beta_{\mathfrak{4}}^+\\
\beta_{\mathfrak{4}}^-
\end{pmatrix}\,,
\qquad
s_1=
\begin{pmatrix}
1 & 0\\
0 & -1
\end{pmatrix}\,,
\quad
s_2=
\frac{1}{2}
\begin{pmatrix}
-1 & \sqrt{3}\\
\sqrt{3} & 1
\end{pmatrix}\,.
\eeq
Notice that the product $s_1 s_2$ generates a $\mathbb{Z}_3$ subgroup.
In order to check this claim it is convenient to observe that
 $(W_{\mathfrak{4}})^{2,\pm}_{\mathfrak{S}}$ transform as  
 $\beta_{\mathfrak{4}}^{\pm}$ under \eqref{S3forD4}.
The relations \eqref{NULLsD4} have definite transformation properties under $S_3$:





\section{Examples of $\cN = 2$ VOA $\mathcal{W}_{\mathcal{\mathsf{G}}}$}
\label{sec:N=2Examples}

\subsection{Rank $1$: $\mathsf{G}=\mathbb Z_p$}
In this section we examine in detail the proposed free-field
realization for the $\cN = 2$ VOA $\cW_{\mathbb Z_p}$
associated to the rank-1 complex reflection
group $\mathbb Z_p$, $p\ge 2$. These algebras have been 
first analyzed by means of direct bootstrap techniques in \cite{Nishinaka:2016hbw}.

\subsubsection{Construction of the generators and OPEs}

According to our prescription, we need one copy of a $\beta \gamma bc$
system. 
For the sake of simplicity, we omit the subscript 1 from the free fields 
in this section.
The $\cN= 2$ SCA algebra is realized according to the formulae
of section \ref{sec:N2SCAFree}. In particular, the central charge is $c = -3(2p-1)$
and the conformal weight of $\beta$ is $p/2$. 

Let $W$, $\overline W$ denote the chiral, antichiral
extra generators of the VOA.
The only non-trivial
task at hand is the construction of the antichiral generator $\overline W$,
of conformal weight $h=p/2$ and charge $q=-p/2$ using the strategy presented in 
section~\ref{sec:N2freeFIELDprescription}.

Before proceeding, let us stress that, for $p=2$,
supersymmetry enhances from $\cN=2$ to small $\cN = 4$,
and the 
 sought-for VOA is nothing but 
the small $\cN =4$ SCA at central charge $c=-9$,
with the identification $\overline W = J^-$.
In this case the free-field realization coincides
with the one of \cite{Adamovic:2014lra}, 
which has been reviewed in section \ref{sec:A1}.

Let us now discuss the case of generic $p$.
The direct analysis of a few examples reveals that
there exists a unique, up to scaling, chiral $\cN = 2$ superVirasoro
primary of conformal weight $\tfrac{p}{2}$ that can be constructed using a single 
$\beta\gamma b c$ of the type given in table \eqref{betagammabc_hmr}. 
For definiteness, we fix the normalization of the 
generators $W$ and $\overline W$ as follows,
\beq
W = \beta \ , \qquad 
\overline W = \beta^{p-1} \, \gamma^p + \dots \ .
\eeq
All omitted terms in $\overline W$ contain at least one derivative or
a pair of fermionic free fields. 

In terms of the map $\mathcal P$ of section \ref{sec:freefieldsclassPOISSON}, $\mathcal P (\overline W) = \beta^{p-1} \, \gamma^p$.
We may regard $\overline W$ as the
unique $\cN = 2$ super-Virasoro primary completion of the monomial
$\beta^{p-1} \, \gamma^p$. 

Once the form of the antichiral generator $\overline W$
is fixed, we may check that the OPE $\overline W \, \overline W$
is regular, and that the singular part of the $W \, \overline W$
OPE is expressed entirely in terms of $\cN = 2$ super-Virasoro
descendants of the identity, as it must be.
In particular, we can verify   that
\beq\label{WWbarrank1N2}
W \times \overline W = g_{W \, \overline W} \,\, [ \text{id} ]\,,
\qquad
g_{W \overline W}=
 (-)^p  \, \frac{(2p-1)!}{p! \,p^{\, p-1 }}\,,
\eeq
where $ [ \text{id} ]$ denotes the contribution for the $\mathcal{N}=2$ 
super-Virasoro family of the identity operator. More explicitly
\beq
[ \text{id} ] =  \text{id}  + \tfrac{3 \,p}{c} \, \cJ
 + \tfrac{p}{c} \, \mathcal{T}
+ \tfrac{3p(3p-1)}{2c (c-1)}\,
 \big((\cJ \, \cJ)_0-\tfrac{2}{3}\cT\big) + \dots
\eeq
Similarly to the $\mathcal{N}=4$ case given in \eqref{idPIECEN4},
 the contribution of the  stress tensor $\mathcal{T}$
to this OPE is split in two parts: the first correspond to the $\mathfrak{osp}(2|2)$
descendant of $\mathcal{J}$, the second to the completion of $(\cJ \, \cJ)_0$ to a 
$\mathfrak{osp}(2|2)$ primary whose norm is given by $\tfrac{2}{9}c(c-1)$.
The relation \eqref{WWbarrank1N2} has been checked explicitly for $p = 2, 3, \dots, 7$.
For $p = 2,3,4,5$ we can 
record the entire content of the singular part of the $W \overline W$
OPE  in terms of quasiprimary fields
\begin{align}
p =2 \ :  \quad [ \text{id} ]    &= \text{id}
 - \tfrac 23 \cJ \ , \nn  \\[1.5 mm]
p =3 \ :  \quad [ \text{id} ]   &= 
 \text{id}
- \tfrac 35 \, \cJ - \tfrac{3}{10} \, \mathcal{T} + \tfrac {3}{20} \, (\cJ \, \cJ)_0
 \ , \nn \\[1.5 mm]
p =4 \ : \quad [ \text{id} ]   &= 
 \text{id}
 - \tfrac 47 \, \cJ  - \tfrac 27 \,  \mathcal{T} + \tfrac 17  \, (\cJ \, \cJ)_0
- \tfrac{2}{105} \, (\cJ (\cJ \, \cJ)_0)_0
+ \tfrac{2}{35} \, (\mathcal G \, \widetilde {\mathcal G})_0
+ \tfrac{6}{35} \, (\cJ \, \mathcal{T})_0 \ , \nn \\[1.5 mm]
p =5 \ : \quad [ \text{id} ]  &=
 \text{id}
 - \tfrac  59 \, \, \cJ
- \tfrac {5}{18} \, \mathcal{T} 
+ \tfrac{5}{36} \, (\cJ \,\cJ)_0
 - \tfrac{5}{252}  \, (\cJ (\cJ \, \cJ)_0)_0 \nn \\
& + \tfrac{5}{126} \, (\mathcal G \, \widetilde{ \mathcal G})_0
+ \tfrac{10}{63} \, (\cJ \, \mathcal{T})_0
+ \tfrac{25}{504} \, (\mathcal{T} \, \mathcal{T})_0
+ \tfrac{5}{3024}  \, (\cJ (\cJ(\cJ \, \cJ)_0)_0)_0 \nn \\
& - \tfrac{5}{126} \,  (\cJ (\cJ \, \mathcal{T})_0 )_0
- \tfrac{5}{252} \,  (\cJ (\mathcal G \, \widetilde { \mathcal G})_0 )_0
- \tfrac{5}{252} \,  (\mathcal G \, \widetilde { \mathcal G})_{-1}
+ \tfrac{95}{3024} \,   (\cJ  \, \cJ)_{-2}
\ .
\end{align}
We are using a compact notation in which all $z_{12}$
factors and $z$ derivatives are omitted,
since they can be straightforwardly recovered from $\mathfrak {sl}(2)$
covariance. 
For further details on the notation, the reader is referred to appendix \ref{OPEandsl2cov}.
We are only
recording quasiprimary operators that enter the singular part of the OPEs.
Clearly, all OPE coefficients on the RHS of the OPE
$W\, \overline W$ can in principle be recovered
from the two-point function coefficient $g_{W \overline W}$
by exploiting the $\cN = 2$ superVirasoro symmetry.

\subsubsection{Null states} 
 
An essential feature of our proposed free-field realization
is that all null states of the VOA are realized manifestly as zero.
First of all, let us verify this claim for the null state corresponding
to the ``Higgs branch'' relation of the 
associated variety $\mathscr M_{\mathbb Z_p} = (\mathbb C \oplus \mathbb C^*)/\mathbb Z_p$. To describe this variety we
introduce a complex coordinate $z$
and its conjugate $\bar z$, and let the generator
of $\mathbb Z_p$ act on $z$,
$\bar z$ as
\beq
z \mapsto e^{2\pi i/p} \, z \ , \qquad
\bar z \mapsto e^{-2\pi i/p} \, \bar z \ .
\eeq
The invariants are clearly
\beq
w = z^p \ , \qquad \bar w  = \bar z ^p \ , \qquad j = z \, \bar z \ ,
\eeq
satisfying one relation,
\beq \label{Zp_Higgs_rel}
w \, \bar w = j^p \ .
\eeq
At the level of the VOA, this invariant motivates us to consider
the $\cN = 2$ superVirasoro primary composite operator
\beq
\mathfrak X_{p,0}^{W \overline W} = (W \, \overline W)_0 + \dots \ .
\eeq 
The dots represent the terms needed to obtain a superVirasoro primary,
and are uniquely determined by the OPEs of the abstract
VOA.
For example, the full expressions of this composite
for $p = 2,3,4$ are
\begin{align}
p = 2 \ : \quad \mathfrak X_{2,0}^{W \overline W} & = (W \, \overline W)_0 
- \tfrac 14 \, (\cJ \, \cJ)_0 + \tfrac 12 \, T
\ , \nn \\
p = 3 \ : \quad \mathfrak X_{3,0}^{W \overline W} & = (W\, \overline W)_0 
- \tfrac{1}{27} \, (\cJ ( \cJ \, \cJ)_0)_0
+ \tfrac 29 \, (\mathcal G \, \widetilde {\mathcal G})_0
+ \tfrac 49 \, (\cJ \, T)_0 \ , \nn \\
p = 4 \ : \quad \mathfrak X_{4,0}^{W \overline W} & = (W \, \overline W)_0
- \tfrac{1}{256} \, (\cJ (\cJ (\cJ \, \cJ)_0 )_0 )_0 
- \tfrac{3}{16} \, (T \,T)_0
+ \tfrac{9}{64} \, (\cJ (\cJ \, T)_0 )_0  \nn \\
&+ \tfrac{3}{32} \,  (\cJ (\mathcal G \, \widetilde {\mathcal G})_0 )_0   
+ \tfrac{3}{32} \, (\mathcal G \, \widetilde { \mathcal G})_{-1} 
- \tfrac{15}{128} \, (\mathcal J \, \mathcal J)_{-2}  \ .
\end{align}
We checked that, in our free-field realization,
these composites are indeed identically zero.
Even though the full expression of the operator
$ \mathfrak X_{p,0}^{W \overline W}$
in the VOA becomes increasingly complex
as we increase $p$, 
the classical counterpart of $ \mathfrak X_{p,0}^{W \overline W}$
via the 
map $\mathcal P$ of section \ref{sec:freefieldsclassPOISSON}
has a very simple structure. 
Indeed, one verifies that
\begin{align}
W_{\rm cl}  &:= \mathcal P(W) = \beta \ , \qquad
\overline W_{\rm cl}  := \mathcal P(\overline W)
= \beta^{p-1} \, \gamma^p  \ , \qquad
\cJ_{\rm cl}  := \mathcal P(\cJ)  = p \, \beta \, \gamma \ ,
\nn \\
(\mathfrak X_{p,0}^{W \overline W})_{\rm cl} &:= \mathcal P(
\mathfrak X_{p,0}^{W \overline W}
)
= W_{\rm cl} \overline W_{\rm cl} - \frac{1}{p^p} \, (\cJ_{\rm cl})^p \ . 
\end{align}
These expressions show that the operator 
$\mathfrak X_{p,0}^{W \overline W}$ is indeed the null
operator associated to the ``Higgs branch'' relation
\eqref{Zp_Higgs_rel}. It is also straightforward
to find the map between the classical
Poisson variables $z$, $\bar z$ and $\beta$, $\gamma$,
\beq
\beta = z^p \ , \qquad \gamma = z^{1-p} \, \bar z \ .
\eeq

Let us now discuss another pair of null states
that are expected in the VOA $\cW_{\mathbb Z_p}$.
They are a pair of non-chiral, $\cN = 2$ superVirasoro primary
operators linear
in $W_p$ and $\overline W_p$ respectively, given by
\beq
\mathfrak X^{\mathcal G W}_{\frac p2 + \frac 32 , -1}  = (\mathcal G \,   W)_0 - \frac 1p \, (\cJ \, 
\mathcal G _{  W}
)_0  \ , \qquad
\mathfrak X_{\frac p2 + \frac 32 , +1}^{\widetilde {\mathcal G} \overline W}  =(\widetilde { \mathcal G} \, \overline W)_0 + \frac 1p \, (\cJ \, 
\widetilde { \mathcal G} _{\overline W}
)_0   \ .
\eeq
The operator $\mathcal G _{  W}$ is the supersymmetry descendant
of $W_p$, defined as $\{ \mathcal G \,   W \}_1$. 
A similar remark applies to $\widetilde {\mathcal G} _{\overline W}$.
Let us stress that, in order to verify that
$\mathfrak X^{\mathcal G W}_{\frac p2 + \frac 32 , -1} $,
$\mathfrak X_{\frac p2 + \frac 32 , +1}^{\widetilde {\mathcal G} \overline W}$
are superVirasoro primaries, we only need to use the OPEs
of the abstract VOA, and not our specific free-field realization.
As already argued in \cite{Nishinaka:2016hbw}, these composite operators 
must be null in order for the VOA to exist.
In our free-field construction one can indeed verify
that both  $\mathfrak X^{\mathcal G W}_{\frac p2 + \frac 32 , -1}$
and
$\mathfrak X_{\frac p2 + \frac 32 , +1}^{\widetilde {\mathcal G} \overline W}$
vanish identically.

Finally, let us comment on the relation between
the ``Higgs branch'' null state $\mathfrak X^{W \overline W}_{p,0}$
and the fermionic null states 
$\mathfrak X^{\mathcal G W}_{\frac p2 + \frac 32 , -1}$,
$\mathfrak X_{\frac p2 + \frac 32 , +1}^{\widetilde {\mathcal G} \overline W}$.
The latter can be obtained from 
$\mathfrak X^{W \overline W}_{p,0}$
by taking a singular OPE with
$\mathcal G_W$, $\widetilde{ \cG}_{\overline W}$.
For example,
\begin{align}
p = 3 \; : \qquad \{ \cG_W \, \mathfrak X^{W \overline W}_{3,0} \}_2 
& = \tfrac 23 \, \mathfrak X^{\cG W}_{3,-1} \ , &
  \{ \widetilde \cG_{\overline W} \, \mathfrak X^{W \overline W}_{3,0} \}_2 
& = \tfrac 43 \, \mathfrak X^{\widetilde \cG \overline W}_{3,+1} \ , \nn \\
p = 4 \; : \qquad \{ \cG_W \, \mathfrak X^{W \overline W}_{4,0} \}_3 
& = \tfrac{3}{8} \, \mathfrak X^{\cG W}_{\frac 72,-1} \ , &
  \{ \widetilde \cG_{\overline W} \, \mathfrak X^{W \overline W}_{4,0} \}_3 
& = \tfrac{27}{8} \, \mathfrak X^{\widetilde \cG \overline W}_{\frac 72,+1} \ .
\end{align}
For general $p$, $\mathfrak X^{\mathcal G W}_{\frac p2 + \frac 32 , -1}$
enters the order $(p-1)$ pole of the 
$\cG_W \, \mathfrak X^{W \overline W}_{p,0}$ OPE,
and similarly  $\mathfrak X^{\widetilde {\mathcal G} \overline W}_{\frac p2 + \frac 32 , -1}$
enters the order $(p-1)$ pole of the 
$\widetilde \cG_{\overline W} \, \mathfrak X^{W \overline W}_{p,0}$ OPE.
In other words, the operators 
$\mathfrak X^{\mathcal G W}_{\frac p2 + \frac 32 , -1}$,
$\mathfrak X^{\widetilde {\mathcal G} \overline W}_{\frac p2 + \frac 32 , -1}$
belong to the ideal of $\cW_{\mathbb Z_p}$
generated\footnote{The ideal is obtained by taking the 
regular and singular part of the  OPE of 
$\mathfrak X^{W \overline W}_{p,0}$ with the element of the the VOA.
 } by $\mathfrak X^{W \overline W}_{p,0}$.
This observation is consistent with the expectation
that the ``Higgs branch'' null state
$\mathfrak X^{W \overline W}_{p,0}$
generates all null states in $\cW_{\mathbb Z_p}$.

\subsubsection{Screening operator}

We have a conjectural characterization of the VOA $\cW_{\mathbb Z_p}$
as a subalgebra of the $\beta \gamma bc$ system 
in terms of the kernel of a screening operator $\mathbb S$. 
Our proposal is a natural generalization
of the screening operator discussed in \cite{Adamovic:2014lra}
in the case $p=2$, {\it i.e.}~the small $\cN = 4$ SCA at $c = -9$.

In order to define $\mathbb S$, we first have to express $\beta$ and $\gamma$
in terms of chiral bosons $\chi$, $\phi$.
The OPEs of the chiral bosons
are recorded in \eqref{chiral_boson_OPEs},
and the expressions of $\beta$, $\gamma$ in terms
 of the chiral bosons are given in \eqref{beta_gamma_chiral_bosons}.
We may now define the screening current
\beq \label{screening_current2}
\mathsf J = b \, e^{(p^{-1} -1) (\chi + \phi)} \ .
\eeq
This operator has conformal dimension 1 and charge 0
under $\cJ$.
We conjecture that the VOA $\cW_{\mathbb Z_p}$
coincides with the kernel of $\mathbb{S}=\int \mathsf J$ acting 
on the $\beta\gamma bc$ system.
 Its action  on any object $X$ is defined as
\beq
\mathbb{S} \cdot X = \{ \mathsf J  \, X\}_1 \ .
\eeq
In the case $p=2$,
the conjecture is proven in \cite{Adamovic:2014lra}.
It is worth pointing out that the operator $\mathsf J$
can be expressed as a supersymmetry descendant
of an operator $\mathsf K$ with dimension and charge $1/2$,
\beq
\mathsf J = \{ \cG \, \mathsf K \}_1 \ , \qquad
\mathsf K = e^{ p^{-1} \, (\chi + \phi)}  \ .
\eeq
The operator $\mathsf K$ is a chiral $\cN=2$ super-Virasoro primary.

It is a matter of straightforward computation
to verify that all generators of the VOA $\cW_{\mathbb Z_p}$
are annihilated by the action of $\mathbb{S}$.
It is then immediate that $\cW_{\mathbb Z_p}$ is contained
in the kernel of $\mathbb{S}$. We do not have a general argument for
 the reversed inclusion for $p \ge 3$, but we have checked our claim in a few examples,
 for some states with low twist $h-m$.
More precisely, we worked at generic $p$ and considered
operators of twist up to 2. A direct computation of the kernel
of $\mathbb{S}$ in the $\beta \gamma bc$ system
shows a perfect agreement with a counting of states in $\cW_{\mathbb Z_p}$.
In order for this match to work, it is essential that in our free-field
construction all null states are (conjecturally) identically zero.

Let us close this section by analyzing a few properties of the classical
counterpart of the screening current $\mathsf J$ of \eqref{screening_current2}.
The map $\cP'$ of section \ref{sec:freefieldsclassPOISSON}
maps the VOA $\cW_{\mathbb Z_p}$ to the Poisson superalgebra of
functions of the variables $\beta$, $\gamma$, $b$, $c$.
The classical counterpart of \eqref{screening_current2} is simply
\beq
\mathsf J_{\rm cl}  = b \, \beta^{\frac 1p -1} \ .
\eeq
Its action on the classical variables $\beta$, $\gamma$, $b$, $c$
is the following,
\begin{align}
\{ \mathsf J_{\rm cl} , b \}_{\rm PB} &= 0 \ , &
\{ \mathsf J_{\rm cl} , \beta \}_{\rm PB} &= 0 \ , \nn \\
\{ \mathsf J_{\rm cl} , c \}_{\rm PB} &= \beta^{\frac 1p-1} \ , &
\{ \mathsf J_{\rm cl} , \gamma \}_{\rm PB} &= \left( 1 - \tfrac 1p \right) 
b \, \beta^{\frac 1p-2} \ .
\end{align}


\subsection{A rank $2$ example: $\mathsf{G}=G(3,1,2)$}
\label{N2rank2Example}

In this section we discuss
the proposed free-field
realization for the $\cN = 2$ VOA associated to the rank-2 complex reflection
group $G(3,1,2)$.

\paragraph{The ring $\mathscr{R}_{\mathsf{G}}$ for $\mathsf{G}=G(3,1,2)$.}
To begin with let us describe the family of complex reflection groups $G(k,1,\mathsf{r})$.
Recall that $\mathsf{r}$ is the rank of $G(k,1,\mathsf{r})$
and the invariants have degrees
\begin{equation}
k,2k,\dots, \mathsf{r} k\,.
\end{equation}
The action of  $G(k,1,\mathsf{r})$ on
 $(z_1,\dots,z_\mathsf{r})\in V_{\mathsf{G}}\simeq \mathbb{C}^\mathsf{r}$
 is generated by permutations $\sigma_i=p_{i,i+1}$ (with an obvious action on the coordinates) together with
\begin{equation}
\tau:\,(z_1,\dots,z_\mathsf{r})\mapsto
( \omega\, z_1,\,z_2\,\dots,\, z_\mathsf{r})\,,
\qquad \omega^k=1.
\end{equation}
The ring of invariants is freely generated by the 
elementary symmetric polynomials
in
 $z_1^k,\dots, z_\mathsf{r}^k$. 
Notice that for  $k=2$ this group is the 
 Coxeter group $B_{\mathsf{r}}$.

In the following we consider in more details the example of $G(3,1,2)$.
In this case the ring $\mathscr{R}_{\mathsf{G}}$ is generated by 
\begin{equation}
\begin{aligned}
&w_3\,=\,z_1^3+z_2^3\,,
\qquad
w_6\,=\,z_1^6+z_2^6\,,
\qquad
O\,=\,z_1^4\bar{z}_1+z_2^4\bar{z}_2\,,\\
&\overline {w}_3\,=\,\bar{z}_1^3+\bar{z}_2^3\,,
\qquad
\overline{w}_6\,=\,\bar{z}_1^6+\bar{z}_2^6\,,
\qquad
\overline{O}\,=\,z_1\bar{z}_1^4+z_2\bar{z}_2^4\,,
\\
&\qquad \qquad \quad\!\!
j\,=\,z_1\bar{z}_1+z_2\bar{z}_2\,,
\qquad
U\,=\,z_1^2\bar{z}_1^2+z_2^2\bar{z}_2^2\,.
\
\end{aligned}
\end{equation}
The relations of  lowest conformal weight, namely  $h=4$ and $h=4+\tfrac{1}{2}$,
take the explicit form
\begin{align}
\label{nullClassicalN2rank2}
U^2
-\overline  {O}\, w_3
-O\, \overline{w}_3+
j\,w_3 \overline {w}_3+
+\tfrac{1}{2}\,j^2\,U
-\tfrac{1}{2}\,j^4=0\,,\\
(w_6+w_3w_3)\overline{w}_3- O\,U-j^2\,O+2\,j\,U\,w_3=0  \ .
\end{align}
The second equation is accompanied by it conjugate.


\paragraph{Free-field realization.}
As explained in section~\ref{sec:free_fields} the free fields 
in this case are two copies of the $\beta\gamma bc$ system, 
with weights determined by the degree of the invariants
of $\mathbb C^2/G(3,1,2)$,
\beq
(p_1 , p_2) = (3, 6) \ .
\eeq
The small $\cN = 2$ algebra with central charge $c=-48$ is realized
according to the formulae in \ref{sec:N2SCAFree},
and the chiral generators $\cW_3$, $\cW_6$ are simply realized
as
\beq
\cW_3 = \beta_1 \ , \qquad \cW_6 = \beta_2 \ .
\eeq
The remaining generators\footnote{
In \cite{Lemos:2016xke} a close relative of this VOA was constructed by bootstrap methods.
In that case the operators  $\mathcal{O}$ and $\overline {\mathcal{O}}$ were not included as generators.
} with their quantum numbers are summarized in the following table
\begingroup
\renewcommand{\arraystretch}{1.3}
\begin{equation}
\begin{tabular}{| c||c | c |c | c|c|c | c|c | }
\hline
 & $\mathcal{W}_3$ &$\mathcal{W}_6$ &$\mathcal{J}$ & $\mathcal{O}$ & $\mathcal{U}$&$\overline{\mathcal{W}}_3$  & 
 $\overline{\mathcal{O}}$ & $\overline{\mathcal{W}}_6$
 \\
 \hline \hline
$h$ & $\tfrac{3}{2}$ & $3$ & $1$ & 
$\tfrac{5}{2}$& $2$   & $\tfrac{3}{2}$ &  $\tfrac{5}{2}$&  $3$
 \\
 \hline
$m$ & $\tfrac{3}{2}$ & $3$ & $0$ & 
$\tfrac{3}{2}$& $0$   & $-\tfrac{3}{2}$ &  $-\tfrac{3}{2}$&  $-3$
 \\
 \hline
$h-m$ & $0$ &  $0$ & $1$ & 
$1$ & $2$  & $3$ & $4$ &  $6$
 \\
 \hline
$\sharp$  & $-$ & $ -$ & $-$ & 
$4$ & $8$  &  $13$ &  $36$& $104$ \\
\hline
\end{tabular}
\label{N2rank2weigths}
\end{equation}
\endgroup
According to the prescription given in section \ref{sec:N2freeFIELDprescription}, the first step is to 
construct all  $\mathcal{N}=2$ superVirasoro primary with the appropriate weights. 
The entry $\sharp$ denotes the number of such objects. The symbol $-$ indicates that the
corresponding entry has already been constructed so we do not need an Ansatz for it.
After imposing that the algebra closes one finds the complete list of OPEs to be 
\begin{subequations}
\begin{align}
&\mathcal{W}_3\times \overline{\mathcal{W}}_3\,\sim\,[\text{id}] + [\mathcal{U}]\,, \\
&\mathcal{W}_6\times \overline{\mathcal{W}}_6\,\sim\,[\text{id}] + 
\tfrac{7}{143}\,[\mathcal{U}]-
\tfrac{268}{429}\,
[\mathcal{W}_3 \overline{\mathcal{W}}_3]+
\dots\,, \\
&\mathcal{O}\times \overline{\mathcal{O}}\,\sim\,
\tfrac{13365}{59584}
\left([\text{id}] -\tfrac{295}{627} [\mathcal{U}]-
\tfrac{347}{285}\,[\mathcal{W}_3 \overline{\mathcal{W}}_3]\right)\,, 
\qquad 
\mathcal{U}\times\mathcal{U}\,\sim\,\tfrac{99}{392}\,[\text{id}] -\tfrac{45}{98}\, [\mathcal{U}]\,, \\
&
\mathcal{W}_3\times \mathcal{U} \,\sim\,\tfrac{99}{392}\,[\mathcal{W}_3] + [\mathcal{O}]\,,
\qquad  
\mathcal{W}_3\times \overline{\mathcal{O}} \,\sim\,-\tfrac{135}{152}\,[\mathcal{U}] 
+
\tfrac{27}{76} [\mathcal{W}_3\overline{\mathcal{W}}_3]\,,
 \\
&\mathcal{W}_3\times \mathcal{O} \,\sim\,
\tfrac{27}{2128}\sqrt{\tfrac{2145}{2}} \,[\mathcal{W}_6] \,,
\quad
\mathcal{W}_6\times \mathcal{O} \,\sim\,
\tfrac{27}{1064}\,[\mathcal{W}_3\mathcal{W}_6]+
\tfrac{81}{266}\sqrt{\tfrac{15}{286}}\,[\mathcal{W}_3\mathcal{W}_3\mathcal{W}_3]\,,
\\
&
\mathcal{W}_6\times \overline{\mathcal{O}} \,\sim\,
-\tfrac{27}{2128}\sqrt{\tfrac{2145}{2}}\,[\mathcal{W}_3]+
\tfrac{43}{133}\sqrt{\tfrac{5}{858}}\,[\mathcal{O}]+
-\tfrac{435}{266}\sqrt{\tfrac{15}{286}}\,[\mathcal{W}_3\mathcal{U}]
+ \dots\,,
\\
&
\mathcal{W}_3\times \overline{\mathcal{W}}_6\,\sim\,
\tfrac{3}{14}\sqrt{\tfrac{165}{26}} \,[ \overline{\mathcal{W}}_3] -
\tfrac{14}{3}\sqrt{\tfrac{26}{165}} \,[\overline{\mathcal{O}}]-
6\sqrt{\tfrac{6}{715}} \,[\mathcal{U} \overline{\mathcal{W}}_3] \,,\\
&
\mathcal{W}_6\times \mathcal{U} \,\sim\,
-\tfrac{117}{931}\,[\mathcal{W}_6]+
\tfrac{27}{266}\sqrt{\tfrac{165}{26}}\,[\mathcal{W}_3\mathcal{W}_3]+
4\sqrt{\tfrac{10}{429}}\,[\mathcal{W}_3\mathcal{O}]\,,
\\
&
 \mathcal{O} \times \mathcal{O} \,\sim\,
 \tfrac{405}{161728}
 \sqrt{\tfrac{2145}{2}}\,[\mathcal{W}_6]-
 \tfrac{40095}{283024}\,
 [\mathcal{W}_3\mathcal{W}_3]\,,
\\
&  \mathcal{U} \times \mathcal{O} \,\sim\,
 \tfrac{13365}{59584} [\mathcal{W}_3]-
 \tfrac{885}{7448}
  [\mathcal{O}]-
\tfrac{27}{56}  [\mathcal{U}\mathcal{W}_3]\,.
\end{align}
\end{subequations}
while the $\mathcal{W}$-$\mathcal{W}$, $\overline{\mathcal{W}}$-$\overline{\mathcal{W}}$ OPEs
 are regular.  The \dots above indicate contributions from super-Virasoro primaries with higher conformal dimensions. They have the same quantum numbers as the relations
\eqref{nullClassicalN2rank2}.  It is easy to check that the expression of the  corresponding operators in the free-field realization is  zero.
 Notice that the  first four OPEs are real,  the remaining  nine are complex so  they are accompanied by their conjugate.
 
A few  additional remarks are in order. First of all, in order to constrain the form of the generators in terms of free fields, it is convenient to  start by imposing all the linear constraints originating form 
having the right structure of the OPE of the chiral generators $\mathcal{W}_3,\mathcal{W}_6$
with everything else. Second of all, observe that once $\overline{\mathcal{W}}_3$ is constructed, all the remaining generators, namely $\mathcal{U}, \mathcal{O}, \overline{\mathcal{O}}$ and
 $\overline{\mathcal{W}}_6$
are generated\footnote{The operator $ \overline{\mathcal{O}}$ and 
$\overline{\mathcal{W}}_6$ are generated in the $\overline{\mathcal{W}}_3 \,\mathcal{U}$
and $\overline{\mathcal{W}}_3 \,\overline{\mathcal{O}}$ respectively which are conjugate to the OPEs appearing above. }
in the OPE.
Finally,  let us point out that the procedure given in section \ref{sec:N2freeFIELDprescription}
lives the freedom of redefining $\overline{\mathcal{W}}_6
\rightarrow  x(\alpha)\,\overline{\mathcal{W}}_6+
\alpha\,\overline{\mathcal{W}}_3\overline{\mathcal{W}}_3$.
We fix this freedom by imposing that the OPEs are manifestly  $\mathbb{Z}_2$ conjugation symmetric.





\section{The $R$-filtration from free fields}

\label{sec:R}

This work was largely motivated by the desire to achieve a better understanding  of the VOAs that arise from four-dimensional    ${\cal N}=4$ 
superconformal field theories  via the map introduced in \cite{Beem:2013sza}.  
Applying the construction of  \cite{Beem:2013sza}
to the ${\cal N}=4$ super Yang-Mills theories, one finds a class of ${\cal N}=4$ VOAs labelled by Weyl groups.  In this paper, we have offered an alternative construction
of ${\cal N}=4$ VOAs labelled by Weyl groups,
and conjectured that they are in fact the {\it same} as the ones that arise from SYM theories.\footnote{To our surprise, we have also found ${\cal N}=4$ VOAs associated to non-crystallographic Coxeter groups. Their four-dimensional interpretation (if any) remains unclear.}  

So far, we have studied these algebras in their own right, as novel and interesting examples of VOAs. In this section we wish to go back to their four-dimensional interpretation. We propose that the
free-field representations that we have introduced allow to solve a longstanding  open problem, the assignment of the  ``$R$-filtration'' \cite{Beem:2017ooy}. Let us briefly review the terms of the problem, for the general
case of VOAs that descend from arbitrary ${\cal N}=2$ $4d$ SCFTs.

According to the map of \cite{Beem:2013sza}, the operators of the VOA are in one-to-one correspondence with the so-called ``Schur operators'' of the parent ${\cal N}=2$ SCFT.
Schur operators belong to certain (semi)short representations of the $\sl(4|2)$ superconformal algebra -- as such, due to the shortening conditions, they are labelled by three out of the five Cartan quantum numbers\footnote{In ${\cal N}=4$ and ${\cal N}=3$ SCFTs, Schur operators are labelled by four out of the six Cartans of the $\psl(4|4)$ and $\sl(4|3)$ superconformal algebras, respectively.}
 of $\mathfrak{sl}(4|2)$. 
 Of these quantum numbers, all except  one  survive as good quantum number of the VOA,  defining  {gradings} respected by the operator product expansion.
The exception is the the $\sl(2)_R$ symmetry Cartan, denoted by  $R$. While Schur operators are  all highest weights of $\sl(2)_R$,  the cohomological construction of \cite{Beem:2013sza} involves the lower components of the  $\sl(2)_R$ multiplet to which a Schur operators belongs, and 
as a result  $R$ does {\it not} descend to a grading of the VOA. It is however easy to argue \cite{Beem:2017ooy} that $R$ can only decrease or remain constant 
in the OPE, and as such it defines a {\it filtration} of the VOA.  It follows that any VOA associated to a $4d$ SCFT by the map of \cite{Beem:2013sza} must admit such an $R$-filtration.
 However, the $R$-filtration 
does not appear to be intrinsic to the VOA -- at least, not in any obvious way. Given an abstract presentation of the VOA (say in terms of strong generators and their singular OPEs) it is {\it a priori} unclear how to determine its $R$-filtration.  This is a severe limitation if one's goal is to use the VOA as a tool to study the parent $4d$ theory, because without knowledge of the $R$ quantum number the identification of $4d$ Schur representations
is ambiguous.\footnote{In
 simple special cases, the $R$-filtration can be assigned by specifying a set of rules designed to reproduce  the known Macdonald index of the parent superconformal field theory \cite{Song:2016yfd}, 
but 
the general  story has remained elusive.}

Our main new observation is that the free-field constructions analyzed in this paper come equipped with a natural filtration, which we call the ``${\cal R}$-filtration''.
We conjecture that  for the ${\cal N}=4$ VOAs  that arise from ${\cal N}=4$ SYM theories, the ${\cal R}$- and $R$-filtrations
{\it coincide}. We  have performed several successful checks of this conjecture.
 We expect this statement to generalize
to the VOAs labelled by complex reflection groups that descend from $4d$ ${\cal N}=3$ SCFTs, 
but we did not perform any check of that more general conjecture.

We begin in the next subsection with a review of the $R$-filtration for VOAs that arise from general ${\cal N}=2$ SCFTs, and indicate its obvious extension
to the ${\cal N}=4$ case.
In section \ref{sec:RR}  we show that the VOAs studied in this paper admit a natural ``${\cal R}$-filtration'', defined in terms of their
free-field realizations. According to the basic dictionary of \cite{Beem:2013sza}, the vacuum character of the VOA reproduces the Schur limit of the superconformal index
of its parent $4d$ SCFT, which is a function of a single superconformal fugacity $q$.
Knowledge of the $R$-filtration allows to refine the vacuum character, so that it yields the full Macdonald index of the $4d$ SCFT,
a function of {\it two} superconformal fugacities $q$ and $t$. The basic check that we have correctly identified the $R$-filtration consists in matching the refined vacuum character computed from our free-field realization
with the well-known Madconald index
of an ${\cal N}=4$ SYM theory. In section \ref{sec:identification}, we perform this check  in several examples, up to some order in an expansion in the conformal weight. 
In the limit $q \to 0$, the Macdonald index reduces to the so-called Hall-Littlewood index, which is much simpler to compute. In section~\ref{sec:HL}
 we collect some further evidence for our proposal, showing that it  correctly reproduces the full Hall-Littlewood index in rank-one and  rank-two examples.

\subsection{The   $R$-filtration}

For the reader's convenience, we start by reviewing the salient facts of the $4d$/$2d$ correspondence introduced in  \cite{Beem:2013sza}.


\subsubsection{The $4d$/$2d$ map and Schur operators}

Given an ${\cal N}=2$ SCFT, the associated VOA is obtained by passing to the cohomology of a certain nilpotent fermionic generator $\qq$
of the ${\cal N}=2$ superconformal algebra $\sl(4|2)$. 
We denote the Cartan quantum numbers of $\sl(4|2)$ by $(E, j_1, j_2, R, r)$, where $E$ is the conformal dimension and $j_1$,  $j_2$, $R$,  $r$ are eigenvalues with respect to the Cartan generators of $\sl(2)_1\oplus \sl(2)_2\oplus \sl(2)_R\oplus  {\frak{gl}}(1)_r$, respectively.
The nontrivial cohomology classes of local operators inserted at the origin  in $\mathbb{R}^4 \cong \mathbb{C}^2$ $(z=\bar z = w = \bar w = 0)$ have canonical representatives which are the \emph{Schur operators} \cite{Gadde:2011uv}.
These are local operators whose quantum numbers satisfy the relations\footnote{The first condition in \eqref{eq:schur_conditions} implies the second for representations of $\sl(4|2)$  that can appear in unitary superconformal theories \cite{Beem:2013sza}.}
\begin{equation}
\begin{split}
\label{eq:schur_conditions}
E - (j_1 + j_2) - 2R &= 0~,\\
r + j_1 - j_2 &= 0~.
\end{split}
\end{equation}
Schur operators are always the highest weight states of their respective $\sl(2)_1\oplus \sl(2)_2\oplus \sl(2)_R$ modules. The various (unitary) supermultiplets that contain Schur operators and the positioning of Schur operators within those multiplets is summarized in table \ref{tab:schurTable}.  

\renewcommand{\arraystretch}{1.5}
\begin{table}
\centering
\begin{tabular}{|l|l|l|l|}
\hline
Multiplet					& $\OO_{\rm Schur}$															& $h$			& $r$								\\
\hline
\hline
$\hat\BB_R$					& $\Psi^{11\dots 1}$ 														& $R$			& $0$											\\
\hline
$\DD_{R(0,j_2)}$			& $\wt{\QQ}^1_{\dot +}\Psi^{11\dots1}_{\dot+\dots\dot+}$ 					& $R+j_2+1$		& $j_2+\frac12$ 		\\
\hline
$\bar{\DD}_{R(j_1,0)}$		& ${\QQ}^1_{+}\Psi^{11\dots 1}_{+\dots+}$ 									& $R+j_1+1$		& $-j_1-\frac12$ 				\\
\hline
$\hat{\CC}_{R(j_1,j_2)}$	& ${\QQ}^1_{+}\wt{\QQ}^1_{\dot+}\Psi^{11\dots1}_{+\dots+\,\dot+\dots\dot+}$	& $R+j_1+j_2+2$	& $j_2-j_1$				\\
\hline
\end{tabular}
\caption{\label{tab:schurTable} Summary of the appearance of Schur operators in short multiplets of the ${\cal N}=2$ superconformal algebra, $\sl(4|2)$. The superconformal primary in a supermultiplet is denoted by $\Psi$. There is a single conformal primary Schur operator ${\OO}_{\rm Schur}$ in each listed superconformal multiplet. The holomorphic dimension $h$ and ${\frak{gl}}(1)_r$ charge $r$ of ${\OO}_{\rm Schur}$ are given in terms of the quantum numbers $(R,j_1,j_2)$ that label the shortened multiplet (left-most column). } 
\end{table}

Finite linear combinations of local operators inserted away from the origin cannot define nontrivial $\qq$-cohomology classes unless $w=\bar w=0$.
A canonical choice of representatives for local operators inserted on the $w=\bar w=0$ plane, $\Cb_{[z,\zb]}$, is given by \emph{twisted translated} Schur operators,
\begin{equation}
\label{eq:twisted_translated}
\OO(z)\equiv e^{zL_{-1}+\bar{z}(\overline{L}_{-1}+R^-)}\OO_{\rm Sch}(0)e^{-zL_{-1}-\bar{z}(\overline{L}_{-1}+R^-)}~,
\end{equation}
where $L_{-1}$ and $\bar{L}_{-1}$ are the generators of holomorphic and antiholomorphic translations in $\Cb_{[z,\zb]}$, $R^{-}$ is the lowering operator of $\mathfrak{su}(2)_R$, and $\OO_{\rm Sch}(z,\zb)$ is a Schur operator. The OPE of twisted-translated Schur operators, taken at the level of $\qq$-cohomology, is $\slf(2)_z$ covariant and $\slf(2)_{\bar z}$ invariant, with the holomorphic dimension of the twisted-translated operator $\OO(z)$ being determined in terms of the quantum numbers of the corresponding Schur operator according to
\begin{equation}
\label{eq:holomorphic_dimensions}
[L_0,\OO(z)] = h\, \OO(z)~,\qquad h = \frac{E + j_1 + j_2}{2} = E - R~.
\end{equation}
This holomorphic OPE endows the vector space  of Schur operators with the structure of a vertex operator algebra.

In this work, we use the notation $\{ AB \}_n$ to denote the coefficient of $z_{12}^{-n}$
in the holomorphic OPE $A(z_1) B(z_2)$, see \eqref{bracket_notation}. We find it convenient to introduce
alternative notations for the special cases $n =0,1$. More precisely, we define
\beq \label{NO_and_bracket}
{\rm NO}[A, B] := \{ AB \}_0 \ , \qquad   \{\!\!\{ A, B  \}\!\!\}  := \{ AB \}_1 \ .
\eeq
We refer to these operations as the normal order product of $A$ and $B$,
and the bracket of $A$ and $B$, respectively.

\subsubsection{Gradings and filtration} \label{gradings_and_filtration}

The vector space ${\cal V}$ of Schur operators has a triple grading by $(h, R, r)\in \hf\Zb_+\times\hf\Zb_+\times\hf\Zb$,
\begin{equation}
\VV=\bigoplus_{h,R,r}\VV_{h,R,r}~.
\end{equation}
The normally-ordered product preserves $h$ and $r$ but not $R$, making the $R$ grading unnatural from the point of view of the VOA structure. However, the specifics of the twisted translation construction implies that $R$-charge violation occurs with a definite sign,
\begin{equation}\label{eq:R-violation}
\NO{\VV_{h_1,R_1,r_1}}{\VV_{h_2,R_2,r_2}}\subseteq \bigoplus_{k\geqslant0} \VV_{h_1+h_2,R_1+R_2-k,r_1+r_2}~.
\end{equation}
Consequently, there is a \emph{filtration} by $R$ that is preserved by the normally-ordered product. That is, if we define,
\begin{equation}\label{eq:R-filtration}
\FF_{h,R,r}=\bigoplus_{k\geqslant 0}\VV_{h,R-k,r}~,
\end{equation}
then we have the following filtered property for normally-ordered multiplication,
\begin{equation}\label{eq:R-filtered-product}
{\rm NO}(\FF_{h_1,R_1,r_1},\FF_{h_2,R_2,r_2})\subseteq \FF_{h_1+h_2,R_1+R_2,r_1+r_2}~.
\end{equation}
In addition, the bracket operation $\{\!\!\{ \cdot, \cdot \}\!\!\}$ defined
in \eqref{NO_and_bracket} obeys
\begin{equation}\label{eq:R-filtered-bracket}
\{\!\!\{    \FF_{h,R,r},\FF_{h',R',r'}   \}\!\!\}   \subseteq \FF_{h+h'-1,R+R'-1,r+r'}~,
\end{equation}
so, the bracket is filtered of tri-degree $(-1,-1,0)$.
The  associated graded of our filtered VOA is defined in the usual fashion,
\begin{equation} \label{associated_graded}
{\rm gr}_\FF\VV=\bigoplus_{h,R,r} \GG_{h,R,r}~,\qquad \GG_{h,R,r}= \FF_{h,R,r}/\FF_{h,R-1,r}~.
\end{equation}
On this space, which is isomorphic as a vector space to $\VV$,  one can show that the normally-ordered product induces a grade-preserving (with respect to all the gradings) commutative, associative product, and the bracket induces an anti-symmetric bracket of tri-degree $(-1,-1,0)$ that obeys the Jacobi identity. %

 Table \ref{tab:schurTable} makes it clear that knowledge of only the $h$ and $r$ quantum numbers state of the VOA is not sufficient to characterize uniquely the $4d$ protected operator from which it descends---the $R$ quantum number is also needed for a precise uplift.


\subsubsection{The  superconformal index}

The superconformal index of a $4d$ $\cN = 2$ SCFT $\mathcal T$ (see, {\it e.g.}, \cite{Gadde:2011uv, Rastelli:2014jja})
is defined as
\beq
\cI^{\mathcal T}(p,q,t) := {\rm STr}_{{\cal H}  }  \left( p^{\frac 12 (E + 2 j_1 - 2 R  - r)} \,
q^{\frac 12 (E - 2 j_1 - 2 R  - r]} \, t^{R+r}
   \right) \ ,
\eeq
where ${\rm STr}$ denotes the supertrace, and $\cH$ is the Hilbert
space of local operators of the SCFT. The index receives
contributions only from the states that lie in short 
representations of the superconformal algebra, 
with the contributions being such that the index is insensitive
to recombinations.
Let us also recall the definition of two
special limits of the full superconformal index:
the Macdonald index and the Schur index.
The Macdonald index 
depends on two fugacities only, and is given by
\beq\label{MacDonaldIndexDEF}
\cI^{\mathcal T}_{\rm Macdonald}(q,t): = {\rm STr}_{{\cal H}_M  }  (  q^{E - 2 R - r} \, t^{R+r}
   ) \ ,
\eeq
where $\cH_M$ denotes the subspace of states in $\cH$ 
satisfying $E + 2j_1 - 2R -r = 0$.
The Schur index depends on
only one fugacity, and reads
\beq
\cI^{\mathcal T}_{\rm Schur}(q): = {\rm STr}_{{\cal H}  }  (  q^{E-R}
   ) \ .
\eeq
Notice that, even if the trace is taken over the entire Hilbert space
$\cH$, only Schur operators contribute to the
Schur limit of the superconformal index.

Under the map of \cite{Beem:2013sza}, the Schur index of a $4d$ $\cN = 2$ SCFT
is mapped to the vacuum character of the associated VOA,\footnote{The usual vacuum character of a VOA should includes the additional overall factor $q^{-c/24}$, where $c$ is the central charge.
This factor is necessary (among other things) to obtain the correct modular  properties, but we are omitting it here since it plays no role in the comparison with the $4d$ superconformal index.
In the conventions used here, both the $2d$ character and the $4d$ superconformal index are normalized to start with ``1'' in their $q$ expansion.}
\beq
\mathcal V = \chi[\cT] \ , \qquad \chi_{\mathcal V} (q) :=  {\rm STr}_{ \mathcal V  }    (q^{L_0 }) \ , \qquad
 \chi_{\mathcal V} (q) = \cI^{\mathcal T}_{\rm Schur}(q) \ .
\eeq
In order to reconstruct the 
Macdonald index \eqref{MacDonaldIndexDEF}
of the parent $4d$ theory from VOA data,
we need control over the $r$ grading and the $R$ filtration
discussed in section \ref{gradings_and_filtration}.
If these pieces of data are successfully identified,
the Macdonald index can be recovered via
a refinement of the VOA vacuum character,
\beq
\mathcal V = \chi[\cT] \ , \qquad 
\cI^{\mathcal T}_{\rm Macdonald}(q,t) = \chi_{\mathcal V}(q,t) :=
{\rm STr}_{ \mathcal V  }    (q^{L_0 - R -r } \, t^{R+r}) \ .
\eeq
In the definition of $ \chi_{\mathcal V}(q,t)$, the supertrace is
implicitly understood to be taken on the associated
graded algebra \eqref{associated_graded},
in which both quantum numbers $R$ and $r$
define gradings.

No simple recipe is known to extract
the $R$-filtration from a presentation of the VOA
in terms of strong generators and their singular OPEs.
In this work, we propose an efficient way to
identify the $R$-filtration in the case in which the
VOA coincides with $\cW_{\mathsf G}$ for some
(crystallographic) complex reflection group 
$\mathsf G$. Our proposal relies on the free-field
realization of the VOA. More precisely, 
the free-field realization of 
$\cW_{\mathsf G}$ allows us to introduce
a filtration on $\cW_{\mathsf G}$,
denoted $\mathcal R$-filtration and defined in section \ref{sec:RR}.
We propose the identification of this new $\mathcal R$-filtration
with the sought-for $R$-filtration.
Several checks of this proposal are discussed in sections \ref{sec:identification}
and \ref{sec:HL}.

\subsubsection{The ${\cal N}=4$ case}

Unitary irreducible highest weight representations of the four dimensional $\mathcal{N}=4$ 
super-conformal algebra 
$\mathfrak{psu}(2,2|4)$ are labelled by six quantum numbers $\{E,(j_2,j_2),[q_1,p,q_2]\}$ where 
$E,(j_1,j_2)$ are as in the previous discussion while $[q_1,p,q_2]$
 are $\mathfrak{su}(4)_R$ Dynkin labels, see \cite{Dolan:2002zh}.
 When these quantum numbers satisfy certain relations the corresponding 
supermultiplet obeys shortening conditions.
One of the properties of the chiral algebra  map  $\chi$ introduced in  \cite{Beem:2013sza} 
specialized to four dimensional $\mathcal{N}=4$ SCFT is that it acts on irreducible representations as
\begin{equation}
\chi\,:\,\, \text{Reps of $\mathfrak{psl}(4|4)$}\,\rightarrow  \text{Reps of $\mathfrak{psl}(2|2)$}\,.
\end{equation}
A little inspection shows that, using the notation of \cite{Dolan:2002zh} for four dimensional multiplets
 \begin{equation}
 \label{chiMAPpsu}
\begin{aligned}
\chi(\mathcal{B}_{[0,p,0]})&=\,\mathfrak{S}_{h=\tfrac{1}{2}p}\\
\chi(\mathcal{B}_{[q,p,q]})&=\,\mathfrak{L}_{(h,j)=(q+\tfrac{1}{2}p,\tfrac{1}{2}p)}\\
\chi(\mathcal{C}_{[q_1,p,q_2],(j_1,j_2)})&=\,\mathfrak{L}_{(h,j)
=(\tfrac{1}{2}(p+q_1+q_2)+j_1+j_2+2,\tfrac{1}{2}p)}\\
\chi(\mathcal{D}_{[q_1,p,q_2],(0,j_2)})&=\,\mathfrak{L}_{(h,j)
=(\tfrac{1}{2}(p+q_1+q_2)+j_2+1,\tfrac{1}{2}p)}\\
\chi(\bar{\mathcal{D}}_{[q_1,p,q_2],(j_1,0)})&=\,\mathfrak{L}_{(h,j)
=(\tfrac{1}{2}(p+q_1+q_2)+j_1+1,\tfrac{1}{2}p)}\\
\chi(\mathcal{A})&=\,0\,.
\end{aligned}
\end{equation}
The action \eqref{chiMAPpsu} can be determined by considering the 
decomposition of $\mathcal{N}=4$ superconformal multiplets in 
$\mathcal{N}=2$ superconformal multiplets and the results of table \ref{tab:schurTable}. 
The $\mathcal{N}=2$ superconformal algebra is embedded as
\beq\label{embedding}
\mathfrak{psl}(4|4)\,\supset\,\mathfrak{sl}(4|2)\oplus \mathfrak{sl}(2)_y\,,
\eeq
with 
\beq
\mathfrak{sl}(4)_R\,\supset\,\mathfrak{sl}(2)_R\oplus \mathfrak{sl}(2)_y\oplus \mathfrak{gl}(1)_r\,,
\qquad
[1,0,0]\mapsto (\tfrac{1}{2},0)_{+\frac{1}{2}} \oplus (0,\tfrac{1}{2})_{-\frac{1}{2}}\,.
\eeq
The decomposition of the short supermultiplets with respect to the embedding
\eqref{embedding} is given by\footnote{A similar decomposition in the $\mathcal{N}=3$ case is given in equations  $(2.6)$-$(2.13)$ of \cite{Lemos:2016xke}.}
\begin{subequations}
\label{N2N4brancing}
\begin{align}
\mathcal{B}_{[q,p,q]}\,& \rightarrow \hat{\mathcal{B}}_{R=\frac{p}{2}+q}\otimes [\tfrac{p}{2}]\,+\,\dots \,
 \\
\mathcal{C}_{[q_1,p,q_2],(j_1,j_2)}\,& \rightarrow\,
\hat{\mathcal{C}}_{R=\frac{p+q_1+q_2}{2},(j_1,j_2)}\otimes [\tfrac{p}{2}]
\,+\,\dots \\
\mathcal{D}_{[q_1,p,q_2],(0,j_2)}\,& \rightarrow\,
\mathcal{D}_{R=\frac{p+q_1+q_2}{2},(0,j_2)}\otimes [\tfrac{p}{2}]
\,+\,\dots \\
\bar{\mathcal{D}}_{[q_1,p,q_2],(j_1,0)}\,& \rightarrow\,
\bar{\mathcal{D}}_{R=\frac{p+q_1+q_2}{2},(j_1,0)}\otimes [\tfrac{p}{2}]
\,+\,\dots
\end{align}
\end{subequations}
where the \dots indicate terms that are either obtained by acting with the $Q,\bar{Q}$ 
generators in  $\mathfrak{psl}(2|2)$ 
or that do not contain Schur operators\footnote{
More precisely, the terms \dots~are obtained by acting with the generators of $\mathfrak{t}$ on the 
first terms in \eqref{N2N4brancing}, where
$\mathfrak{psl}(4|4)\simeq \mathfrak{sl}(4|2)\oplus \mathfrak{sl}(2)_y\oplus\,\mathfrak{t}$
as a vector space.  The $\qq$-cohomology of $\mathfrak{t}$  is spanned by the
fermionic generators in $\mathfrak{psl}(2|2)$. These generate Schur operators when acting on the generalized highest weight state. The elements of $\mathfrak{t}$ that are not in the $\qq$-cohomology  on the other hand do not generate Schur operators.
}.
The chiral algebra map \eqref{chiMAPpsu} follows: 
$h$ is the same as in the $\mathcal{N}=2$ case, see   table \ref{tab:schurTable} 
and $j=\tfrac{p}{2}$.
%
To be precise, the image of the  map $\chi$ is a representation of $\mathfrak{pl}(2|2)$,
which is  the extension of 
$\mathfrak{psl}(2|2)$ by the outer automorphism $GL(1)_r\subset SL(2)$. 
The  $GL(1)_r$ quantum number of the  $\mathfrak{pl}(2|2)$ primary is obtained combining \eqref{N2N4brancing} with 
the content of  table \ref{tab:schurTable}.
 Similarly the $R$ quantum number is found to be 
\begin{subequations}
\label{Runderchimap}
\begin{align}
R[\mathcal{B}_{[q,p,q]}]&=q+\tfrac{1}{2}p\,,
&
R[\mathcal{C}_{[q_1,p,q_2],(j_1,j_2)}]&=\tfrac{1}{2}(p+q_1+q_2)+1\,,\\
R[\mathcal{D}_{[q_1,p,q_2],(0,j_2)}]&=\tfrac{1}{2}(p+q_1+q_2)+\tfrac{1}{2}\,,
&
R[\bar{\mathcal{D}}_{[q_1,p,q_2],(j_1,0)}]&=\tfrac{1}{2}(p+q_1+q_2)+\tfrac{1}{2}\,,
\end{align}
\end{subequations}
where the first formula can be applied to all values of $q$ including $q=0$.
An important feature of the map $\chi$ is that in general it cannot be inverted since 
different types of four dimensional multiplets correspond to the same $\mathfrak{pl}(2|2)$ multiplet.
The pair of maps $( \chi,R )$, on the other hand, can be inverted.

\subsection{The $\mathcal{R}$-filtration }
\label{sec:RR}
According to our free-field construction,
$\cW_\Gamma$ is realized as a subalgebra of the 
total $\beta\gamma bc$ system $\mathbb{M}^{(\Gamma)}_{ \beta\gamma bc }$
generated by
 $\beta_\ell$, $\gamma_\ell$, $b_\ell$, $c_\ell$, with $\ell = 1, \dots, \mathsf{r}$,
where
$\mathsf{r} = {\rm rank}(\Gamma)$.
 We first define a filtration $\widetilde \cR$
at the level of the $\beta \gamma bc$ system.
To this end, it is sufficient to assign an additive 
weight, referred to as $\cR$-weight in the following, to the 
fundamental `letters'  $\beta_\ell$, $\gamma_\ell$, $b_\ell$, $c_\ell$,
according to the following table,
\vspace{1mm}
\begingroup
\renewcommand{\arraystretch}{1.3}
\begin{equation}
\begin{tabular}{| c||c | c |c | c|c | }
\hline
 & $\beta_\ell$ & $\gamma_\ell$ & $b_\ell$ & $c_\ell$ & $\partial$  \\
 \hline \hline
${\cal R}$ & $\frac 12 \, p_\ell$ & $1- \frac 12 \, p_\ell$ & $\frac 12 \, p_\ell$ & 
$1- \frac 12 \, p_\ell$  & $0$  
 \\
 \hline
$h-{\cal R}$ & $\,\,0$ &  $\,\,0$ & $+\frac 12$ & 
$- \frac 12$ & $1$  \\
\hline
\end{tabular}
\label{tableRweightsbetagammabc}
\end{equation}
\endgroup

\vspace{1mm} \noindent
For convenience we reported here also the combination $h- \cR$.
Together with
\eqref{betagammabc_hmr}, we have
thus  assigned quantum numbers $h$, $m$, $r$, $\cR$ to the free fields.
Given any monomial in derivatives of 
$\beta_\ell$, $\gamma_\ell$, $b_\ell$, $c_\ell$, 
its $\cR$-weight is simply the sum of the $\cR$-weights of its letters.
Given a polynomial, we define its $\cR$-weight as the maximal
$\cR$-weight of its monomials.
Given  $\cR\in\tfrac{1}{2}\mathbb{Z}$, we may then 
introduce the vector space
\beq\label{FtildeMORE}
\widetilde \cV_\cR :=
 \text{linear span of states in $\mathbb{M}^{(\Gamma)}_{ \beta\gamma bc }$ with $\cR$-weight $\cR-k$, $k \in \mathbb Z_{\ge 0}$ }
 \ .
\eeq
The collection of vector spaces $\{ \widetilde \cV_{\cR} \}_{\cR \in \frac 12 \mathbb Z}$
defines the filtration $\widetilde \cR$
of $\mathbb{M}^{(\Gamma)}_{ \beta\gamma bc }$,
\beq
\mathbb{M}^{(\Gamma)}_{ \beta\gamma bc } =   \bigcup_{R \in \frac 12 \mathbb Z} \widetilde
\cV_\cR 
 \ .
\eeq
To verify the compatibility of the filtration with the VOA structure
of $\mathbb{M}^{(\Gamma)}_{ \beta\gamma bc }$,
we have to ensure that, 
in the OPE of two operators of $\cR$-weights $\cR_1$ and $\cR_2$,
only operators with $\cR$-weight $\le \cR_1 + \cR_2$  appear,
both in the singular and in the regular part.
This property is easily verified by
noticing that 
 OPEs in  $\mathbb{M}^{(\Gamma)}_{ \beta\gamma bc }$
are computed via Wick contractions.
Since 
the $\cR$-weights of a pair $\beta_\ell \, \gamma_\ell$
or $b_\ell c_\ell$ is one, every Wick contraction decreases the 
$\cR$-weight by one unit. 
The same observation implies that in the singular OPEs the 
$\cR$-weights are strictly smaller than $\cR_1+\cR_2$.

The filtration $\widetilde \cR$ descends to a filtration 
$\cR$ of the VOA $\cW_\Gamma$,
since the latter is a subalgebra of $\mathbb{M}^{(\Gamma)}_{ \beta\gamma bc }$.
The filtration $\cR$ is specified by the collection
of vector spaces
\beq\label{Wgammafilt1}
\cV_\cR := \widetilde \cV_\cR \cap \cW_{\Gamma} \,  , \qquad \cR \in \tfrac 12 \mathbb Z \ .
\eeq
Notice that in $\mathbb{M}^{(\Gamma)}_{ \beta\gamma bc }$
 we can construct operators with arbitrarily
low $\cR$-weight. In contrast, all operators in $\cW_\Gamma$
have $\cR \ge 0$.
As a result, we can write
\beq\label{Wgammafilt2}
\cW_\Gamma = \bigcup_{\cR \in \frac 12 \mathbb Z_{\ge 0}} \cV_\cR 
 \ .
\eeq
Given any filtered algebra such as the pair $(\cW_\Gamma, \cR)$,
there is a standard  coset
construction that yields a graded algebra,
the associated graded algebra  $\mathscr G(\cW_\Gamma)$,
\beq\label{associated_graded2}
\mathscr G(\cW_\Gamma) = \bigoplus_{\cR \in \frac 12 \mathbb Z_{\ge 0}}
\cH_\cR \ , \qquad \cH_0 = \cV_0\simeq \mathbb{C} \,,\qquad  \cV_{\frac{1}{2}}=0\ , \qquad
\cH_\cR = \cV_\cR/ \cV_{\cR - 1} \ , \quad \cR \ge 1  \ .
\eeq
Notice that  $\mathscr G(\cW_\Gamma)$
and  $\cW_\Gamma$ are isomorphic as vector spaces.
The graded algebra $\mathscr G(\cW_\Gamma)$
will be useful in the next paragraph for the definition
of the refined vacuum character.

It should be noticed that 
all generators of the small
$\cN = 4$ algebra have  $\cR=1$ and they actually 
exhaust the $\cR=1$ component, {\it i.e.},  $\cV_1\simeq \mathfrak{S}_1$ 
as  $\mathfrak{psl}(2|2)$ modules.
 In more generality one can argue 
 that the action of  $\mathfrak{psl}(2|2)$ does not change
 the  $\cR$-weight.
 A particular example is given by operators transforming in short representations of 
  $\mathfrak{psl}(2|2)$. In this case the weight of the super-multiplet can be determined unambiguously since the corresponding  quasi-primary, due to the $h=m$ condition, is necessarily a function of
   $\{\beta_1,\dots,\beta_{\mathsf{r}}\}$   so its  $\cR$-weight is equal to its conformal dimension, $\cR=h$. As we will see in the next subsection, this fact has a clear four-dimensional interpretation. 

\paragraph{Refined vacuum character.}
By means of the filtration $\cR$,
we can refine the vacuum character of the 
VOA $\cW_\Gamma$.
This is achieved in a standard way via the 
associated graded algebra $\mathscr G(\cW_\Gamma)$
defined in  \eqref{associated_graded2}:
\beq\label{refined_vacuum_char}
\chi_{\cW_\Gamma}(q,\xi, z)\,=\,
\sum_{h,r,m,\cR}\, 
 \text{sdim} (\cH_{h,r,m,\cR} ) \, q^h \, \xi^{\cR+r} \, z^{m}  \,,
\qquad
\cH_{\cR}=\bigoplus_{h,r,m}\,\cH_{h,r,m, \cR}\,.
\eeq
Some comments are in order. The variables $q$, $\xi$, $z$
are fugacities. The quantum numbers $h$, $m$, $r$ were defined in 
section \ref{sec:prelliminaries}, their characterization is reported
 here for convenience:
$h$ is the conformal dimension,
$r$ is  associated to the outer automorphisms of the small
$\cN = 4$ and $\cN = 2$ SCA normalized as  $r[G]= 1/2$ and $r[\cG]= 1/2$ respectively,
finally $m \in \tfrac{1}{2}\mathbb{Z}$ is the  Cartan of the $\mathfrak{sl}(2)_y$ 
 of the small $\cN = 4$ SCA or the $\mathfrak{gl}(1)$
of the $\cN = 2$ SCA.
The super-dimension $\text{sdim}(\mathcal{H})$ is defined as 
 $\dim(\mathcal{H}^{(\text{bos})})-\dim(\mathcal{H}^{(\text{ferm})})$.
Notice that the character \eqref{refined_vacuum_char} can
 be further refined by applying the substitution
 $\xi^{\cR+r} \mapsto \hat{\xi}_1^{\cR}\,y^{2r}$
 in \eqref{refined_vacuum_char}.

\paragraph{On the computation of $\text{sdim} (\cH_{h,r,m,\cR} )$.}
Let us comment briefly on the computation of 
the refined character \eqref{refined_vacuum_char} level by level.
The VOA $\cW_\Gamma$ is triply graded as
\beq
\mathcal{W}_{\Gamma}\,=\,
\bigoplus_{h\geq 0,r,m}\, V_{h,r,m}   \ , \qquad
V_{h,r,m}   =  V_{h,r,m}^{(\rm bos)}  \oplus  V_{h,r,m}^{(\rm ferm)}    \,,
\eeq
where each $V_{h,r,m}^{\rm (bos)}$ and $V_{h,r,m}^{\rm (ferm)}$ are 
finite-dimensional.
Given any base of $V_{h,r,m}^{\rm (bos)}$,
we can determine the $\cR$-weights of each element
of the base using the free-field construction.
Let us   select a basis of $V_{h,r,m}^{(\rm bos)}$
that is minimal, in the sense that the
$\cR$-weights of its elements are the lowest possible.
Let $\cR_1< \dots < \cR_k$ be the distinct
$\cR$-weights of the elements of a minimal basis,
occurring with multiplicities $n_1$, \dots, $n_k$.
In an analogous fashion, we can determine a minimal basis of
$V_{h,r,m}^{(\rm ferm)}$, in which the
distinct $\cR$-weights are 
  $\cR'_1< \dots < \cR'_{k'}$,
occurring with multiplicities
$n'_1$, \dots, $n'_{k'}$.
The contribution of $V_{h,r,m}$ to the refined
character \eqref{refined_vacuum_char} is then
\beq \label{operational_R_grading}
\sum_{i = 1}^k \,  n_i \, q^h  \, \xi^{\cR_i+r} \, z^{m}
- \sum_{i = 1}^{k'} \,  n'_i \, q^h  \, \xi^{\cR'_i+r} \, z^{m}  \ .
\eeq
\vspace{1mm}

\subsection{Identification of $\cR$-filtration and $R$-filtration}

\label{sec:identification}

We propose the following identification:\footnote{We
expect that the obvious generalization of this statement 
to the case of 
 complex reflection groups is true, 
but we did not perform any check of this conjecture.
}

\vspace{0.4cm}
\noindent
\emph{Suppose $\cW_\Gamma$ is the VOA associated to a four-dimensional
 $\mathcal{N}=4$ SCFT $\cT$,
 $\cW_\Gamma = \chi[\cT]$.
Then the $\cR$-filtration of $\cW_\Gamma$,
defined in \eqref{tableRweightsbetagammabc}, \eqref{FtildeMORE}, \eqref{Wgammafilt1}, \eqref{Wgammafilt2}, coincides with the
$R$-filtration 
canonically attached to $\cW_\Gamma$
by its definition in terms of four-dimensional CFT data
of $\cT$.
}
\vspace{0.4cm}

\noindent This proposed identification implies in particular the 
equality of the Macdonald index of $\cT$
with the refined vacuum character \eqref{refined_vacuum_char} of $\cW_\Gamma$,
\beq \label{macdonald_conjecture2}
\chi_{
\cW_{
\Gamma
}
} (q,\xi,z) 
= \cI^{\cT}_{\rm Macdonald} (q,\xi,z) \ ,  \qquad
\cW_\Gamma = \chi[\cT] \ ,
\eeq
where we have refined the $4d$ 
Macdonald index  
using the $\mathfrak{sl}(2)_y$ flavor fugacity $z$.
To provide evidence for \eqref{macdonald_conjecture2}, we followed two approaches.
Firstly, we performed a direct match of the Macdonald
index of $4d$ $\cN =4$ SYM with
gauge algebra $\mathfrak a_1$, $\mathfrak a_2$,
 and the refined vacuum character 
of $\cW_\Gamma$ for $\Gamma = A_1, A_2$, 
 working up to and including 
terms $q^5$, $q^3$ respectively.
Secondly, we analyzed the  Hall-Littlewood limits
of the Macdonald index of $4d$ $\cN = 4$ SYM with gauge algebra 
$\mathfrak g$ and of the refined vacuum
character of $\cW_{\rm Weyl (\mathfrak g)}$, and we provided general arguments for their equality,
for any simple Lie algebra $\mathfrak g$.

\paragraph{The Macdonald  index of $4d$ $\cN = 4$ SYM.}
The definition of the Macdonald index was recalled in \eqref{MacDonaldIndexDEF}.
For $4d$ $\mathcal{N}=4$ SYM with  gauge group $G$, 
the Macdonald index can be written as an integral over the maximal torus 
of $G$ as follows
\beq \label{g_Macdonald}
 \cI^{\text{SYM}(\mathfrak{g})}_{\rm Macdonald} (q,\xi,z)
 = \int [du] \, {\rm P.E.} \bigg[ \xi^{\frac 12}\,
 \chi^{\mathfrak{pl}(2|2)}_{\mathfrak{S}_{\frac{1}{2}}} (q,\xi,z)\,
  \chi^{\mathfrak{g}}_{\text{Adj}}(u)  \bigg] \, .
\eeq
The notation deserves some explanations:
 $\mathfrak{g}=\text{Lie}(G)$ and 
$[du]$ is the corresponding normalized Haar measure, the $\mathfrak{pl}(2|2)$
character of the extra-short representation $\mathfrak{S}_{\frac{1}{2}}$ is given by
\beq
\chi^{\mathfrak{pl}(2|2)}_{\mathfrak{S}_{\frac{1}{2}}} (q,\xi,z)\,=\,
 \frac{q^{\frac 12} \!} {1-q}\, (z^{\frac{1}{2}}-z^{-\frac{1}{2}})
-\frac{q}{1-q} (\xi^{\frac{1}{2}}-\xi^{-\frac{1}{2}}) \ .
\eeq
Finally the plethystic exponential is defined as 
\beq
{\rm P.E.}[ f(q,\xi,z,u)] = \exp \bigg[ \sum_{m = 1}^\infty \frac 1m  f(q^m,\xi^m,z^m,u^m) \bigg] \ .
\eeq

\vspace{1mm}

\paragraph{Match between index and character for $\Gamma = A_1$.}
In the case $\Gamma = A_1$, the VOA $\cW_\Gamma$ is  
(the simple quotient of)
the small $\cN = 4$ SCA with central charge $c = -9$.
To compute the refined vacuum  character \eqref{refined_vacuum_char} up to and including $q^5$
terms, we performed a counting of states with definite
quantum numbers $h$, $j$, $r$, $\cR$, up to $h =5$. The results
are collected in table \ref{su2_table}. Clearly, all states with integer $h$ are bosons,
and all states with half-integer $h$ are fermions.
Let us stress that this counting differs from the analogous counting
at generic central charge $c$, because of the states
that become null upon setting $c  = -9$.
Fortunately, the counting of states is facilitated by the fact that,
in the free-field realization, all null states are automatically
zero, as proven in \cite{Adamovic:2014lra}.

\begin{table}[H]
\centering
\begin{tabular}{| c || l |}
\hline 
$h$ & $\mathfrak{su}(2)$ multiplets with quantum numbers $(j)_r^{\cR}$ \\
\hline \hline
0 & $(0)^0_0$  \\ \hline 
$\frac 12$ & --- \\ \hline
1 & $(1)_0^1$ \\ \hline
$\frac 32$ &  $(\frac 12)_{\pm 1/2}^1$  \\ \hline
2 & $(2)^2_0  + (1)^1_0  + (0)^1_0$ \\ \hline
$\frac 52$ &  $(\frac 32)_{\pm 1/2}^2 +(\frac 12)^1_{\pm 1/2}$ \\ \hline
3 & $(3)^3_0 + (2)^2_0 + 2\, (1)^2_0 + (1)^1_0 + (0)^1_0$ \\ \hline
$\frac  72$ & 
$(\frac 52)_{\pm 1/2}^3  + 2  \, (\frac 32)^2_{\pm 1/2} 
+ (\frac 12)^2_{\pm 1/2}  + (\frac 12)^1_{\pm 1/2}$    \\ \hline
4 & 
$(4)_0^4  + (3)^3_0 + 2 \, (2)^3_0  + 2 \, (2)^2_0 
+ 4 \, (1)^2_0  + (1)^1_0  
+ 2 \, (0)^2_0  + (0)^1_0  + (1)^2_{\pm 1} $
  \\  \hline
$\frac 92$ & 
$(\frac 72)_{\pm 1/2}^4  + 2 \, (\frac 52)_{\pm 1/2}^3 
+ (\frac 32)^3_{\pm 1/2} + 3 \, (\frac 32)^2_{\pm 1/2} 
+ 3 \, (\frac 12)^2_{\pm 1/2}  + (\frac 12)^1_{\pm 1/2}$ 
\\ \hline
$5$ & 
$(5)_{0}^5 + (4)_{0}^4   + 2 \, (3)_0^4
  + 2 \, (3)_0^3
+ 5\, (2)_0^3
+ 2\, (2)_0^2 
+ 2\, (1)_0^3
+ 7\, (1)_0^2
+   (1)_0^1 $
\\
& $   
{}+ 3\, (0)_0^2
+    (0)_0^1 
+  (2)_{\pm1}^3
+  (1)_{\pm 1}^2
+  (0)_{\pm 1}^2
$
\\
\hline
\end{tabular}
\caption{States in the VOA $\cW_\Gamma$ with $\Gamma =A_1$ with $h \le 5$.
The notation $(j)^\cR_r$ encodes the spin $j$ under
the $\mathfrak{sl}(2)_y$ symmetry of the small $\cN = 4$ SCA,
the outer automorphism quantum number $r$,
and the $\cR$-weight. When we write, for instance, $(\frac 12)^1_{\pm 1/2}$,
we mean that multiplets with $r = 1/2$ and $r = -1/2$
are found with the same multiplicity.
}
\label{su2_table}
\end{table}

Upon assembling the refined vacuum character
from the data of table \ref{su2_table}, we found a perfect match
with the Macdonald index of $4d$ $\cN = 4$ SYM with gauge algebra
$\mathfrak{a}_1$, up to $q^5$ terms. The Macdonald index is computed using
\eqref{g_Macdonald}.

The content of table \ref{su2_table} can also be encoded more compactly
in 
the expression
 \beq\label{A1MacDcharacter}
 \chi_{A_1}(q,\hat{\xi},y,z)=
1+\chi^{(A_1)}_{\text{S}}
 +
 \hat{\xi}^{2}\,\mathfrak{L}_{(3,1)}
 +  \hat{\xi}^{3}\,\mathfrak{L}_{(4,2)}
 +  \hat{\xi}^{2}\,\mathfrak{L}_{(4,0)}
 + \hat{\xi}^{4}\,\mathfrak{L}_{(5,3)}
  + \hat{\xi}^{3}\,\mathfrak{L}_{(5,2)}
   + (\hat{\xi}^{3}+\hat{\xi}^{2})\,\mathfrak{L}_{(5,1)}+\dots
 \eeq
The RHS is obtained from \eqref{refined_vacuum_char}
by means of the further refinement
 $\xi^{R+r} \mapsto \hat{\xi}_1^{R}\,y^{2 r}$. 
 On the LHS, 
 $\chi^{(A_1)}_{\text{S}}=\sum_{R=1}^{\infty}\,\hat{\xi}^{R}\,\mathfrak{S}_R$
 and $\mathfrak{S}_R(q,y,z)$, $\mathfrak{L}_{(h,j)}(q,y,z)$ denote the  $\mathfrak{pl}(2|2)$
 character of the
 corresponding representations introduced in  table \ref{table_reps}.

\paragraph{Match between index and character for $\Gamma = A_2$.}
The counting of states up to $h=3$ in the case $\Gamma = A_2$
is summarized in table \ref{su3_table}. In contrast to the case $\Gamma = A_1$,
this time we have a non-trivial interplay between bosons and
fermions with the same weight $h$. Once again, it is essential to take into
account the states that become null for $c = -24$, but luckily this task is
efficiently performed with the help of the free-field realization.

Constructing the refined vacuum character \eqref{refined_vacuum_char}
up to $q^3$
from the data in table \eqref{su3_table}, we find a perfect
match with the Macdonald index of $4d$ $\cN = 4$ SYM
with gauge algebra $\mathfrak a_2$, as given by
\eqref{g_Macdonald}.

We can repackage the content of table \ref{su3_table} in the 
compact expression
 \beq\label{A2MacDcharacter}
 \chi_{A_2}(q,\hat{\xi},y,z)=
1+\chi^{(A_2)}_{\text{S}}
 +
 \hat{\xi}^{2}\,\mathfrak{L}_{(2,0)}
 +
( \hat{\xi}^{2}+\hat{\xi}^{3})\,\mathfrak{L}_{(3,1)}
 +  \hat{\xi}^{5/2}\,\mathfrak{L}_{(\frac{5}{2},\frac{3}{2})}+\dots
 \eeq
where we are using the same notation as is \eqref{A1MacDcharacter}, and on the RHS
 $\chi^{(A_2)}_{\text{S}}=\chi^{(A_1)}_{\text{S}}+\hat{\xi}^{3/2}\,\mathfrak{S}_{\frac{3}{2}}+\hat{\xi}^{5/2}\,\mathfrak{S}_{\frac{5}{2}}+\hat{\xi}^{3}\,\mathfrak{S}_{3}+\dots$.

\begin{table}[H]
\centering
\begin{tabular}{|   l ||  l  |  l  |}
\hline
 $h$  & bosons & fermions \\ \hline \hline
$0$   &\small $(0)^0_0$ & --- \\ \hline
$\frac 12$   & --- &  --- \\ \hline
$1$   & \small $(1)^1_0$ &  --- \\ \hline
$\frac 32$   & \small $\left( \frac32 \right)_0^{ 3/2} $ &
  $\left( \frac 12 \right)_{\pm1/2}^{ 1}$ \\ \hline
$2$ &  \small $(2)^2_0 
+ (1)^1_0 + (0)_0^2 + (0)_0^1    $  &
$(1)^{3/2}_{\pm 1/2}$ \\ \hline
$\frac 52$ &
\small $  (\frac 52)_0^{5/2} 
+  (\frac 32)_0^{5/2}
+ (\frac 32)_0^{3/2} 
+  (\frac 12)_0^{3/2}$ &
\small $ (\frac 32)_{\pm1/2}^2
+ (\frac 12)_{\pm1/2}^2 +  (\frac 12)_{\pm1/2}^1 $ \\ \hline
$3$ &
\small $
2\, (3)_0^3
+ (2)_0^2
+ (1)_0^3 +   3\, (1)_0^2 + (1)_0^1
+ 2\,(0)_0^2 + (0)_0^1 +  (0)^2_{\pm 1}  $
& 
\small $2\, (2)_{\pm1/2}^{5/2}
+ (1)_{\pm1/2}^{5/2} 
+ (1)_{\pm1/2}^{3/2}$
\\
\hline
\end{tabular}
\caption{
States in the VOA $\cW_\Gamma$ with $\Gamma = A_2$ with $h \le 3$.
The notation $(j)^\cR_r$ encodes the spin $j$ under
the $\mathfrak{sl}(2)_y$ symmetry of the small $\cN = 4$ SCA,
the outer automorphism quantum number $r$,
and the $\cR$-weight.
}
\label{su3_table}
\end{table}

\subsection{The Hall-Littlewood limit}
\label{sec:HL}

In this section we give additional evidence that the $R$-filtration,
 defined in terms of the 4d parent theory SCFT data in section~\ref{sec:R},
 and the  $\mathcal{R}$-filtration, 
 defined in terms of the free-field realization in section~\ref{sec:RR}, coincide 
 by focusing on a subsector of operators known as  Hall-Littlewood (HL) chiral ring \cite{Gadde:2011uv}.
This ring is obtained  by restricting to operators  satisfying the condition $h=R+r=\cR+r$.
As the Higgs branch chiral ring, the HL chiral ring carries the structure of a  Poisson algebra, 
see \cite{Beem:2017ooy}.
At the level of the character
and of the Macdonald index, the HL limit
 corresponds to 
\begin{align}\label{HL}
\chi_{\cW_\Gamma}^{\rm HL} (\tau,x)& = 
\lim_{q\rightarrow 0}\,\chi_{\cW_\Gamma}(q,q^{-1}\tau^2,x^2)\,, 
\nn \\
 \cI^{\text{SYM}(\mathfrak{g})}_{\rm HL} (\tau,x)
& = \lim_{q\rightarrow 0}\, \cI^{\text{SYM}(\mathfrak{g})}_{\rm Macdonald} (q,q^{-1} \tau^2,x^2) \ ,
\end{align}
where the character $\chi_{\cW_\Gamma}$ is defined in 
 \eqref{refined_vacuum_char} and the index in \eqref{g_Macdonald}. 
 
 In the following we  describe more explicitly the HL chiral ring as obtained from the 
 free-field description by applying the map  $\mathcal{P}'$ introduced in  \eqref{PprimeDEF} and  the  four dimensional HL chiral ring denoted by  $\mathscr{R}_{HL}[\text{SYM}_{\mathfrak{g}}]$. 
As a consistency check of our proposed equality 
 $\mathcal{W}_{\text{Weyl}(\mathfrak{g})}=\chi[\text{SYM}_{\mathfrak{g}}]$ and equivalence  
 of $\mathcal{R}$- and $R$-filtrations we have 
  \beq\label{threeareequal_HL}
 \mathcal{P}'(\mathcal{W}_{\Gamma})
 \,
\simeq 
\mathscr{R}_{HL}[\text{SYM}_{\mathfrak{g}}]\,,
\qquad 
\Gamma=\text{Weyl}(\mathfrak{g})\,,
 \eeq
 where  the right hand side is given in \eqref{actualRHL}.
We verify this isomorphism explicitly  for\footnote{In 
the non-crystallographic cases $p\neq 3,4,6$ the right hand side in the identity 
 \eqref{threeareequal_HL} should be replaced with $\mathscr{R}'_{\Gamma}$, see
 \eqref{actualRHL}, \eqref{MprimeANDRprime}.
} $\Gamma=A_1,I_{2}(p)$ with $p=3,4,6$.

\paragraph{The HL chiral ring from the free-field description.}
Restricting to HL operators, defined by the condition $h=\cR+r$,
 in the free-field description is straightforward.
According to the weight assignments given in 
\eqref{betagammabc_hmr}, \eqref{tableRweightsbetagammabc}
all the constituent $\{\beta_{\ell},\gamma_{\ell},b_{\ell},c_{\ell}\}$ satisfy the condition 
$h=\cR+r$, while adding a derivative will violate this condition\footnote{One
could also consider the conjugate ring consisting of operators satisfying 
$h=\cR-r$. The two rings must be isomorphic, but the description of the latter in terms of the given 
free-field realization is more complicated.  This can be seen by looking at the weight assignment
%
\begingroup
\renewcommand{\arraystretch}{1.3}
\begin{equation}
\begin{tabular}{| c  ||  c | c |c | c|c   |}
\hline
 & $\beta_\ell$ & $\gamma_\ell$ & $b_\ell$ & $c_\ell$ & $\partial$  \\
 \hline \hline
$h-(\cR+r)$ & $0$ &$0$ & $0$ & 
$0$  & $1$  
 \\
 \hline
$h-(\cR-r)$ & $0$ &  $0$ & $+1$ & 
$- 1$ & $1$  
\\ \hline
\end{tabular}
\end{equation}
\endgroup
as follows from 
\eqref{betagammabc_hmr} and  \eqref{tableRweightsbetagammabc}.
 }.
This implies that the (candidate) HL
chiral ring is obtained by setting to zero derivatives
 in the the free-field realization of the  strong generators of $\mathcal{W}_{\Gamma}$.
 This is the definition of the map 
 $\mathcal{P}'$ introduced in  \eqref{PprimeDEF} so that the candidate HL chiral ring is
  $\mathcal{P}'(\mathcal{W}_{\Gamma})$ with the Poisson ring structure defined in section~\ref{sec:freefieldsclassPOISSON}.

 Interestingly, this ring (conjecturally) admits an alternative description as the subring of 
 $\mathbb{C}[\beta,\gamma,b,c]$ annihilated by a certain set of nilpotent operators which are interpreted as the image of the screening charges of the VOA under the map $\mathcal{P}'$.
 In equation
 \beq\label{PprimeKernel}
\mathcal{P}'(\mathcal{W}_{\Gamma})=\text{Kernel}(\mathsf{J}^{(1)}_{\text{cl}})\,
\cap\dots\,
\cap\text{Kernel}(\mathsf{J}^{(\mathsf{r})}_{\text{cl}})\,,
 \eeq
 where $\mathsf{J}^{(\ell)}_{\text{cl}}$ acts by Poisson
  brackets $\{ \mathsf{J}^{(\ell)}_{\text{cl}},\cdot\}_{\text{PB}}$.
 Some examples are given in  the end of this section.
\paragraph{The HL chiral ring 
$\mathscr{R}_{HL}[\textrm{SYM}_{\mathfrak{g}}]$.}
Let us recall that the HL chiral ring is a consistent truncation of the usual $\mathcal{N}=1$ chiral ring obtained by restricting to Schur operators.
The $\mathcal{N}=1$ chiral ring  $\mathscr{R}_{\mathcal{N}=1}[\textrm{SYM}_{\mathfrak{g}}]$ of
 $\mathcal{N}=4$ SYM is given by
 \begin{equation}
 \mathscr{R}_{\mathcal{N}=1}[\textrm{SYM}_{\mathfrak{g}}]=
 \mathbb{C}[\mathscr{M}''_{\Gamma}]\,,
 \qquad \mathscr{M}''_{\Gamma}\,=\,
 \frac{\mathbb{C}^{(3|2)}
  \otimes V_{\Gamma}}{ 
 \Gamma}\,,
\eeq
with   $V_{\Gamma}\simeq \mathbb{R}^{\text{rank}(\Gamma)}$,
 generalizing 
 \eqref{MGammavarietyintro}
 and \eqref{wannabeHiggsBrChiralRing}, 
see \cite{Cachazo:2002ry}, \cite{Kinney:2005ej}.
By restricting to Schur operators we conclude that
\beq
\label{actualRHL}
\mathscr{R}_{HL}[\textrm{SYM}_{\mathfrak{g}}]=
\mathscr{R}'_{\Gamma}:=
 \mathbb{C}[\mathscr{M}'_{\Gamma}]\,,
 \eeq
 \begin{equation}
 \label{MprimeANDRprime}
 \mathscr{R}'_{\Gamma}\,:=\,
\mathbb{C}[z_1^+,\dots,z_{\mathsf{r}}^+,z_1^-,\dots,z_{\mathsf{r}}^-,
\theta_1,\dots, \theta_{\mathsf{r}}]^{\Gamma}\,,
\qquad   \mathscr{M}'_{\Gamma}\,=\,
 \frac{\mathbb{C}^{(2|1)}
  \otimes V_{\Gamma}}{ 
 \Gamma}\,.
\eeq
As $\mathscr{R}_{\Gamma}$ the Hall-Littlewood chiral ring $\mathscr{R}'_{\Gamma}$
carries the action of $GL(2)$.
This is actually extended to the action of  $SL(2|1)$.

It should be remarked that the HL chiral ring admits an alternative description which involve 
solving 
a certain BRST cohomology problem. 
This is the truncation of the BRST definition of 
 the VOA $\chi[\text{SYM}_{\mathfrak{g}}]$, see \cite{Beem:2013sza}.
It is a non-trivial fact\footnote{In
fact this is the main conjecture in
  \cite{Felder:2014}.} that this definition reproduces the HL chiral ring \eqref{actualRHL}.
Further evidence of this equivalence can be obtained by matching the corresponding Hilbert series.
The Hilbert series of \eqref{actualRHL} can be computed using the Molien formula\footnote{The
super-determinant is defined in the standard way. The basic building block is 
 \beq
 \frac{1}{\text{sdet}_{\mathbb{C}^{(2|1)}}(1-h)}\,=\,
 \frac{1-\tau^2}{(1-x \tau)(1-x^{-1} \tau)}\,.
 \eeq
 It is possible to further refine \eqref{Molien2} by keeping track of the $\mathfrak{gl}(1)_r$ 
quantum number. This is achieved by taking $h=\text{diag}(\tau x,\tau x^{-1},z\tau^2)$, where
$z$ is a  $\mathfrak{gl}(1)_r$ fugacity. This refinement of the Hilbert series cannot be obtained as a specialization of the four dimensional index.
 }
 \beq\label{Molien2}
\mathsf{HS}'_{\Gamma}(\tau,x)
\,=\,
\frac{1}{|\Gamma|}\sum_{g\in\Gamma}\,
\frac{1}{\text{sdet}_{\mathbb{C}^{(2|1)}\otimes V_{\Gamma}}(1-h\otimes g)}\,,
\qquad 
h=
\begin{psmallmatrix}
\tau x & \!0 & 0 \\
0 & \,\tau x^{-1} & 0 \\
0 &0 & \tau^2 
\end{psmallmatrix}\,.
\eeq
The BRST definition of the HL chiral ring on the other hand implies that its Hilbert series is 
 obtained by taking the 
HL limit given in \eqref{HL} of  the integral \eqref{g_Macdonald} and gives 
\beq \label{g_HL}
 \cI^{\text{SYM}(\mathfrak{g})}_{\rm HL} (\tau,x)
 = \int [du] \, {\rm P.E.} \bigg[ \left((x+x^{-1})\,\tau-\tau^2\right)\,
  \chi^{\mathfrak{g}}_{\text{Adj}}(u)  \bigg] \, .
\eeq
Since the argument of the plethystic exponential
is a Laurant polynomial, the integrand is a rational function, see \eqref{g_HLappendix}.
The equivalence of the two descriptions implies that
\beq\label{IndexisMolien2}
 \cI^{\text{SYM}(\mathfrak{g})}_{\rm HL} (\tau,x)=\mathsf{HS}'_{\Gamma}(\tau,x)\,,
 \qquad
 \Gamma=\text{Weyl}(\mathfrak{g})\,,
\eeq
See appendix \ref{app:index} for more details.

\subsubsection{Examples}
\paragraph{Rank $1$ example: $\Gamma=A_1$.} 
Let us describe the entries of \eqref{threeareequal_HL} in this example.
In this case the VOA $\mathcal{W}_{A_1}$ is isomorphic to the simple quotient of the 
small $\mathcal{N}=4$ super-Virasoro algebra at $c=-9$. Applying the map $\mathcal{P}'$
to the free-field realization of the generators of $\mathcal{W}_{A_1}$  gives 
\beq\label{generatorsHLZ2}
j^+=\beta\,,
\quad
j^0=2\beta\gamma+b c\,,
\quad
j^-=\gamma(\beta \gamma +b c)\,,
\qquad
g^+=b\,,
\quad
g^-=b\gamma\,,
\eeq
where we used the notation $\mathcal{P}'(X)=x$ when acting on the generators. 
Above  $(\beta,\gamma,b,c)$ are super-commuting variables, 
they can be thought of as coordinates of $\mathbb{C}^{2|2}$. 
It should be 
noticed that the image of $T$ and $\widetilde{G}^{\pm}$ is zero
 since derivatives are set to zero. 
 It is easy to verify that the following combinations
\beq\label{NullsA1HL}
\text{Nulls}_{A_1}=\Big{\{}
j^+j^--\tfrac{1}{4}j^0j^0\,,\,\,
j^{\pm}g^{\mp}-\tfrac{1}{2}j^0g^{\pm}
\,,\,\,g^{+}g^{-}\Big{\}}\,
\eeq
are identically zero. Recall that the relations
 $g^{\pm}g^{\pm}=0$ and $g^{+}g^{-}+g^{-}g^{+}=0$  hold by  definition in 
 a supercommutative ring.
We conclude that 
  \beq\label{HLA1case_gen_relations}
\mathcal{P}'(\mathcal{W}_{A_1})\,\simeq\, \frac{\mathbb{C}[j^+,j^0,j^-,g^+,g^-]}{\text{Nulls}_{A_1}=0}\,.
 \eeq
 The Hilbert series of this ring can be computed by standard methods, the result is given by 
 \eqref{HLA1} below.
 Next let us consider the ring \eqref{MprimeANDRprime} with $\Gamma=A_1$.
 It is rather clear the it is generated by 
 \beq\label{generatorsHLZ2_quotientVAR}
j^+=z^+z^+\,,
\quad
j^0=2 z^+z^-\,,
\quad
j^-=z^-z^-\,,
\qquad
g^+=z^+\theta\,,
\quad
g^-=z^-\theta\,.
\eeq
The character of this ring is given by the Molien formula \eqref{Molien2} and in this case gives
\beq\label{HLA1}
\mathsf{HS}'_{A_1}(\tau,x)=
\frac{1-(x+x^{-1})\tau^3-\tau^4+(x+x^{-1})\tau^5}{(1-\tau^2\,x^{-2})(1-\tau^2)(1-\tau^2\,x^{+2})}\,.
\eeq
This expression is a generalization of  \eqref{MolienZ2text}.
 Similarly to the Higgs branch chiral ring, 
 the most efficient method to show the equivalence  \eqref{threeareequal_HL}
   is to establish a relation between the $\beta\gamma b c$ coordinates and the
  quotient coordinates by equating the genetators
 \eqref{generatorsHLZ2}  with \eqref{generatorsHLZ2_quotientVAR}. 
 This gives
 \beq\label{HLcosetVFbetagammaA1}
 \beta=z^+z^+\,,
 \qquad
 \gamma-\tfrac{1}{2}\beta^{-1}bc=\frac{z^-}{z^+}\,,
 \qquad
 b=z^+\theta\,.
 \eeq
 Notice that the expressions 
 \eqref{generatorsHLZ2} are invariant under the transformation
  $\gamma \mapsto \gamma +\eta b, c\mapsto c +2\eta \beta$ where $\eta$ is a fermionic parameter.
  For this reason only the invariant combinations under this transformation, given in  \eqref{HLcosetVFbetagammaA1}, 
  can be determined. As already anticipated  in \eqref{PprimeKernel},
   $\mathcal{P}'(\mathcal{W}_{A_1})$ can be identified with the kernel of
    $\mathsf{J}_{\text{cl}}=b\beta^{-1/2}$, 
    which generates the fermionic symmetry described above, 
   in $\mathbb{C}[\beta,\gamma,b,c]$.

\paragraph{Rank $2$ example: $\Gamma=I_2(p)$.} 
Let us start from describing the generators of $\mathscr{R}'_{I_2(p)}$. They are given by $j(y), w(y)$
defined in \eqref{jwgenerators} together with their $\mathfrak{sl}(2|1)$ fermionic partners
\beq\label{ggwgenerators} 
g(y)=z_1(y)\theta_2+z_2(y)\theta_1\,,
\qquad
g_w(y)=p\,\left(z_1(y)^p \,\theta_1 + z_2(y)^p\,\theta_2\right)\,.
\eeq
These generators should be compared to the image of the $\mathcal{W}_{I_2(p)}$
 generators, namely \eqref{allVIrpartW2p} and the $\mathfrak{psl}(2|2)$ descendants of $W^{\text{h.w.}}=\beta_2$, under $\mathcal{P}'$.
 The resulting expressions are rather involved  but one can show that they coincide with 
\eqref{jwgenerators},  \eqref{ggwgenerators} upon using the identification
  \eqref{canonical_transf} and 
 \beq\label{HLcosetVFbetagammaI2p}
b_1=z_1^+\theta_2+z_2^+\theta_1\,,
 \qquad
 b_2=p\,\left((z_1^+)^{p-1}\theta_1+(z_2^+)^{p-1}\theta_2\right)\,.
\eeq
%
This concludes the proof of the isomorphism  \eqref{threeareequal_HL} for
 $\Gamma=I_2(p)$.

As in the case $\Gamma=A_1$, see \eqref{HLcosetVFbetagammaA1}, also in this case the
 expression of $\gamma_1, \gamma_2$ can be determined only up to certain nilpotent quantities.
 This is due to the fact that  $\mathcal{P}'(x)$ with $x\in \mathcal{W}_{I_2(p)}$
 are invariant under the transformations
 \begin{align}
\nonumber
 \gamma_1 \,\mapsto & \,\gamma_1+\Lambda\,\eta\, \beta_1^{p-2} b_1+\tilde{\eta}\,b_2\,,\\
\nonumber
 \gamma_2 \,\mapsto &  \,\gamma_2+\,\eta\, b_2+\tilde{\eta}\,b_1\,,\\
\nonumber
c_1 \,\mapsto &\,c_1+2\,\Lambda\,\eta\, \beta_1^{p-1}+p\,\tilde{\eta}\,\beta_2\,,  \\
\nonumber
c_2 \,\mapsto&  \,c_2+\tfrac{p}{p-1}\,\eta\,\beta_2+\tfrac{2}{p-1}\,\tilde{\eta}\,\beta_1\,,
    \end{align} 
where $\eta,\tilde{\eta}$ are fermionic parameters. 
These transformations are generated  by\footnote{Notice that $\eta$, $\tilde{\eta}$ have a fixed non-trivial dependence on
 $\beta \gamma b c$.}
  $\mathsf{J}^{\pm}_{\text{cl}}$ given in \eqref{JclfromK} 
with the suffix $\pm$ referring to the  two solutions of the differential equation \eqref{diffeqscreening}.




\section*{Acknowledgments}
The authors are grateful to Philip Argyres, Chris Beem, Mario Martone for useful conversations and correspondence.
The work of CM is supported in part by grant \#494786 from the Simons Foundation.
The work of LR is supported in part by NSF Grant PHY-1620628.
We thank the Galileo Galilei Institute for
Theoretical Physics for the hospitality and the INFN for partial support.
\vfill\eject


\appendix
\section{Some basic facts on OPEs and VOAs}
\label{OPEandsl2cov}

For the convenience of the reader, in this appendix
we collect some standard facts about OPEs
in VOAs, with emphasis on the cases with 
$\cN =2$ and small $\cN = 4$  supersymmetry.

\paragraph{Covariance under $\mathfrak{sl}(2)_z$.}
Given the operators $A(z)$, $B(z)$, we adopt the notation
\beq
A(z_1)  \, B(z_2) = \sum_{n \in \mathbb Z} \, \frac{\{ A  \, B\}_n (z_2)}{z_{12}^n} \ , \qquad
z_{12}:= z_1 - z_2 \ .
\eeq
An operator $\cO(z)$ is an $\mathfrak{sl}(2)_z$ primary
(or quasi-primary)
if it transforms tensorially under the global part of the
conformal group on the Riemann
sphere, denoted $SL(2)_z$,
\beq
\cO'(z') = \left( \frac{\partial z'}{\partial z}\right)^{-h_\cO}  \,
 \cO(z) \ , 
\eeq
where $h_\cO$ is the conformal dimension of $\cO$ and
\beq
z' = \frac{a \, z + b}{c \, z + d} \ , \qquad
\begin{pmatrix}
a & b \\
c & d
\end{pmatrix} \in SL(2)_z \ .  
\eeq
 An  $\mathfrak{sl}(2)_z$  primary operator $\cO$
satisfies
\beq
\{ T \cO \}_3 = 0 \ , \qquad
\{ T \cO \}_2 = h_\cO \, \cO \ , \qquad
\{ T \cO \}_1 = \partial \cO \ ,
\eeq
where $T$ is the stress tensor.

The OPE of $\mathfrak{sl}(2)_z$ primary operators $\cO_1$, $\cO_2$
with dimensions $h_1$, $h_2$  
is constrained by $\mathfrak{sl}(2)_z$
covariance to take the form
\beq \label{general_z_OPE}
\cO_1(z_1) \, \cO_2(z_2)
= \sum_{\cO \in B_h} 
\lambda_{\cO_1 \cO_2}{}^\cO \,
\frac{1}{z_{12}^{h_1 + h_2 - h}} \,
\cD_{h_1, h_2 ; h}(z_{12} , \partial_{z_2}) \,
\cO(z_2) \ ,
\eeq
where $B_h$ is a basis in the 
space of $\mathfrak{sl}(2)_z$ primary operators with dimension $h$,
 $\lambda_{\cO_1 \cO_2}{}^\cO$
are OPE coefficients, and the differential operator
$\cD_{h_1, h_2 ; h}(z_{12} , \partial_{z_2})$
is given by
\beq \label{z_diff_op}
\cD_{h_1, h_2 ; h}(z_{12} , \partial_{z_2})
= \sum_{k = 0}^\infty  \frac{(h + h_1 - h_2)_k}{k! \, (2h)_k} \, z_{12}^k \, \partial_{z_2}^k 
\ .
\eeq
The quantity $(x)_k$ is the ascending Pochhammer symbol,
$(x)_k = \prod_{i=0}^{k-1} (x+i)$.
 
Notice that, if $\cO_1$, $\cO_2$ are $\mathfrak{sl}(2)_z$ primaries,
the operators $\{ \cO_1 \, \cO_2 \}_n$, $n \in \mathbb Z$ are not
necessarily $\mathfrak{sl}(2)_z$ primaries.
As a consequence of \eqref{general_z_OPE},
however, $\{ \cO_1 \, \cO_2 \}_n$ is generically given
as the sum of an $\mathfrak{sl}(2)_z$ primary of weight $h_1 + h_2 -n$
and derivatives
of other $\mathfrak{sl}(2)_z$ primaries of lowest weight.
There is a standard formula for extracting
the $\mathfrak{sl}(2)_z$ primary
with  $h=h_1 + h_2 -n$ from $\{ \cO_1 \cO_2 \}_n$.
In our normalization conventions, the formula reads
\begin{align}\label{quasiPrimary_extraction}
(\cO_1 \, \cO_2)_n(z) &= \sum_{p \ge 0}
\cK_{h_1, h_2, n, p} \,
\partial^p_{z} \{ \cO_{1} \cO_{2} \}_{n+p}(z) \ , 
\end{align}
where
\begin{align} \label{K_coeffs}
\cK_{h_1, h_2, n, p} &= 
\frac{(-)^p \,(2h_1 - n - p)_p}{p! \, (2h_1 + 2h_2 - 2n - p -1)_p}  \ .
\end{align}
Our normalization is chosen in such a way that 
\beq 
\cO_1(z_1) \, \cO_2(z_2)
= \sum_{n\in \mathbb Z} 
\frac{1}{z_{12}^{h_1 + h_2 - h}} \,
\cD_{h_1, h_2 ; h}(z_{12} , \partial_{z_2}) \,
( \cO_1 \cO_2)_n(z_2) \ .
\eeq

\paragraph{Covariance under $\mathfrak{sl}(2)_y$}
The operator content of VOAs with small $\cN =4$ superconformal
symmetry falls into representations of $\mathfrak{sl}(2)$ R-symmetry.
We find it convenient to study irreps of $\mathfrak{sl}(2)$
by means of a standard index-free formalism,
based on the introduction of an auxiliary variable $y$.
The group $SL(2)$ acts on $y$ via M\"obius transformations.
We often denote this $SL(2)$ group as $SL(2)_y$,
in order to distinguish it 
from the global part of the conformal group on the Riemann
sphere, which is  denoted $SL(2)_z$.
An object $\cO(y)$ transforms in the irrep of $\mathfrak{sl}(2)_y$
with spin $j$ if it satisfies
\beq \label{sl2y_transformation}
\cO'(y') =  
\left( \frac{\partial y'}{\partial y}\right)^{j}  \, \cO(y) \ , 
\eeq
where we suppressed the $z$ dependence, and
\beq
y' = \frac{\hat a \, y + \hat b}{\hat c \, y + \hat d} \ , \qquad
\begin{pmatrix}
\hat a & \hat b \\
\hat c & \hat d
\end{pmatrix} \in SL(2)_y \ .
\eeq
As a function of $y$, $\cO$ is a polynomial of degree $2j$.

Consider any objects $\cO_1(y)$, $\cO_2(y)$
transforming under $SL(2)_y$ according to \eqref{sl2y_transformation}
with spins $j_1$, $j_2$. Let $\cB$ denote any bilinear
operation. We are mainly interested in the cases
$\cB(\cdot , \cdot ) = \{ \cdot \; \cdot \}_n$
or $\cB(\cdot , \cdot ) = ( \cdot \;  \cdot  )_n$ in a VOA,
but the following considerations also apply
if $\cB$ is, for instance, the commutative product in a polynomial
ring, or the Poisson bracket in a Poisson algebra.
It is useful to have a formula to 
decompose the product $\cB(\cO_1(y_1), \cO_2(y_2))$
into contributions of definite spin $j$.
Such a formula reads
\beq
\cB (\cO_1(y_1) , \cO_2(y_2) ) = \sum_j y_{12}^{j_1 + j_2 - j} \, \widehat \cD_{j_1, j_2 ; j}(y_{12} , \partial_{y_2}) \,
\cB(\cO_1, \cO_2)^j (y_2) \ .
\eeq
Some comments are in order. 
The range of the summation over $j$ is determined by
the usual rules for composing angular momenta,
\beq
j \in \{ |j_1 - j_2|  , |j_1 - j_2| + 1 \ , \dots , j_1 + j_2 \} \ .
\eeq
The differential operator
$\widehat \cD_{j_1, j_2 ; j}(y_{12} , \partial_{y_2})$ is given by
\beq \label{diff_oper_for_y}
\widehat \cD_{j_1, j_2 ; j}(y_{12} , \partial_{y_2})
= \sum_{k = 0}^\infty  \frac{(-j -j_1 + j_2)_k}{k! \, (-2j)_k} \, 
y_{12}^k \, \partial_{y_2}^k \ ,
\eeq
and can be thought of as the continuation of the operator
\eqref{z_diff_op} to $h = -j$. As usual, $y_{12} = y_1 - y_2$.
Notice that the sum over $k$  
always truncates to a finite sum.
The object $\cB(\cO_1, \cO_2)^j$ is the projection
onto the part with definite spin $j$. In order to define it more precisely,
we introduce the notation
\beq
\cO_1(y) = \sum_{k_1 = 0}^{2j_1} \, \cO_{1, k_1} \, y^{k_1} \ , \qquad
\cO_2(y) = \sum_{k_2 = 0}^{2j_2} \, \cO_{1, k_2} \, y^{k_2} \ .
\eeq
We may then write
\beq
\cB(\cO_1, \cO_2)^j(y_2) = \sum_{k_1=0}^{2j_1} \sum_{k_2 = 0}^{2j_2} \,
\cC_{j_1, j_2, j, k_1, k_2} \, \cB(\cO_{1, k_1} , \cO_{2, k_2}) \ ,
\eeq
where the coefficients $\cC$ are given by
\begin{align} \label{C_coeffs}
\cC_{j_1, j_2, j, k_1, k_2}  &= \sum_{r = 0}^{j_1 + j_1 -j} 
\frac{ (-)^r \, (  j_1 - j_2 + j +r )^\downarrow_r}{r! \,
(    2 j + r  + 1)_r^\downarrow} 
\, \frac{1}{(j_1 + j_2 - j-r)!} \,
\sum_{s = 0}^r \binom{p}{s}  \, (k_1)^\downarrow_{j_1 + j_2 - j - s} \, 
(k_2)^\downarrow_{s} \ .  
\end{align}
We have used the descending Pochhammer symbol,
$(x)^\downarrow_k = \prod_{i=0}^{k-1} (x-i)$.

\paragraph{A compact notation for OPEs.}
As discussed above, covariance under $\mathfrak{sl}(2)_z$
and $\mathfrak{sl}(2)_y$
completely fixes the way the $\partial_z$ and $\partial_y$
derivative of a  $\mathfrak{sl}(2)_z$,  $\mathfrak{sl}(2)_y$
primary operator enter the OPE of two
$\mathfrak{sl}(2)_z$,  $\mathfrak{sl}(2)_y$
primary operators. This allows us to use a compact notation
in which all factors $z_{12}$, $y_{12}$ and all
terms with $\partial_z$ and/or $\partial_y$
derivatives are omitted. If needed, such elements can be easily
reconstructed unambiguously 
with the formulae recorded above. 
For example, the non-trivial OPEs of the small $\cN = 4$ algebra 
at level $k$ in compact notation read
\begin{align}
J \, J & = - k \, {\rm id}  + 2 \, J \ , &
J \, G & = G \ , &
J \, \widetilde G & = \widetilde G \ , \nn \\
T \, J & = J \ , &
T \, G & = \frac 32 \, G  \ , &
T \, \widetilde G & = \frac 32 \, \widetilde G  \ , \nn \\
T \, T & =3 \, k \, {\rm id} + 2 \, T \ , &
G \, \widetilde G & = - 2 \, k \, {\rm id} + 2 \, J - T \ .
\end{align}
To make contact with the OPEs is section \ref{smallN4recap}
in the main text, one uses the parametrization
\beq
J(y) = J^+ + J^0 \, y + J^- \, y^2 \ , \qquad
G(y) = G^+ + G^- \, y \ , \qquad
\widetilde G(y) = \widetilde G^+ + \widetilde G^- \, y \ .
\eeq
In a completely analogous
fashion, the full $\cN =2$ SCA at level $k$
is encoded in the non-trivial OPEs
\begin{align}
\cJ \, \cJ & = 2 \, k \, {\rm id} \ , &
\cJ \, \cG & = - \cG \ , &
\cJ \, \widetilde \cG & = \widetilde \cG \ , \nn \\
\cT \, \cJ & = \cJ \ , &
\cT \, \cG & = \frac 32 \, \cG \ , &
\cT \, \widetilde \cG & = \frac 32 \, \widetilde \cG \ , \nn \\
\cT \, \cT & = 3 \, k \, {\rm id} + 2 \, \cT \ , &
\cG \, \widetilde \cG & =  -2 \, k \, {\rm id} + \cJ - \cT \ .
\end{align}

\paragraph{Primary conditions.}
Let us summarize 
the different notions of primary operators 
encountered in this work in the case of VOAs
with small $\cN = 4$ supersymmetry.
\begin{itemize}
\item $\mathfrak{sl}(2)_z$ primary of dimension $h$:
\beq
\{ T \, \cO(y) \}_3= 0 \ , \qquad 
\{ T \, \cO(y) \}_2= h \, \cO(y) \ , \qquad 
\{ T \, \cO(y) \}_1= \partial_z \cO(y) \ . \qquad 
\eeq
\item Virasoro primary of dimension $h$:
\beq
\{ T \, \cO(y) \}_{n \ge 3}= 0 \ , \qquad 
\{ T \, \cO(y) \}_2= h \, \cO(y) \ , \qquad 
\{ T \, \cO(y) \}_1= \partial_z \cO(y) \ . \qquad 
\eeq
\item Operator with definite $\mathfrak{sl}(2)_y$ spin $j$:
\beq
\{ J(y_1) \, \cO(y_2) \}_1 = 2 \, j \, y_{12} \, \widehat D_{1,j;j} \cO(y_2) \ .
\eeq
\item AKM primary with definite $\mathfrak{sl}(2)_y$ spin $j$:
\beq
\{ J(y_1) \, \cO(y_2) \}_{n \ge 2} =0 \ , \qquad
\{ J(y_1) \, \cO(y_2) \}_1 = 2 \, j \, y_{12} \, \widehat D_{1,j;j} \cO(y_2) \ .
\eeq
\item $\mathfrak{psl}(2|2)$ primary with quantum numbers $(h,j)$:
\begin{align}
&\{ T \, \cO(y) \}_3 = 0 \ , \qquad 
\{ T \, \cO(y) \}_2  = h \, \cO(y) \ , \qquad 
\{ T \, \cO(y) \}_1  = \partial_z \cO(y) \ , \\
& \{ J(y_1) \, \cO(y_2) \}_1  = 2 \, j \, y_{12} \, \widehat D_{1,j;j} \cO(y_2) \ ,   \\
& \{ G(y_1) \, \cO(y_2) \}_2 = 0 \ , \qquad
\{ \widetilde G(y_1) \, \cO(y_2) \}_2 = 0  \ .
\end{align}
\item small $\cN = 4$ super-Virasoro primary with quantum numbers $(h,j)$:
\begin{align}
&\{ T \, \cO(y) \}_{n \ge 3} = 0 \ , \qquad 
\{ T \, \cO(y) \}_2  = h \, \cO(y) \ , \qquad 
\{ T \, \cO(y) \}_1  = \partial_z \cO(y) \ , \\
& \{ J(y_1) \, \cO(y_2) \}_{n \ge 2} = 0 \ , \qquad
 \{ J(y_1) \, \cO(y_2) \}_1  = 2 \, j \, y_{12} \, \widehat D_{1,j;j} \cO(y_2) \ ,   \\
& \{ G(y_1) \, \cO(y_2) \}_{n \ge 2} = 0 \ , \qquad
\{ \widetilde G(y_1) \, \cO(y_2) \}_{n \ge 2} = 0  \ .
\end{align}
\end{itemize}
The differential operator $ \widehat D_{1,j;j}$ was defined in 
\eqref{diff_oper_for_y} and takes the simple form
\beq
 \widehat D_{1,j;j} (y_{12}, \partial_{y_2}) = 1
 + \frac{1}{2 \,j} \, y_{12} \, \partial_{y_2} \ .
\eeq
It is useful to notice that the AKM primary condition, combined
with the $\mathfrak{psl}(2|2)$ primary condition,
is equivalent to the super-Virasoro primary condition.
 
The analogous notions of primary operators in the case
with $\cN = 2$ supersymmetry are obtained with minimal modifications.
The R-symmetry of the small $\cN =2$ SCA is $\mathfrak{gl}(1)$,
and therefore we do not need the auxiliary variable $y$.
\begin{itemize}
\item Operator with definite $\mathfrak {gl}(1)$ charge $m$:
\beq
\{ \cJ \, \cO \}_1 = 2 \, m \, \cO \ .
\eeq
\item AKM primary with definite $\mathfrak {gl}(1)$ charge $m$:
\beq
\{ \cJ \, \cO \}_{n \ge 2} = 0 \ , \qquad
\{ \cJ \, \cO \}_1 = 2 \, m \, \cO \ .
\eeq
\item $\mathfrak{osp}(2|2)$ primary with quantum numbers $(h,m)$:
\begin{align}
&\{ T \, \cO \}_3 = 0 \ , \qquad 
\{ T \, \cO \}_2  = h \, \cO \ , \qquad 
\{ T \, \cO \}_1  = \partial_z \cO \ , \\
& \{ \cJ \, \cO \}_1  = 2 \, m \,  \cO  \ ,   \\
& \{ \cG  \, \cO  \}_2 = 0 \ , \qquad
\{ \widetilde \cG  \, \cO \}_2 = 0  \ .
\end{align}
\item $\cN = 2$ super-Virasoro primary with quantum numbers $(h,m)$:
\begin{align}
&\{ T \, \cO \}_{n \ge 3} = 0 \ , \qquad 
\{ T \, \cO \}_2  = h \, \cO \ , \qquad 
\{ T \, \cO \}_1  = \partial_z \cO \ , \\
& \{ \cJ \, \cO \}_{n \ge 2} = 0 \ , \qquad
 \{ \cJ \, \cO \}_1  = 2 \, m \,  \cO  \ ,   \\
& \{ \cG  \, \cO  \}_{n \ge 2} = 0 \ , \qquad
\{ \widetilde \cG  \, \cO \}_{n \ge 2} = 0  \ .
\end{align}
\end{itemize}
In analogy with the previous case, the AKM primary condition, combined with the
$\mathfrak{osp}(2|2)$ primary condition, is equivalent to the
$\cN = 2$ super-Virasoro primary condition.

\paragraph{Superconformal multiplets.}
The action of the fermionic generators of $\mathfrak{psl}(2|2)$
is encoded in the OPE of an operator $\cO$ with the
supersymmetry currents $G$, $\widetilde G$.
The fermionic generators of $Q$ type are encoded in the
order-one pole of the OPE, while the fermionic
generators of $S$ type are encoded in the order-two pole.
We are mainly interested in the action of generators of $Q$ type.
To describe it efficiently, we introduce the notation
\beq
G^\uparrow \cO := (G \, \cO)^{j+\frac 12}_1 \ , \quad
G^\downarrow \cO := (G \, \cO)^{j-\frac 12}_1 \ , \quad
\widetilde G^\uparrow \cO := (\widetilde G \, \cO)^{j+\frac 12}_1 \ , \quad
\widetilde G^\downarrow \cO := (\widetilde G \, \cO)^{j-\frac 12}_1 \ ,
\eeq
where $\cO$ has spin $j$.
Suppose $W$ is a $\mathfrak{psl}(2|2)$ primary
operator with $h=j$. The corresponding supersymmetry
multiplet is a short multiplet, denoted $\mathfrak S_h$.
The content of $\mathfrak S_h$ is the following
\beq
\label{Sncontent}
\begin{array}{ccc}
& W & \\
G_W := G^\downarrow W & & \widetilde G_W := \widetilde G^\downarrow W \\
& T_W := - G^\downarrow \widetilde G^\downarrow W \ .
\end{array}
\eeq
Let us now consider a $\mathfrak{psl}(2|2)$ primary operator $X$
with generic $h >j$. In this case, the relevant supersymmetry
multiplet is a long multiplet denoted $\mathfrak L_{h,j}$.
Its content can be presented in the following way,
\footnotesize
\beq \!\!\!\!\!\!\!\!\!
\begin{array}{cccccc}
& j -1  & j-  \frac 12  & j  & j + \frac12 & j + 1\\[3mm]
h &   &   & X \\
h +\frac 12 &   & G^\downarrow X \ , \  \tilde G^\downarrow X  &  & 
G^\uparrow X \ , \  \tilde G^\uparrow X  \\
h +1 & G^\downarrow \tilde G^\downarrow X &   &   
G^\downarrow   G^\uparrow X \ , \
\tilde G^\downarrow \tilde G^\uparrow X \ , \ 
G^\downarrow \tilde G^\uparrow X \ , \
\tilde G^\downarrow   G^\uparrow X
 &   & 
G^\uparrow \tilde G^\uparrow X       \\
h +\frac 32 &   &   
G^\downarrow \tilde G^\downarrow   \tilde G^\uparrow X    \ , \
\tilde G^\downarrow   G^\downarrow     G^\uparrow X
& & 
G^\uparrow \tilde G^\downarrow   \tilde G^\uparrow X    \ , \
\tilde G^\uparrow   G^\downarrow     G^\uparrow X   \\
h +2 &   &   &  G^\downarrow \tilde G^\downarrow G^\uparrow \tilde G^\uparrow X   \\
\end{array}  \nn
\eeq
\normalsize
The cases $j = 1/2$ and $j=0$ deserve special attention.
If $j=1/2$, the state $G^\downarrow \widetilde G^\downarrow X$,
which would have spin $-1/2$, is identically zero.
In the case $j=0$, all states that would have negative spin in the above table
are identically zero. Moreover, the spin-0 states 
$G^\downarrow \tilde G^\uparrow X$ and $\tilde G^\downarrow   G^\uparrow X$
become linearly dependent
because of the identity
\beq
G^\downarrow \tilde G^\uparrow X = \tilde G^\downarrow   G^\uparrow X \ .
\eeq

Similar considerations apply to the case with $\cN = 2$ supersymmetry.
In that case, we introduce the notation
\beq
\cG \cdot \cO = (\cG \cO)_1 \ , \qquad
\widetilde \cG \cdot \cO = (\widetilde \cG \cO)_1 \ .
\eeq
If we start with an $\mathfrak{osp}(2|2)$ primary $X$
with generic $h \neq m$, the relevant supersymmetry multiplet
is non-chiral and denoted $\mathfrak X_{h,m}$.
Its content is
\beq
\begin{array}{ccc}
& X & \\
\cG_X := \cG \cdot X & & \widetilde \cG_X := \widetilde \cG \cdot X \\
& \cT_X := - \cG \cdot (\widetilde \cG \cdot X) \ .
\end{array}
\eeq
In the special cases $h = \pm m$, the primary $X$ is annihilated 
by the action of $\cG$ or $\widetilde \cG$, and therefore we obtain 
a chiral or antichiral multiplet $\mathfrak C_h$, $\bar {\mathfrak C}_h$,
which only contains two states.

\section{Hilbert series and indices}

\subsection{Molien series}
\label{appMOLIEN}

Recall the definition of the plethystic logarithm applied to the Hilbert series  \eqref{Molien}:
\beq
\mathsf{PL}_{\Gamma}(\tau,x)=\sum_{k=1}^{\infty}\,\frac{\mu(k)}{k}\,
\log\left(
\mathsf{Molien}_{\Gamma}(\tau^k,x^k)
\right)\,,
\eeq
where $\mu(k)$ is the M{\"o}bius function.
It is convenient to remove the contribution of short generators from $\mathsf{PL}_{\Gamma}$
and define
\beq
\mathsf{X}_{\Gamma}(\tau,x):=\mathsf{PL}_{\Gamma}(\tau,x)-\sum_{\ell=1}^{\mathsf{r}}
\chi_{\frac{p_{\ell}}{2}}(x)\,\tau^{p_{\ell}}\,,
\eeq
where $\mathsf{r}=\text{rank}(\Gamma)$ and the degrees of the invariants $\{p_1,\dots,p_{\mathsf{r}}\}$ 
given in table~\ref{tableinvariants}
and $\chi_j$ are  $SL(2)$ characters  
\beq
\chi_j:=\chi_j(x)=\frac{x^{2j+1}-x^{-2j-1}}{x-x^{-1}}\,.
\eeq
Lets collect the $\tau$ expansion of 
$\mathsf{X}_{\Gamma}(\tau,x)$ for all Coxeter groups 
up to the first relation, {\it i.e.}, the first negative sign in the expansion in $GL(1)\times SL(2)$ characters:
\begin{subequations}
\label{ALLplethysticLOGS}
\begin{align}
A_1\,&:\qquad 
-\tau^4\,.\\
A_{N-1}\,&:\qquad 
-\chi_{\frac{N}{2}-1}\tau^{N+2}+\dots\,,\\
B_{N}/C_{N}\,&:\qquad 
-\tau^{2N+2}\,\sum_{\ell=1}^{
\left \lfloor{\frac{N+1}{2}}\right \rfloor
}\chi_{N-2\ell+1}+\dots\,,\\
& \!\!\!\!\!\!\!\!\!\!\!\!\!\!\!
\!\!\!\!\!
\!\!\!
 \begin{cases}
\,\,\,\,\,D_4\,&:\qquad 
\tau^6\,-\tau^8 \left(\chi_{2}+\chi_{1}+2\right)+\dots
\\
\,\,\,\,\,D_5\,&:\qquad 
\tau^7\chi_{1/2}\,-\tau^9 \chi_{1/2}+\dots
\\
\,\,\,\,\,D_6\,&:\qquad 
\tau^8\chi_{1}\,-\tau^{12} \left(\chi_{4}+2\chi_{2}+\chi_{1}+2\right)+\dots
\\
\,\,\,\,\,D_7\,&:\qquad 
\tau^9\chi_{3/2}+
\tau^{11}\chi_{1/2}
\,-\tau^{13} \left(\chi_{3/2}+\chi_{1/2}\right)+\dots
\\
\,\,\,\,\,D_8\,&:\qquad 
\tau^{10}\chi_{2}+
\tau^{12}(\chi_{1}+1)\,
-\tau^{14} \chi_{1}+\dots
\\
\,\,\,\,\,D_9\,&:\qquad 
\tau^{11}\chi_{5/2}+
\tau^{13}(\chi_{3/2}+\chi_{1/2})\,
-\tau^{17} \left(\chi_{5/2}+\chi_{3/2}+2\chi_{1/2}\right)+\dots
\end{cases}
\\
E_6\,&:\qquad 
\tau^{8}+
\tau^{9}\,\chi_{\frac{3}{2}}+
\tau^{12}\,\chi_{3}-
\tau^{11}\,\chi_{\frac{1}{2}}
+\dots\,,\\
E_7\,&:\qquad 
\tau^{10}\,\chi_{1}+
\tau^{12}\,\chi_{3}+
\tau^{14}\,\chi_{3}+
\tau^{16}\,\chi_{5}-
\tau^{16}\,\chi_{2}
+\dots\,,\\
E_8\,&:\qquad 
\tau^{12}+
\tau^{14}\,\chi_{3}+
\tau^{18}\,(\chi_{6}+\chi_{4}+\chi_{3})+
\tau^{20}\,(\chi_{6}+1)
-\tau^{22}\,\,\chi_{4}+\dots\\
F_4\,&:\qquad 
\tau^{8}
+\tau^{12}\,\chi_{3}
-\tau^{12}\,\chi_{2}
-\tau^{14}\,(\chi_{5}+\chi_{4}+\chi_{3}+\chi_{2}+\chi_{1})
+\dots\\
H_3\,&:\qquad 
-\tau^{12}\,(\chi_{4}+\chi_{2}+1)+\dots
\\
H_4\,&:\qquad
\tau^{12}+
\tau^{20}\,(\chi_{6}+1) 
-\tau^{24}\,(\chi_{8}+\chi_{6}+\chi_{4}+\chi_{2}+1)+\dots
\\
I_2(p)\,&:\qquad 
-\chi_{\frac{p}{2}-1}\tau^{p+2}+\dots\,,
\end{align}
\end{subequations}
The Weyl group $G_2=I_2(6)$.
 From the expressions in \eqref{ALLplethysticLOGS}
  one immediately reads off the long generators  from table~\ref{tableinvariants}
  as well as the quantum number of the lightest relation by identifying terms of the form
  $\tau^{2h}\chi_j(x)$ with quantities with quantum numbers $(h,j)$.
Notice that for $E_{6,7,8}, H_4$ there are short generators with smaller $h$-weight than the lightest relation.

The Molien series for all Coxeter groups  are recorded in an ancillary Mathematica file.
They are obtained either by direct summation over the elements of the Coxeter group\footnote{The
Weyl groups of classical types are easy to describe.
The Weyl group
 $A_{n-1}$  consists of permutations of $n$
  variables $x_1,\dots,x_n$ constrained by $\sum x_i=0$.
The Weyl group of type $B_n/C_n$
 is the semi-direct product of the symmetric groups $S_n$ with $\mathbb{Z}_2^n$.
The Weyl group of type  $D_n$ is a subgroup of type $B_n/C_n$
corresponding to the restriction to $\{\sigma_1,\dots, \sigma_n \}\in \mathbb{Z}_2^n$
such that $\prod \sigma_i=1$.
} 
or by summing over conjugacy classes and using the results of \cite{Carter1972}.
The second approach is  particularly convenient in type $E_{6,7,8}$ for which 
 the order of  the Weyl group
  $ |\Gamma|=\prod_{\ell}\,p_{\ell}$
 is very large.
 
 For the classical cases $A$ and $B/C$ the Molien series can be extracted from a simple generating function in the following way.
 Let
\begin{subequations}
\begin{align}
\mathcal{Z}_{A}(p,v_1,v_2) & :=\frac{1}{\prod_{n,m=0}^{\infty}(1-p\,v_1^nv_2^m)}\,=\,
\exp\left(\sum_{k=1}^{\infty} \frac{p^k}{k}\frac{1}{(1-v_1^k)(1-v_2^k)}\right)\,,
\\
\mathcal{Z}_{B/C}(p,v_1,v_2) & :=
\mathcal{Z}_{A}(p,v_1^2,v_2^2)
\mathcal{Z}_{A}(p\, v_1 v_2,,v_1^2,v_2^2)\,.
\end{align}
\end{subequations}
The following relations hold
 \begin{subequations}
\begin{align}
\mathcal{Z}_{A}(p,\tau x,\tau x^{-1}) &=1+z_{u(1)}\, \sum_{k=1}^{\infty}\,p^k\,
\mathsf{Molien}_{A_{k-1}}(\tau,x)\,,
\\
\mathcal{Z}_{B/C}
(p,\tau x,\tau x^{-1}) &=1+\sum_{k=1}^{\infty}\,p^k\,
\mathsf{Molien}_{B/C_{k}}(\tau,x)\,,
\end{align}
\end{subequations}
where $z_{u(1)}=(1-x^{-1}\tau)^{-1}(1-x\tau)^{-1}$ and  $\mathsf{Molien}_{A_{0}}=1$.

\paragraph{The ring $\mathscr{R}_{\Gamma}$ as a finitely generated  $\mathscr{I}_{\Gamma}$-module.}
It is interesting to observe that the Hilbert series of  $\mathscr{R}_{\Gamma}$ defined in \eqref{HS}
 can be rewritten as
\beq\label{MolienFREEmodule}
\mathsf{Molien}_{\Gamma}(\tau,x)\,=\,
\mathcal{Z}_{\text{CB}}(\tau x)\,
\mathcal{Z}_{\text{CB}}(\tau x^{-1})\,
\left(\mathsf{P}_{\Gamma}(\tau,x)+\tau^{|\Phi_{\Gamma}|}\,\mathsf{P}_{\Gamma}(\tau^{-1},x) \right)\,,
\eeq
where
\beq
\mathcal{Z}_{\text{CB}}(y)=\prod_{\ell=1}^{\mathsf{r}}\frac{1}{1-y^{p_{\ell}}}\,,
\eeq
and $|\Phi_{\Gamma}|$ is the cardinality of the root system associated to $\Gamma$.
$\mathsf{P}_{\Gamma}(\tau,x)$ is a polynomial in $\tau$ and
satisfies $2\mathsf{P}_{\Gamma}(1,1)=|\Gamma|$.
For example, for $\mathsf{r}=1$ one has  $\mathsf{P}_{A_1}(\tau,x)=1$.
In rank two
\beq
\mathsf{P}_{I_2(p)}(\tau,x)+\tau^{2p}\,\mathsf{P}_{I_2(p)}(\tau^{-1},x) \,=\,
\frac{1-\tau^{2p+2}}{1-\tau^2}\,+\,\frac{x^{p-1}-x^{1-p}}{x-x^{-1}}\,\tau^p\,.
\eeq
The presentation \eqref{MolienFREEmodule}  can be interpreted as follows.
Let
\begin{equation}
\mathscr{I}_{\Gamma}
:=\mathbb{C}[z_1^+,\dots,z_{\mathsf{r}}^+]^{\Gamma}
 \otimes \mathbb{C}[z_1^-,\dots,z_{\mathsf{r}}^-]^{\Gamma}\,=\,
\mathbb{C}[\mathcal{E}^{+}_1,\dots,\mathcal{E}^{+}_{\mathsf{r}},
\mathcal{E}^{-}_1,\dots,\mathcal{E}^{-}_{\mathsf{r}}]\,,
\end{equation}
where $\mathcal{E}^{\pm}_{\ell}$ are the $\Gamma$-invariants and are algebraically independent.
The ring of invariants \eqref{wannabeHiggsBrChiralRing} is a finitely generated free 
$\mathscr{I}_{\Gamma}$-module. The Hilbert series  \eqref{MolienFREEmodule} makes this 
fact manifest.
Let us finally observe that  
\beq
\mathsf{Molien}_{\Gamma}(\tau^{-1},x)\,=\,
\tau^{2\,\text{rank}(\Gamma)}\,\mathsf{Molien}_{\Gamma}(\tau,x)\,.
\eeq
This can be easily verified from the form \eqref{MolienFREEmodule} and \eqref{cmassagedform}.

\subsection{The index in some limits}
\label{app:index}

In this appendix we include some details about the integral representations of the 
index \eqref{g_Macdonald} in two limits.

\paragraph{Coulomb branch index.}
As a warm up lets consider a limit of the index  \eqref{g_Macdonald}  which reproduced the 
Hilbert series of $\mathbb{C}[z_1,\dots,z_{\mathsf{r}}]^{\Gamma}$.
 It is given by
\beq \label{g_CB}
 \cI^{\text{SYM}(\mathfrak{g})}_{\rm CB} (t)
 =\! \int [du] \, {\rm P.E.} \bigg[ t\,
  \chi^{\mathfrak{g}}_{\text{Adj}}(u)  \bigg]=
 \frac{1}{|\Gamma|} \frac{1}{(1-t)^{\mathsf{{\mathsf{r}}}}}\,
 \int\frac{d^{\mathsf{r}} u}{(2\pi i u)^{\mathsf{r}}}\prod_{\alpha\in \Phi_{\Gamma}}
 \left(
 \frac{1-u^{\alpha}}{1-t \,u^{\alpha}}\right)
 =\prod_{\ell=1}^{{\mathsf{r}}}\frac{1}{1-t^{p_{\ell}}}
\eeq
where ${\mathsf{r}}=\text{rank}(\Gamma)$, $\Gamma=\text{Weyl}(\mathfrak{g})$ and $p_{\ell}$ are the degrees of the invariants, see table~\ref{tableinvariants}.
The explicit evaluation of the integral \eqref{g_CB} is non-trivial.
The simplest example is 
\beq \label{g_CBsu2}
 \cI^{\text{SYM}(\mathfrak{su}(2))}_{\rm CB} (t)
 =
 \frac{1}{2} \frac{1}{(1-t)}\,
 \int_{0}^{2\pi}\frac{d\theta}{2\pi}
 \frac{(1-e^{i\theta})(1-e^{-i\theta})}{(1-t\,e^{i\theta})(1-t\,e^{-i\theta})}
 =\frac{1}{1-t^{2}}\,.
\eeq
This expression is valid for $|t|<1$.
The integral above can be computed by summing the two residues.
\paragraph{Hall-Littlewood index.}
The HL index given in \eqref{g_HL} can be massaged to the form
\beq \label{g_HLappendix}
 \cI^{\text{SYM}(\mathfrak{g})}_{\rm HL} (\tau,x)
=
 \frac{1}{|\Gamma|} \left(\frac{(1-\tau^2)}{(1-x\tau)(1-x^{-1}\tau)}\right)^{\mathsf{r}}\,
\int\frac{d^{\mathsf{r}} u}{(2\pi i u)^{\mathsf{r}}}
\prod_{\alpha\in \Phi_{\Gamma}}
 \frac{(1-u^{\alpha})(1-\tau^2 u^{\alpha})}{(1-x\tau  \,u^{\alpha})(1-x^{-1}\tau  \,u^{\alpha})}\,.
\eeq
In this example one should take $|\tau x|, |\tau x^{-1}|<1$.
 We checked in various examples that the integral \eqref{g_HLappendix} coincides with the
Molien series of $\mathscr{R}'_{\Gamma}$ defined in \eqref{Molien2}.
It is likely that the equality could be proven by showing that  
the residues in \eqref{g_HLappendix} are in one to one correspondence with elements of the Weyl group.

A trained eye might recognize that for $\mathfrak{g}=\mathfrak{su}(N)$ the index \eqref{g_HLappendix}
is related to the Nekrasov instanton partition function, 
see e.g.~equation (14) in \cite{Felder:2017}.
 Introduce $a_N$ via the generating function
\beq
\sum_{N=0}^{\infty}\,a_N(q_1,q_2)\,z^N\,=\,
\exp\left(\sum_{N=1}^{\infty}\,\frac{1-q_1^Nq_2^N}{(1-q_1^N)(1-q_2^N)}\frac{z^N}{N}\right)\,.
\eeq
The expressions \eqref{g_HLappendix}  can be integrated to 
\beq
\cI^{\text{SYM}(\mathfrak{su}(N))}_{\rm HL} (\tau,x)=
\frac{(1-x\,\tau)(1-x^{-1}\tau)}{(1-\tau^2)}\,a_N(x\,\tau,x^{-1}\,\tau)\,.
\eeq
The relative factor in this equations is interpreted as a 
$\cI^{\text{SYM}(\mathfrak{u}(1))}_{\rm HL} (\tau,x)$.
Similarly, the generating series for the $B_{\mathsf{r}}$ family reads
\beq
1+\sum_{\mathsf{r}=1}^{\infty}\,\cI^{\text{SYM}(\mathfrak{b}_r)}_{\rm HL} (\tau,x)=
\exp\left(\sum_{k=1}^{\infty}\,\frac{1+q_1^kq_2^k
-q_1^kq_2^k(q_1^k+q_2^k)}{(1-q_1^{2k})(1-q_2^{2k})}\frac{z^k}{k}\right)\,,
\eeq
with $q_1=x^{+1}\tau$, $q_2=x^{-1}\tau$.

\

\bibliographystyle{./aux/ytphys}
\bibliography{./aux/refs}

\providecommand{\href}[2]{#2}\begingroup\raggedright\begin{thebibliography}{10}

\bibitem{Beem:2013sza}
C.~Beem, M.~Lemos, P.~Liendo, W.~Peelaers, L.~Rastelli, and B.~C. van Rees,
  ``{Infinite Chiral Symmetry in Four Dimensions},''
  \href{http://dx.doi.org/10.1007/s00220-014-2272-x}{{\em Commun. Math. Phys.}
  {\bfseries 336} no.~3, (2015) 1359--1433},
\href{http://arxiv.org/abs/1312.5344}{{\ttfamily arXiv:1312.5344 [hep-th]}}.

\bibitem{Beem:2014rza}
C.~Beem, W.~Peelaers, L.~Rastelli, and B.~C. van Rees, ``{Chiral algebras of
  class S},'' \href{http://dx.doi.org/10.1007/JHEP05(2015)020}{{\em JHEP}
  {\bfseries 05} (2015) 020},
\href{http://arxiv.org/abs/1408.6522}{{\ttfamily arXiv:1408.6522 [hep-th]}}.

\bibitem{Lemos:2014lua}
M.~Lemos and W.~Peelaers, ``{Chiral Algebras for Trinion Theories},''
  \href{http://dx.doi.org/10.1007/JHEP02(2015)113}{{\em JHEP} {\bfseries 02}
  (2015) 113},
\href{http://arxiv.org/abs/1411.3252}{{\ttfamily arXiv:1411.3252 [hep-th]}}.

\bibitem{Lemos:2015orc}
M.~Lemos and P.~Liendo, ``{$\mathcal{N}=2$ central charge bounds from $2d$
  chiral algebras},'' \href{http://dx.doi.org/10.1007/JHEP04(2016)004}{{\em
  JHEP} {\bfseries 04} (2016) 004},
\href{http://arxiv.org/abs/1511.07449}{{\ttfamily arXiv:1511.07449 [hep-th]}}.

\bibitem{Cecotti:2015lab}
S.~Cecotti, J.~Song, C.~Vafa, and W.~Yan, ``{Superconformal Index, BPS
  Monodromy and Chiral Algebras},''
  \href{http://dx.doi.org/10.1007/JHEP11(2017)013}{{\em JHEP} {\bfseries 11}
  (2017) 013},
\href{http://arxiv.org/abs/1511.01516}{{\ttfamily arXiv:1511.01516 [hep-th]}}.

\bibitem{Arakawa:2016hkg}
T.~Arakawa and K.~Kawasetsu, ``{Quasi-lisse vertex algebras and modular linear
  differential equations},''
\href{http://arxiv.org/abs/1610.05865}{{\ttfamily arXiv:1610.05865 [math.QA]}}.

\bibitem{Bonetti:2016nma}
F.~Bonetti and L.~Rastelli, ``{Supersymmetric localization in AdS$_{5}$ and the
  protected chiral algebra},''
  \href{http://dx.doi.org/10.1007/JHEP08(2018)098}{{\em JHEP} {\bfseries 08}
  (2018) 098},
\href{http://arxiv.org/abs/1612.06514}{{\ttfamily arXiv:1612.06514 [hep-th]}}.

\bibitem{Song:2016yfd}
J.~Song, ``{Macdonald Index and Chiral Algebra},''
  \href{http://dx.doi.org/10.1007/JHEP08(2017)044}{{\em JHEP} {\bfseries 08}
  (2017) 044},
\href{http://arxiv.org/abs/1612.08956}{{\ttfamily arXiv:1612.08956 [hep-th]}}.

\bibitem{Fredrickson:2017yka}
L.~Fredrickson, D.~Pei, W.~Yan, and K.~Ye, ``{Argyres-Douglas Theories, Chiral
  Algebras and Wild Hitchin Characters},''
  \href{http://dx.doi.org/10.1007/JHEP01(2018)150}{{\em JHEP} {\bfseries 01}
  (2018) 150},
\href{http://arxiv.org/abs/1701.08782}{{\ttfamily arXiv:1701.08782 [hep-th]}}.

\bibitem{Cordova:2017mhb}
C.~Cordova, D.~Gaiotto, and S.-H. Shao, ``{Surface Defects and Chiral
  Algebras},'' \href{http://dx.doi.org/10.1007/JHEP05(2017)140}{{\em JHEP}
  {\bfseries 05} (2017) 140},
\href{http://arxiv.org/abs/1704.01955}{{\ttfamily arXiv:1704.01955 [hep-th]}}.

\bibitem{Song:2017oew}
J.~Song, D.~Xie, and W.~Yan, ``{Vertex operator algebras of Argyres-Douglas
  theories from M5-branes},''
  \href{http://dx.doi.org/10.1007/JHEP12(2017)123}{{\em JHEP} {\bfseries 12}
  (2017) 123},
\href{http://arxiv.org/abs/1706.01607}{{\ttfamily arXiv:1706.01607 [hep-th]}}.

\bibitem{Buican:2017fiq}
M.~Buican, Z.~Laczko, and T.~Nishinaka, ``{$ \mathcal{N} $ = 2 S-duality
  revisited},'' \href{http://dx.doi.org/10.1007/JHEP09(2017)087}{{\em JHEP}
  {\bfseries 09} (2017) 087},
\href{http://arxiv.org/abs/1706.03797}{{\ttfamily arXiv:1706.03797 [hep-th]}}.

\bibitem{Beem:2017ooy}
C.~Beem and L.~Rastelli, ``{Vertex operator algebras, Higgs branches, and
  modular differential equations},''
  \href{http://dx.doi.org/10.1007/JHEP08(2018)114}{{\em JHEP} {\bfseries 08}
  (2018) 114},
\href{http://arxiv.org/abs/1707.07679}{{\ttfamily arXiv:1707.07679 [hep-th]}}.

\bibitem{Pan:2017zie}
Y.~Pan and W.~Peelaers, ``{Chiral Algebras, Localization and Surface
  Defects},'' \href{http://dx.doi.org/10.1007/JHEP02(2018)138}{{\em JHEP}
  {\bfseries 02} (2018) 138},
\href{http://arxiv.org/abs/1710.04306}{{\ttfamily arXiv:1710.04306 [hep-th]}}.

\bibitem{Fluder:2017oxm}
M.~Fluder and J.~Song, ``{Four-dimensional Lens Space Index from
  Two-dimensional Chiral Algebra},''
  \href{http://dx.doi.org/10.1007/JHEP07(2018)073}{{\em JHEP} {\bfseries 07}
  (2018) 073},
\href{http://arxiv.org/abs/1710.06029}{{\ttfamily arXiv:1710.06029 [hep-th]}}.

\bibitem{Choi:2017nur}
J.~Choi and T.~Nishinaka, ``{On the chiral algebra of Argyres-Douglas theories
  and S-duality},'' \href{http://dx.doi.org/10.1007/JHEP04(2018)004}{{\em JHEP}
  {\bfseries 04} (2018) 004},
\href{http://arxiv.org/abs/1711.07941}{{\ttfamily arXiv:1711.07941 [hep-th]}}.

\bibitem{Arakawa:2017fdq}
T.~Arakawa, ``{Representation theory of W-algebras and Higgs branch
  conjecture},'' in {\em {International Congress of Mathematicians (ICM 2018)
  Rio de Janeiro, Brazil, August 1-9, 2018}}.
\newblock 2017.
\newblock
\href{http://arxiv.org/abs/1712.07331}{{\ttfamily arXiv:1712.07331 [math.RT]}}.
\newblock

\bibitem{Niarchos:2018mvl}
V.~Niarchos, ``{Geometry of Higgs-branch superconformal primary bundles},''
  \href{http://dx.doi.org/10.1103/PhysRevD.98.065012}{{\em Phys. Rev.}
  {\bfseries D98} no.~6, (2018) 065012},
\href{http://arxiv.org/abs/1807.04296}{{\ttfamily arXiv:1807.04296 [hep-th]}}.

\bibitem{Feigin:2018bkf}
B.~Feigin and S.~Gukov, ``{VOA[$M_4$]},''
\href{http://arxiv.org/abs/1806.02470}{{\ttfamily arXiv:1806.02470 [hep-th]}}.

\bibitem{Creutzig:2018lbc}
T.~Creutzig, ``{Logarithmic W-algebras and Argyres-Douglas theories at higher
  rank},''
\href{http://arxiv.org/abs/1809.01725}{{\ttfamily arXiv:1809.01725 [hep-th]}}.

\bibitem{Argyres:2018zay}
P.~C. Argyres, C.~Long, and M.~Martone, ``{The Singularity Structure of
  Scale-Invariant Rank-2 Coulomb Branches},''
  \href{http://dx.doi.org/10.1007/JHEP05(2018)086}{{\em JHEP} {\bfseries 05}
  (2018) 086},
\href{http://arxiv.org/abs/1801.01122}{{\ttfamily arXiv:1801.01122 [hep-th]}}.

\bibitem{Argyres:2018urp}
P.~C. Argyres and M.~Martone, ``{Scaling dimensions of Coulomb branch operators
  of 4d N=2 superconformal field theories},''
\href{http://arxiv.org/abs/1801.06554}{{\ttfamily arXiv:1801.06554 [hep-th]}}.

\bibitem{Caorsi:2018zsq}
M.~Caorsi and S.~Cecotti, ``{Geometric classification of 4d $\mathcal{N}=2$
  SCFTs},'' \href{http://dx.doi.org/10.1007/JHEP07(2018)138}{{\em JHEP}
  {\bfseries 07} (2018) 138},
\href{http://arxiv.org/abs/1801.04542}{{\ttfamily arXiv:1801.04542 [hep-th]}}.

\bibitem{Caorsi:2018ahl}
M.~Caorsi and S.~Cecotti, ``{Special Arithmetic of Flavor},''
  \href{http://dx.doi.org/10.1007/JHEP08(2018)057}{{\em JHEP} {\bfseries 08}
  (2018) 057},
\href{http://arxiv.org/abs/1803.00531}{{\ttfamily arXiv:1803.00531 [hep-th]}}.

\bibitem{Beem:2019tfp}
C.~Beem, C.~Meneghelli, and L.~Rastelli, ``{Free Field Realizations from the
  Higgs Branch},''
\href{http://arxiv.org/abs/1903.07624}{{\ttfamily arXiv:1903.07624 [hep-th]}}.

\bibitem{Adamovic:2014lra}
D.~Adamovic, ``{A realization of certain modules for the $N=4$ superconformal
  algebra and the affine Lie algebra $A_2 ^{(1)}$},''
\href{http://arxiv.org/abs/1407.1527}{{\ttfamily arXiv:1407.1527 [math.QA]}}.

\bibitem{2003math11012G}
M.~{Geck} and G.~{Malle}, ``{Reflection Groups. A Contribution to the Handbook
  of Algebra},'' {\em arXiv Mathematics e-prints} (Nov, 2003) math/0311012,
  \href{http://arxiv.org/abs/math/0311012}{{\ttfamily arXiv:math/0311012
  [math.RT]}}.

\bibitem{MR2358376}
I.~V. Dolgachev, ``Reflection groups in algebraic geometry,''
  \href{https://doi.org/10.1090/S0273-0979-07-01190-1}{{\em Bull. Amer. Math.
  Soc. (N.S.)} {\bfseries 45} no.~1, (2008) 1--60}.

\bibitem{Argyres:2018wxu}
P.~C. Argyres and M.~Martone, ``{Coulomb branches with complex
  singularities},'' \href{http://dx.doi.org/10.1007/JHEP06(2018)045}{{\em JHEP}
  {\bfseries 06} (2018) 045},
\href{http://arxiv.org/abs/1804.03152}{{\ttfamily arXiv:1804.03152 [hep-th]}}.

\bibitem{Nishinaka:2016hbw}
T.~Nishinaka and Y.~Tachikawa, ``{On 4d rank-one $ \mathcal{N}=3 $
  superconformal field theories},''
  \href{http://dx.doi.org/10.1007/JHEP09(2016)116}{{\em JHEP} {\bfseries 09}
  (2016) 116},
\href{http://arxiv.org/abs/1602.01503}{{\ttfamily arXiv:1602.01503 [hep-th]}}.

\bibitem{Lemos:2016xke}
M.~Lemos, P.~Liendo, C.~Meneghelli, and V.~Mitev, ``{Bootstrapping
  $\mathcal{N}=3$ superconformal theories},''
  \href{http://dx.doi.org/10.1007/JHEP04(2017)032}{{\em JHEP} {\bfseries 04}
  (2017) 032},
\href{http://arxiv.org/abs/1612.01536}{{\ttfamily arXiv:1612.01536 [hep-th]}}.

\bibitem{Bourton:2018jwb}
T.~Bourton, A.~Pini, and E.~Pomoni, ``{4d $\mathcal{N}=3$ indices via discrete
  gauging},''
\href{http://arxiv.org/abs/1804.05396}{{\ttfamily arXiv:1804.05396 [hep-th]}}.

\bibitem{bootstrappaper}
F.~Bonetti, C.~Meneghelli, and L.~Rastelli, ``{Bootstrapping $\mathcal{N}=4$
  VOA},'' {\em {\tt to appear}} .

\bibitem{Arakawa2012}
T.~Arakawa, ``A remark on the c 2-cofiniteness condition on vertex algebras,''
  \href{https://doi.org/10.1007/s00209-010-0812-4}{{\em Mathematische
  Zeitschrift} {\bfseries 270} no.~1, (Feb, 2012) 559--575},
  \href{http://arxiv.org/abs/1004.1492}{{\ttfamily arXiv:1004.1492 [math]}}.

\bibitem{Aharony:2015oyb}
O.~Aharony and M.~Evtikhiev, ``{On four dimensional N = 3 superconformal
  theories},'' \href{http://dx.doi.org/10.1007/JHEP04(2016)040}{{\em JHEP}
  {\bfseries 04} (2016) 040},
\href{http://arxiv.org/abs/1512.03524}{{\ttfamily arXiv:1512.03524 [hep-th]}}.

\bibitem{Argyres:2007tq}
P.~C. Argyres and J.~R. Wittig, ``{Infinite coupling duals of N=2 gauge
  theories and new rank 1 superconformal field theories},''
  \href{http://dx.doi.org/10.1088/1126-6708/2008/01/074}{{\em JHEP} {\bfseries
  01} (2008) 074},
\href{http://arxiv.org/abs/0712.2028}{{\ttfamily arXiv:0712.2028 [hep-th]}}.

\bibitem{Shapere:2008zf}
A.~D. Shapere and Y.~Tachikawa, ``{Central charges of N=2 superconformal field
  theories in four dimensions},''
  \href{http://dx.doi.org/10.1088/1126-6708/2008/09/109}{{\em JHEP} {\bfseries
  09} (2008) 109},
\href{http://arxiv.org/abs/0804.1957}{{\ttfamily arXiv:0804.1957 [hep-th]}}.

\bibitem{Garcia-Etxebarria:2015wns}
I.~Garc\'ia-Etxebarria and D.~Regalado, ``{$ \mathcal{N}=3 $ four dimensional
  field theories},'' \href{http://dx.doi.org/10.1007/JHEP03(2016)083}{{\em
  JHEP} {\bfseries 03} (2016) 083},
\href{http://arxiv.org/abs/1512.06434}{{\ttfamily arXiv:1512.06434 [hep-th]}}.

\bibitem{Aharony:2016kai}
O.~Aharony and Y.~Tachikawa, ``{S-folds and 4d N=3 superconformal field
  theories},'' \href{http://dx.doi.org/10.1007/JHEP06(2016)044}{{\em JHEP}
  {\bfseries 06} (2016) 044},
\href{http://arxiv.org/abs/1602.08638}{{\ttfamily arXiv:1602.08638 [hep-th]}}.

\bibitem{Garcia-Etxebarria:2016erx}
I.~Garc\'ia-Etxebarria and D.~Regalado, ``{Exceptional $ \mathcal{N}=3 $
  theories},'' \href{http://dx.doi.org/10.1007/JHEP12(2017)042}{{\em JHEP}
  {\bfseries 12} (2017) 042},
\href{http://arxiv.org/abs/1611.05769}{{\ttfamily arXiv:1611.05769 [hep-th]}}.

\bibitem{KAC2004400}
V.~G. Kac and M.~Wakimoto, ``Quantum reduction and representation theory of
  superconformal algebras,''
  \href{http://www.sciencedirect.com/science/article/pii/S0001870804000337}{{\em
  Advances in Mathematics} {\bfseries 185} no.~2, (2004) 400 -- 458}.

\bibitem{Thielemans:1994er}
K.~Thielemans, {\em {An Algorithmic approach to operator product expansions, W
  algebras and W strings}}.
\newblock PhD thesis, Leuven U., 1994.
\newblock
\href{http://arxiv.org/abs/hep-th/9506159}{{\ttfamily arXiv:hep-th/9506159
  [hep-th]}}.
\newblock

\bibitem{stanley1979}
R.~P. Stanley, ``Invariants of finite groups and their applications to
  combinatorics,''
  \href{https://projecteuclid.org:443/euclid.bams/1183544328}{{\em Bull. Amer.
  Math. Soc. (N.S.)} {\bfseries 1} no.~3, (05, 1979) 475--511}.

\bibitem{reflectiongroups}
G.~I. Lehrer and D.~E. Taylor, {\em Unitary Reflection Groups}.
\newblock Cambridge University Press, 2009.

\bibitem{Gadde:2011uv}
A.~Gadde, L.~Rastelli, S.~S. Razamat, and W.~Yan, ``{Gauge Theories and
  Macdonald Polynomials},''
  \href{http://dx.doi.org/10.1007/s00220-012-1607-8}{{\em Commun. Math. Phys.}
  {\bfseries 319} (2013) 147--193},
\href{http://arxiv.org/abs/1110.3740}{{\ttfamily arXiv:1110.3740 [hep-th]}}.

\bibitem{Rastelli:2014jja}
L.~Rastelli and S.~S. Razamat,
  \href{http://dx.doi.org/10.1007/978-3-319-18769-3_9}{``{The Superconformal
  Index of Theories of Class $\mathcal {S}$},''} in {\em New Dualities of
  Supersymmetric Gauge Theories}, J.~Teschner, ed., pp.~261--305.
\newblock 2016.
\newblock \href{http://arxiv.org/abs/1412.7131}{{\ttfamily arXiv:1412.7131
  [hep-th]}}.
\newblock
\url{https://inspirehep.net/record/1335343/files/arXiv:1412.7131.pdf}.
\newblock

\bibitem{Dolan:2002zh}
F.~A. Dolan and H.~Osborn, ``{On short and semi-short representations for
  four-dimensional superconformal symmetry},''
  \href{http://dx.doi.org/10.1016/S0003-4916(03)00074-5}{{\em Annals Phys.}
  {\bfseries 307} (2003) 41--89},
\href{http://arxiv.org/abs/hep-th/0209056}{{\ttfamily arXiv:hep-th/0209056
  [hep-th]}}.

\bibitem{Cachazo:2002ry}
F.~Cachazo, M.~R. Douglas, N.~Seiberg, and E.~Witten, ``{Chiral rings and
  anomalies in supersymmetric gauge theory},''
  \href{http://dx.doi.org/10.1088/1126-6708/2002/12/071}{{\em JHEP} {\bfseries
  12} (2002) 071},
\href{http://arxiv.org/abs/hep-th/0211170}{{\ttfamily arXiv:hep-th/0211170
  [hep-th]}}.

\bibitem{Kinney:2005ej}
J.~Kinney, J.~M. Maldacena, S.~Minwalla, and S.~Raju, ``{An Index for 4
  dimensional super conformal theories},''
  \href{http://dx.doi.org/10.1007/s00220-007-0258-7}{{\em Commun. Math. Phys.}
  {\bfseries 275} (2007) 209--254},
\href{http://arxiv.org/abs/hep-th/0510251}{{\ttfamily arXiv:hep-th/0510251
  [hep-th]}}.

\bibitem{Felder:2014}
Y.~Berest, G.~Felder, S.~Patotski, A.~C. Ramadoss, and T.~Willwacher,
  ``{Representation Homology, Lie Algebra Cohomology and Derived Harish-Chandra
  Homomorphism},'' \href{http://arxiv.org/abs/1410.0043}{{\ttfamily
  arXiv:1410.0043 [math]}}.

\bibitem{Carter1972}
R.~W. Carter, ``Conjugacy classes in the weyl group,''
  \href{http://eudml.org/doc/89111}{{\em Compositio Mathematica} {\bfseries 25}
  no.~1, (1972) 1--59}.

\bibitem{Felder:2017}
G.~Felder and M.~Mueller-Lennert, ``{Analyticity of Nekrasov Partition
  Functions },'' \href{http://arxiv.org/abs/1709.05232}{{\ttfamily
  arXiv:1709.05232 [math-ph]}}.

\end{thebibliography}\endgroup


\providecommand{\href}[2]{#2}\begingroup\raggedright\endgroup

\end{document}